\documentclass[a4paper,12pt]{article}

\usepackage{amssymb}
\usepackage{amsmath}

\usepackage{graphicx}
\usepackage{geometry}
\usepackage{txfonts}
\usepackage[]{hyperref}
\usepackage{fixltx2e}
\usepackage[numbers]{natbib}

\usepackage{xcolor}
\usepackage[ragged]{sidecap}
\usepackage{rotating}

\usepackage[version=4]{mhchem}

\usepackage{authblk}

\usepackage{multirow}

\def\deg{^\circ}
\def\kms{\mathrm{km\,s}^{-1}}
\def\e#1{\times 10^{#1}}

\def\msol{\mathrm{M}_\odot}
\def\lsol{L_\odot}

\def\h2{$\mathrm{H}_2$}
\def\so2{$\mathrm{SO}_2$}

\def\spy{\;\msol~\mathrm{ yr}^{-1}}
\def\um{\;\mu\mathrm{m}}
\def\arcsec{^{\prime\prime}}
\def\hc3n{HC$_3$N}

\newcommand*\edits[1]{#1}
\newcommand*\editstwo[1]{#1}

\begin{document}

\title{Chemical tracers of a highly eccentric AGB-main sequence star binary}

\date{} 

\author[1,2,3]{T.~Danilovich}
\author[1]{J.~Malfait}
\author[4]{M.~Van de Sande}
\author[5]{M.~Montarg\`es}
\author[5]{P.~Kervella}
\author[1]{F.~De~Ceuster}
\author[1]{A.~Coenegrachts}
\author[6]{T.~J.~Millar}
\author[7]{A.~M.~S.~Richards}
\author[1,8]{L.~Decin} 
\author[9]{C.~A.~Gottlieb}
\author[2,10]{C.~Pinte}
\author[11]{E.~De~Beck}
\author[2]{D.~J.~Price}
\author[12,13]{K.~T.~Wong}
\author[14,1]{J.~Bolte} 
\author[15]{K.~M.~Menten}
\author[16]{A.~Baudry}
\author[17,1]{A.~de~Koter}
\author[7]{S.~Etoka}
\author[18]{D.~Gobrecht}
\author[7,19]{M.~Gray}
\author[16]{F.~Herpin}
\author[15]{M.~Jeste}
\author[20]{E.~Lagadec}
\author[1]{S.~Maes}
\author[5,21]{I.~McDonald}
\author[16]{L.~Marinho}
\author[22]{H.~S.~P.~M\"uller}
\author[7,19]{B.~Pimpanuwat}
\author[8]{J.~M.~C.~Plane}
\author[23]{R.~Sahai}
\author[1]{S.~H.~J.~Wallstr\"om}
\author[24]{J.~Yates}
\author[7]{A.~Zijlstra}

\affil[1]{Institute of Astronomy, KU Leuven, Celestijnenlaan 200D,  3001 Leuven, Belgium}
\affil[2]{School of Physics \& Astronomy, Monash University, Wellington Road, Clayton 3800, Victoria, Australia}
\affil[3]{ARC Centre of Excellence for All Sky Astrophysics in 3 Dimensions (ASTRO 3D), Clayton 3800, Australia}
\affil[4]{School of Physics and Astronomy, University of Leeds, Leeds LS2 9JT, UK}
\affil[5]{LESIA, Observatoire de Paris, Universit\'e PSL, CNRS, Sorbonne Universit\'e, Universit\'e Paris Cit\'e, 5 place Jules Janssen, 92195 Meudon, France}
\affil[6]{Astrophysics Research Centre, School of Mathematics and Physics, Queen's University Belfast, University Road, Belfast BT7 1NN, UK}
\affil[7]{JBCA, Department Physics and Astronomy, University of Manchester, Manchester M13 9PL, UK}
\affil[8]{School of Chemistry, University of Leeds, Leeds LS2 9JT, UK}
\affil[9]{Harvard-Smithsonian Center for Astrophysics, 60 Garden Street, Cambridge, MA 02138, USA}
\affil[10]{Univ. Grenoble Alpes, CNRS, IPAG, Grenoble, France}
\affil[11]{Department of Space, Earth and Environment, Chalmers University of Technology, Onsala Space Observatory, 43992 Onsala, Sweden}
\affil[12]{Theoretical Astrophysics, Department of Physics and Astronomy, Uppsala University, Box 516, SE-751 20 Uppsala, Sweden} 
\affil[13]{Institut de Radioastronomie Millim\'etrique, 300 rue de la Piscine, F-38406 Saint-Martin-d'H\`eres, France}
\affil[14]{Department of Mathematics, Kiel University, Heinrich-Hecht-Platz 6, 24118 Kiel, Germany}
\affil[15]{Max-Planck-Institut f{\" ur} Radioastronomie, Auf dem H{\" u}gel 69, 53121 Bonn, Germany}
\affil[16]{Universit\'e de Bordeaux, Laboratoire d'Astrophysique de Bordeaux, 33615 Pessac, France}
\affil[17]{University of Amsterdam, Anton Pannekoek Institute for Astronomy, 1090 GE Amsterdam, The Netherlands}
\affil[18]{Department of Chemistry and Molecular Biology, University of Gothenburg, Medicinaregatan 7 B 41390 Gothenburg, Sweden} 
\affil[19]{National Astronomical Research Institute of Thailand, Chiangmai 50180, Thailand}
\affil[20]{Universit\'e C\^ote d'Azur, Laboratoire Lagrange, Observatoire de la C\^ote d'Azur, F-06304 Nice Cedex 4, France}
\affil[21]{School of Physical Sciences, The Open University, Walton Hall, Milton Keynes, MK7 6AA, UK}
\affil[22]{Universit\"at zu K\"oln, I. Physikalisches Institut, 50937 K\"oln, Germany}
\affil[23]{California Institute of Technology, Jet Propulsion Laboratory, Pasadena CA 91109, USA}
\affil[24]{University College London, Department of Computer Science, London WC1E 6BT, United Kingdom}

\maketitle

\section*{Abstract} 

Binary interactions have been proposed to explain a variety of circumstellar structures seen around evolved stars, including asymptotic giant branch (AGB) stars and planetary nebulae. Studies resolving the circumstellar envelopes of AGB stars have revealed spirals, discs and bipolar outflows, with shaping attributed to interactions with a companion. For the first time, we have used a combined chemical and dynamical analysis to reveal a highly eccentric and long-period orbit for W~Aquilae, a binary system containing an AGB star and a main sequence companion. Our results are based on anisotropic SiN emission, the first detections of NS and SiC towards an S-type star, and density structures observed in the CO emission. These features are all interpreted as having formed during periastron interactions. Our astrochemistry-based method can yield stringent constraints on the orbital parameters of long-period binaries containing AGB stars, and will be applicable to other systems.


\section*{Main} 


The asymptotic giant branch (AGB) is a late evolutionary stage of low and intermediate mass stars ($\sim 1$ to 8 solar masses, $\msol$). This stage is characterised by mass-losing stellar winds, rich in molecular gas and dust, which form an extended, expanding circumstellar envelope (CSE) around the star \cite{Hofner2018}. AGB stars eventually transition through the planetary nebula phase and end as white dwarf stars, having chemically enriched their host galaxies through their mass loss \cite{Kobayashi2020}. Binary companions can have a significant impact on this process, potentially affecting mass-loss rates and chemistry \cite{Decin2019,Van-de-Sande2022}, and are thought to shape both the eventual planetary nebula \cite{De-Marco2022} and the CSE during the AGB phase \cite{Decin2020}. 
Binary stars with an AGB component are also the progenitors of various exotic objects, including Barium stars, CH stars, extrinsic S-stars, and novae \cite{Jorissen2004}. 
Hence, understanding binary systems containing AGB stars, especially through observations, is important for understanding their overall evolutionary progress, the initial-final mass relation, and the evolution of their host galaxies.


Recent observations of some AGB stars have identified the signatures of binary companions imprinted in the structure of the CSE. In only a few cases, however, is the precise nature of the companion and its effects on the CSE known, thereby limiting the study of such systems. 
Systems with directly detected companions include Mira, which comprises an oxygen-rich AGB star and a white dwarf, in which the companion has contributed to the shaping of the CSE structure \cite{Karovska1997,Ramstedt2014a}, and L$_2$~Pup, an oxygen-rich AGB star surrounded by a disc with a planetary companion \cite{Kervella2016}.
Bipolar structures around $\pi^1$~Gru have been attributed to a recently detected close companion \cite{Homan2020}, adding to the small number of AGB systems with directly detected companions.
The spiral structures observed around the carbon stars AFGL 3068 and R~Scl \cite{Mauron2006,Kim2017,Maercker2012}, and the bipolar structures around the carbon star V~Hya \cite{Sahai2022}, indicate the possible presence of binary companions that have not been directly detected. 
A more complete understanding of circumstellar structures will come from knowing both cause (e.g. a stellar or planetary companion) and effect (the CSE structure) and should allow us to draw more direct links between AGB stars and planetary nebulae, which have been observed to display a multitude of complex asymmetric structures \cite{Ramos-Larios2016,Decin2020}.

W~Aquilae (W~Aql) is a binary system at a distance of 395~pc (Methods \ref{dist-notes}). It contains an S-type AGB star, which has a mixed carbon-oxygen chemistry (C/O $\sim 1$) \edits{and may be transitioning from being oxygen-rich to carbon-rich}, and an F9 main sequence star \cite{Herbig1965,Danilovich2015} located to the southwest of the AGB star at a projected separation of $\sim 0.5\arcsec$ \cite{Ramstedt2011}.
W~Aql has been extensively studied through observations taken with a variety of telescopes \cite{Ramstedt2011,Mayer2013,Danilovich2014,Ramstedt2017,Brunner2018,Danilovich2021}. Spatially resolved observations of the polarised dust \cite{Ramstedt2011} and CO \cite{Ramstedt2017} around the AGB star have shown a large-scale asymmetry in the direction of the F9 companion, a sign that binary interactions may be shaping the CSE. However, the asymmetry exists at larger scales than the present separation of the two stars, from $\sim 10\arcsec$ to $\sim100\arcsec$ \cite{Ramstedt2011,Ramstedt2017,Mayer2013}. Some indications of spiral structure in the CSE were seen in observations taken by the Atacama Large Millimetre/submillimetre Array (ALMA) at a resolution of $\sim 0.4\arcsec$ \cite{Ramstedt2017} but it was unclear whether these could be caused by the F9 star.




\section{Results} 

We have analysed new, high resolution ALMA observations of the W~Aql system with spatial resolutions to $\sim 0.024\arcsec$, i.e. approximately twice the K-band stellar diameter \cite{Richichi2005}, and 40\% larger than the millimetre stellar diameter (Methods \ref{Udisc}). We combined these with photometric observations, new smooth particle hydrodynamics models, and chemical kinetics models to put new constraints on the orbit of the system. 
\editstwo{We have shown that all the observations are consistent with the hypothesis of a highly eccentric orbit, based primarily on the distributions of molecular species which formed during periastron passage and the structures seen in the CO observations, making such an interpretation highly probable.}

\subsection{Species formed during periastron passage} 

\begin{figure}[t]
\begin{center}
\includegraphics[width=0.49\textwidth]{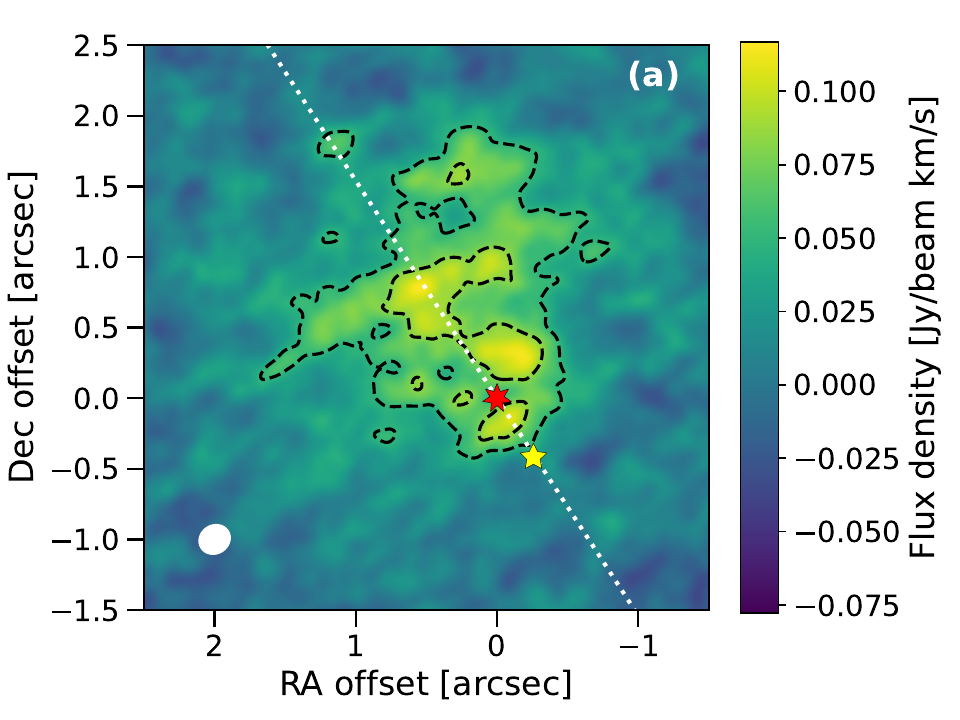}
\includegraphics[width=0.49\textwidth]{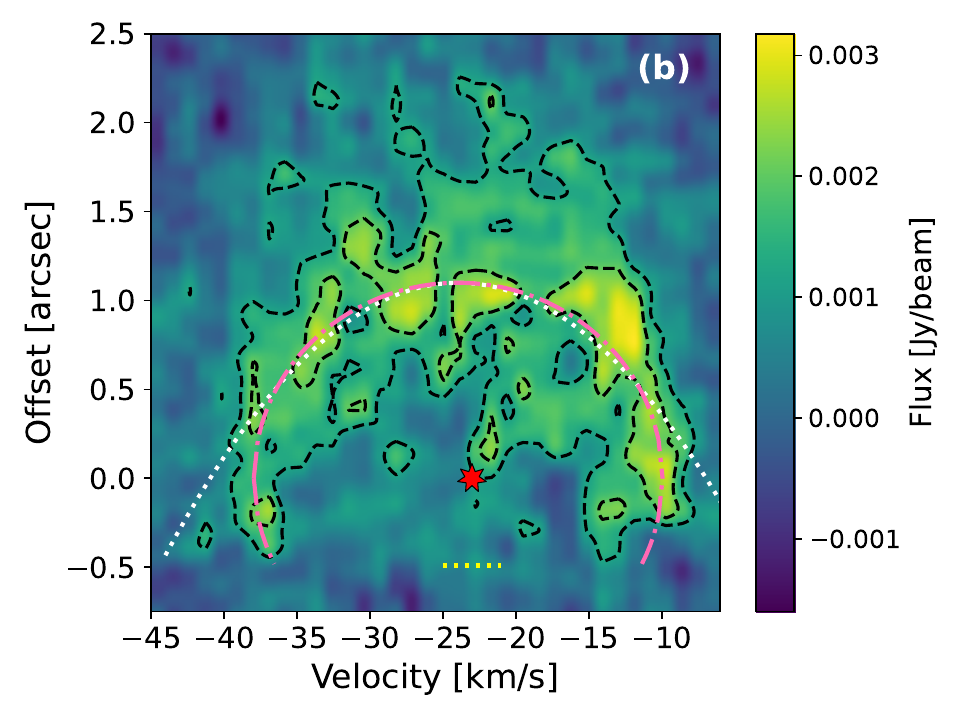}
\caption{\textbf{(a)} Zeroth moment map of SiN ($N,J=6,{13/2}\to5,{11/2}$) towards W~Aql with contours at levels of 3 and $5\sigma$. The position of the AGB star is indicated by the red star at (0,0) and the current location of the F9 star is indicated by the yellow star to the southwest. North is up and east is to the left. The dotted white line indicates the axis used for the PV diagram in (b) and the white ellipse in the bottom left corner indicates the size of the synthesised beam. \textbf{(b)} Position-velocity diagram of SiN towards W~Aql, taken at a position angle of north $33\deg$ east, as indicated by the dotted white line in (a). Dashed black contours are at levels of 3 and $5\sigma$, a dotted white parabola is fit to the data (see Methods \ref{sinsic}), and a dash-dotted pink ellipse is plotted to emphasise the shape of the emission in the PV diagram. The position and LSR velocity ($\upsilon_\mathrm{LSR} = -23~\kms$) of the AGB star is indicated by the red star, and the horizontal yellow dotted line indicates the present offset of the F9 star.}
\label{sin}
\end{center}
\end{figure}

From a detailed examination of the ALMA data (Methods \ref{almadata}), we identified several molecules exhibiting spatially asymmetric emission. Most notable was SiN, which has only been detected towards one other AGB CSE \cite{Turner1992}. In Fig. \ref{sin}a we plot a zeroth moment (integrated intensity) map of SiN, which shows emission in a roughly triangular wedge mainly to the northeast of the AGB star. To further understand the spatial origin of the emission, we constructed a position-velocity diagram (Fig. \ref{sin}b), which reveals an arc of SiN emission that lies side-on ($90\deg$), \edits{i.e. perpendicular} to the plane of the sky (Methods \ref{sinsic}).

The absence of (approximate) spherical symmetry in the emission suggests a spatial and/or temporal dependence for the formation of SiN around W~Aql. Chemical kinetics models indicate that the production of SiN is higher in the presence of UV photons --- such as can be provided by a main sequence companion \cite{Van-de-Sande2022} like the F9 component of W~Aql, but only in sufficiently dense regions of the CSE (see Methods \ref{chemmethods} for further details of the chemistry initiated by the companion's UV field). 
We posit that: (1) the binary orbit is highly eccentric and inclined $i\sim90\deg$; (2) the formation of the arc of SiN was triggered close to periastron (Figs. \ref{orb-phot} and \ref{SiNsketch}), when the F9 star passed close to the AGB star and irradiated part of the dense inner AGB wind; and (3) this temporarily drove chemical reactions through increased \editstwo{(but not complete)} photodissociation and photoionisation, including those reactions which led to the formation of SiN (Methods \ref{chem:sin}).
We used radiative transfer modelling to estimate the abundance of SiN in the arc and found a peak abundance of $1.5\e{-7}$ relative to \h2 (Methods \ref{rtmod}), which is in general agreement with the expectations from chemical models containing an F9-like companion (Methods \ref{chem:sin}). 
Further evidence in support of this formation mechanism is provided by the presence of SiC and NS emission towards W~Aql. These are the first detections of SiC and NS towards an S-type AGB star and their emission is also asymmetric (with a weaker signal to noise ratio (SNR) than SiN; see Methods \ref{sinsic} and \ref{NSdet}, and Figs \ref{sic} and \ref{nsplot} in the Extended Data and Supplementary Materials). The presence of SiC and NS is consistent with chemical model predictions \cite{Van-de-Sande2022} for the effect of the periastron passage of the F9 star on the chemistry of the CSE (Methods \ref{chem:sin} and \ref{nschem}).


\subsection{Photodissociation of common species} 

Farther from the AGB star, such as where the F9 star is presently located, the wind is less dense ($\sim3\e{5}$~cm$^{-3}$ compared with $\sim10^{9}$~cm$^{-3}$ at 10~au from the AGB star) and the chemistry tends to be initiated by photodissociation by the interstellar radiation field. The density in this region is too low for species like SiN to form, however, we see evidence of the F9 star driving additional photodissociation in the zeroth moment maps of SiO, SiS, CS and HCN (Fig.~\ref{othermol} in the Extended Data), all of which show extended emission to the northeast and truncated emission to the southwest, in the direction of the present position of the F9 star. \edits{The central channels of SiS and CS, in particular, show significantly lower molecular emission around the F9 star (Fig.~\ref{cssissinglechans} in the Supplementary Materials).} Spectra centred on the current position of the F9 star show very few detected molecular lines and the line profiles of CS, SiO and HCN show less emission around the LSR velocity compared with spectra centred on the AGB star or at the same distance from the AGB star but on the opposite side of the CSE (see Methods \ref{sec:F9spec} and Fig. \ref{F9spec} in the Extended Data). 

\begin{figure}[t]
\begin{center}
\includegraphics[height=6cm]{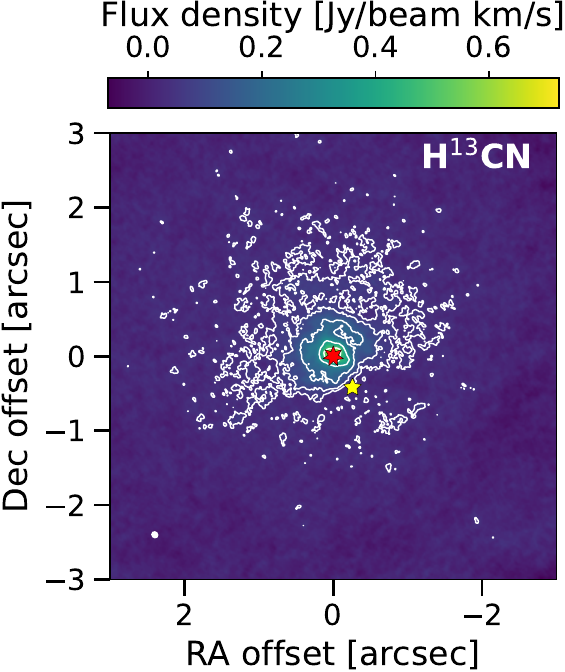}
\includegraphics[height=6cm]{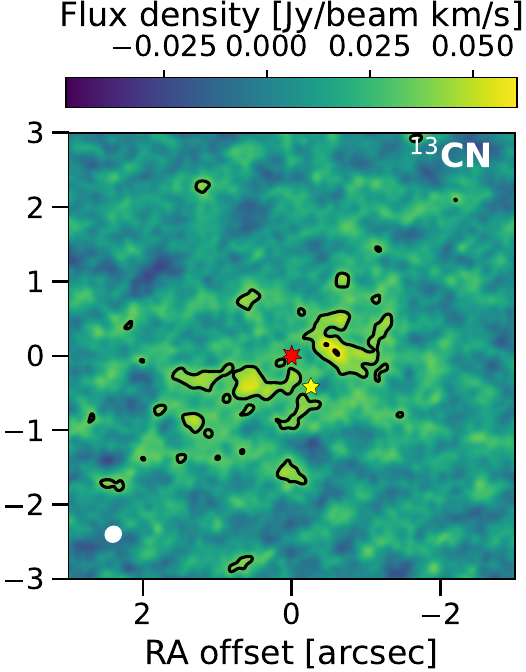}
\includegraphics[height=6cm]{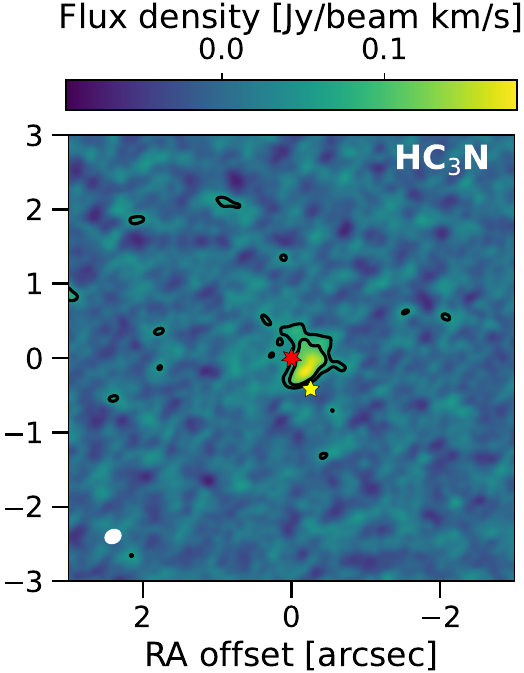}
\caption{Zeroth moment maps of H$^{13}$CN (left), $^{13}$CN (centre), and HC$_3$N ($J=27\to26$, right) towards W Aql. Full transition details are given in Table \ref{resolutions}. Contours are at levels of 3 and 5$\sigma$, and additionally 10, 20, and 30$\sigma$ for H$^{13}$CN. The position of the AGB star is indicated by the red star at (0,0) and the location of the F9 companion is indicated by the yellow star to the southwest. North is up and east is left. The white ellipses in the bottom left corners indicate the sizes of the synthesised beams.}
\label{fig:cn}
\end{center}
\end{figure}

Additional evidence of the F9 star driving photodissociation is found by comparing the distribution of \ce{H^13CN} with the distribution of \ce{^13CN} (note, \ce{^12CN} was not covered by our observations), because CN is a photodissociation product of HCN \cite{Huggins1984}. As shown in Fig.~\ref{fig:cn}, \ce{^13CN} is found to be present mainly in the region in which the \ce{H^13CN} emission is truncated. This is consistent with the F9 star driving the photodissociation of \ce{H^13CN} and hence creating \ce{^13CN}. We also plot the zeroth moment map of the $J=27-26$ transition of HCCCN (the next member in the cyanopolyyne family, hereafter \ce{HC3N}, see Fig.~\ref{fig:cn}), which shows emission on the same side of the AGB star as \ce{^13CN}, albeit over a much smaller region. The other observed transitions of \ce{HC3N} show a similar distribution (Fig.~\ref{hc3n} in the Supplementary Materials). Because \ce{HC3N} forms from CN (Methods \ref{HCNchem}), its asymmetric distribution indicates an asymmetric CN distribution and hence provides further evidence of anisotropic photo-processes in the CSE.

\subsection{Structures in CO emission} 

\begin{figure}[h!p]
\begin{center}
\includegraphics[width=0.49\textwidth]{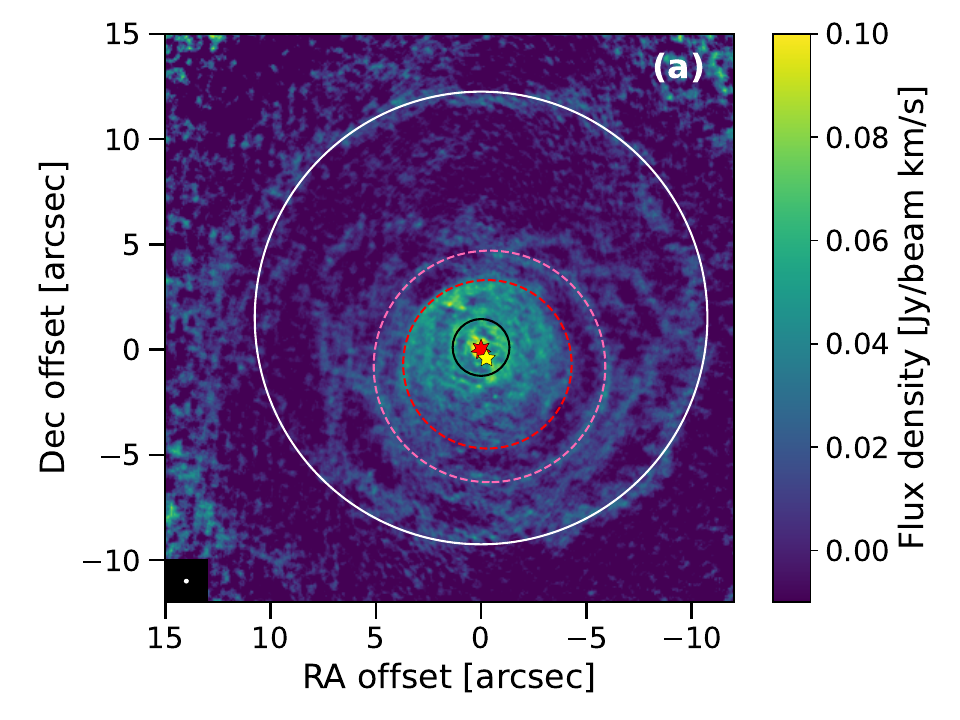}
\includegraphics[width=0.49\textwidth]{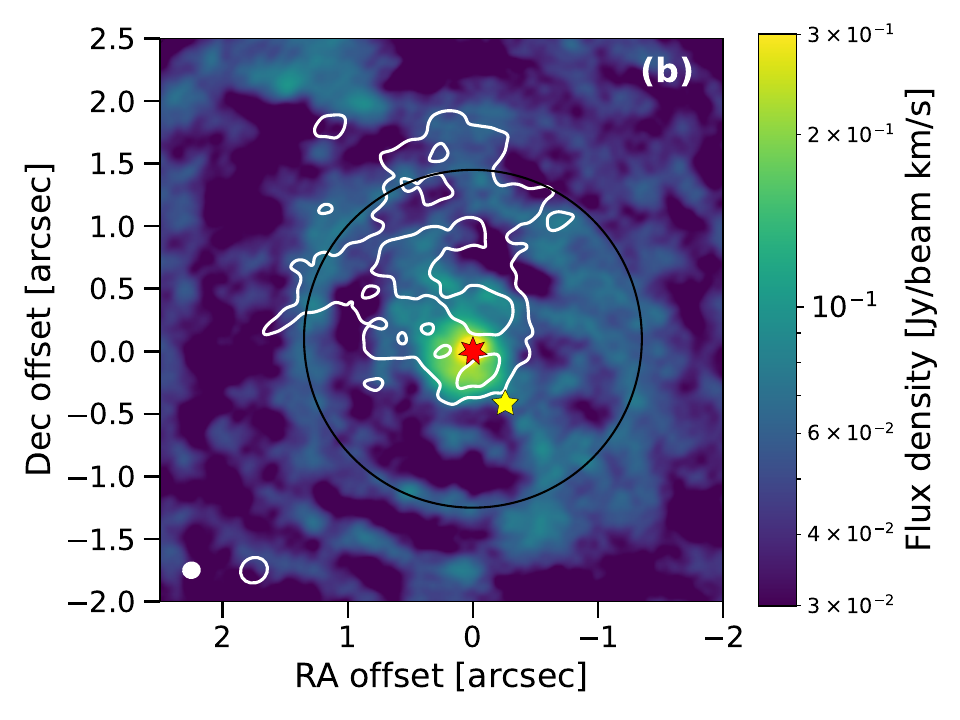}
%
\includegraphics[height=5.0cm]{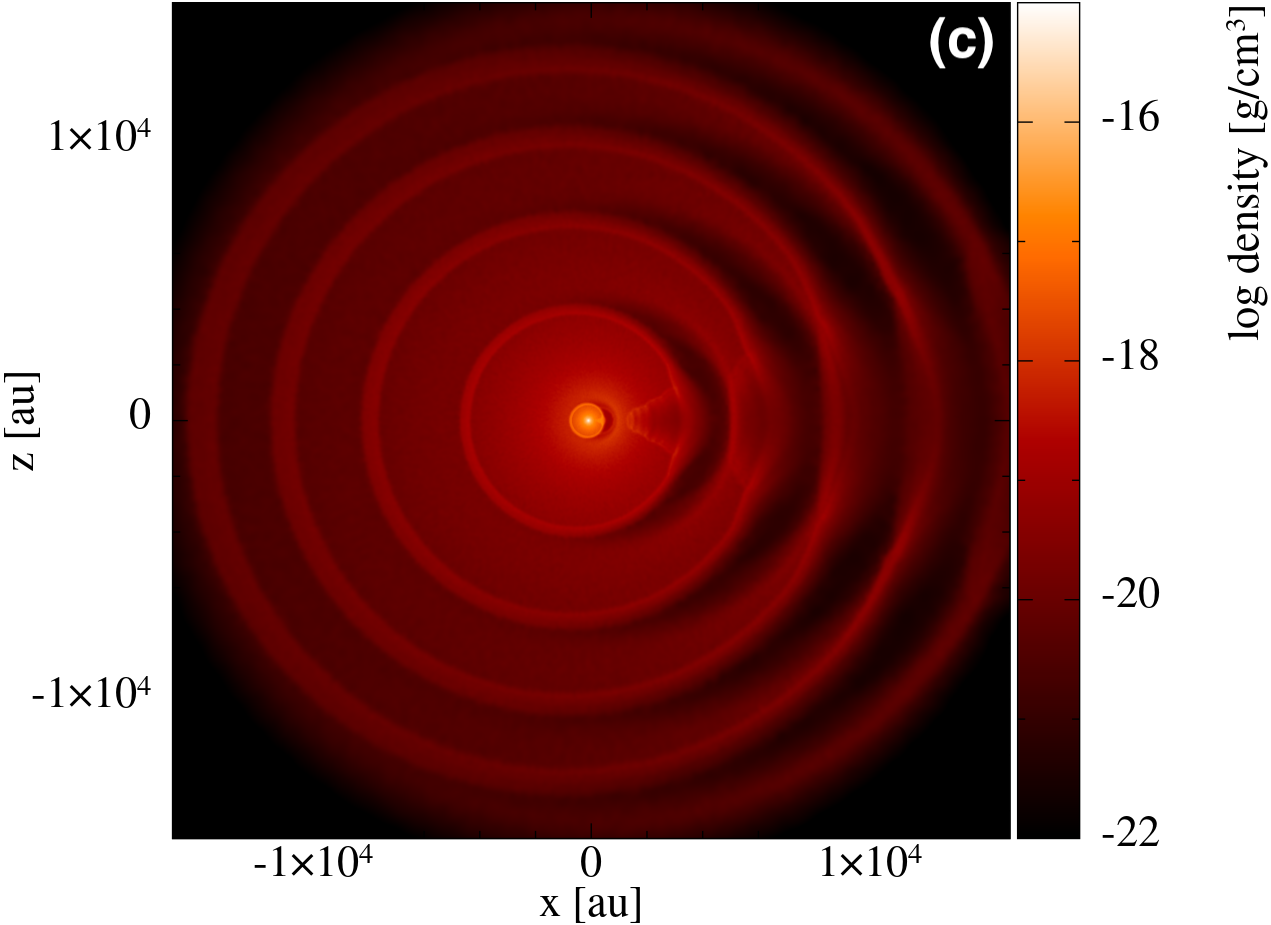}
\includegraphics[width=0.49\textwidth]{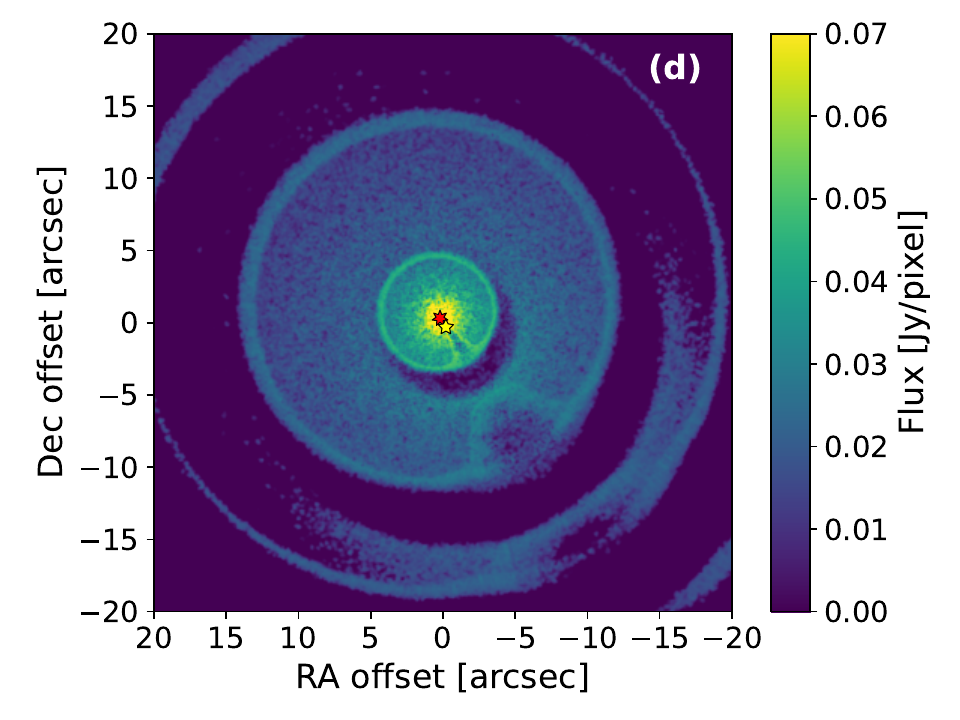}
\caption{\textbf{(a)} A plot of the three central CO channels observed with ALMA summed together (see channels highlighted in Fig.~\ref{cochan}). We include circles (white, black, red, pink)
to guide the eye to structures in the emission. The location of the AGB star is shown as a red star and the present location of the companion is shown as a yellow star. North is up and east is left. The synthetic beam size is shown as a white ellipse inside a black square in the bottom left corner. 
\textbf{(b)} Same as the central part of (a), including the black circle, but plotted on a logarithmic colour scale to emphasise structure. The white contours are the SiN zeroth moment map as shown in Fig. \ref{sin}(a). The filled ellipse in the bottom left corner shows the synthetic beam for the CO data, while the unfilled ellipse is the synthetic beam of the SiN data. 
\textbf{(c)} Density distribution in a 2D slice through a plane perpendicular to the orbital plane ($y=0$), similar to the edge-on orientation of the W~Aql system, from a 3D SPH model with masses $M_\mathrm{AGB}=1.6~\msol$ and $M_\mathrm{2}=1.06~\msol$, eccentricity $e=0.92$, and semimajor axis $a=125$~au. The barycentre of the system is located at 0,0 and at the scale plotted ($1\times10^4$~au $\approx 25\arcsec$) the AGB and F9 stars cannot be distinguished. See Methods \ref{hydro} for more details. \textbf{(d)} The central channel of (c) after processing with a radiative transfer model to convert the model density to CO ($2\to1$) intensity, taking photodissociation into account (see Methods \ref{hydro} for details). Star positions are taken from the model in (c).
\label{coarcs}}
\end{center}
\end{figure}

CO is an abundant stable molecule, commonly used as a density tracer in CSEs. We plot high resolution ($0.132\arcsec\times0.123\arcsec$) channel maps of CO emission in Fig.~\ref{cochan} in the Extended Data and first focus on the central three channels closest to the AGB stellar velocity $\upsilon_\text{LSR}=-23~\kms$ (Fig.~\ref{coarcs}a).
With the aid of angle-radius plots (Fig. \ref{angrad}), we identified two key circular structures in the CO emission, with radii of $1.35\arcsec$ and $10.75\arcsec$, with centres offset from the present position of the AGB star by $0.1\arcsec$ and $1.5\arcsec$ to the north. These are shown in black and white in Fig. \ref{coarcs}a. Other circular structures are highlighted in red and pink and, because these are offset to the southwest, we presume they were formed through different processes to the black and white circles and focus on the latter first.

To better understand the origin of the circular structures, we performed hydrodynamic simulations for highly eccentric systems based on the W~Aql system (details in Methods \ref{hydro}). From these we found that highly elliptical orbits ($e \gtrsim 0.8$) result in almost spherical structures in the wind, which appear circular and slightly offset away from the present position of the companion when viewed edge-on ($i=90\deg$) relative to the plane of the orbit (Fig. \ref{coarcs}c). These structures are generated during periastron passages and are very similar to the black and white circles seen in the ALMA CO data, even more so when the hydrodynamical model is processed with a radiative transfer code (Fig. \ref{coarcs}d). The fact that the \edits{outer} edge of the SiN emission overlaps with the inner circular structure (Fig. \ref{coarcs}b) also suggests they were formed contemporaneously, i.e. during the most recent periastron passage. We also determined that the different emission distributions seen in blue (elongated) and red (circular) channels of our ALMA observations are reproduced in the hydrodynamic model (Fig.~\ref{otherCOchans}).
Based on all of these results, we can constrain the orbital parameters of the W~Aql system.

\subsection{Orbital parameters} 

\begin{figure}[t]
\begin{center}
\includegraphics[height=4.8cm]{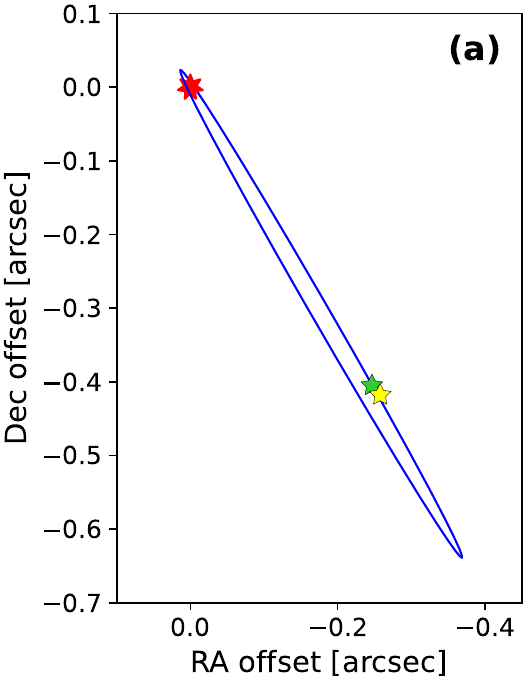}
\includegraphics[height=4.8cm]{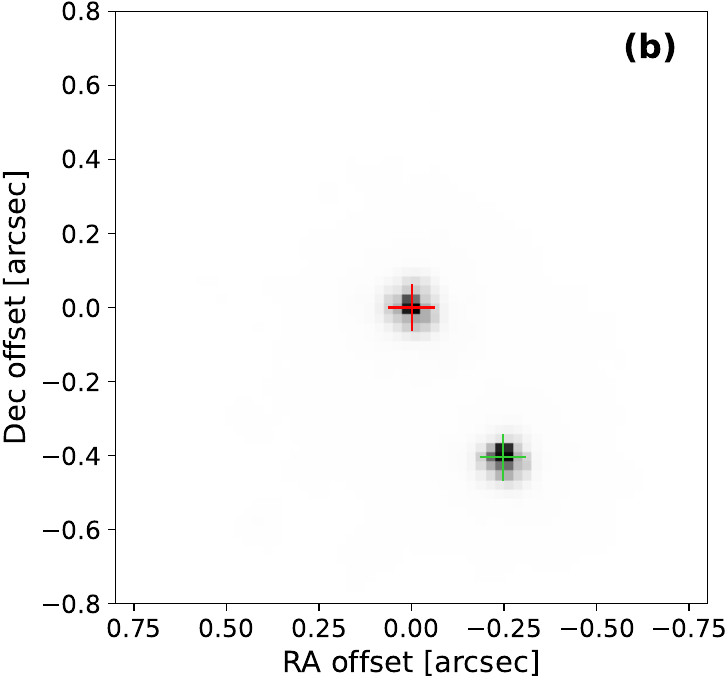}
\includegraphics[height=4.8cm]{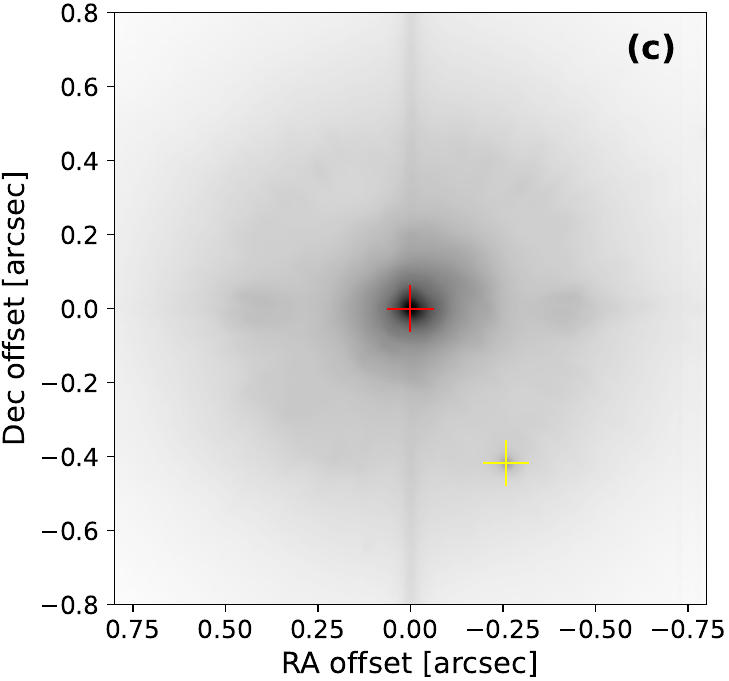}
\caption{Plots of the orbit of the W~Aql system as seen in the plane of the sky. In all panels, north is up and east is left. The orbital parameters shown are for eccentricity $e=0.93$ and periastron separation $r_p= 1.5\e{14}$~cm. Although we find the inclination to be $i=90\pm7\deg$, we plot the orbit with $i=85\deg$ so that it is possible to see the ellipse.  \textbf{(a)} Plot of the orbit of the W~Aql system in the frame of the AGB star. The location of the AGB star is shown as a red star at (0,0), the position of the F9 star from the SPHERE observation is shown as a yellow star and from the HST observation as a green star. \textbf{(b)} B-band image of W~Aql observed with HST in 2004 \cite{Ramstedt2011}, plotted with a linear intensity scale. The measured centres of the AGB and F9 stars are indicated with the red and green crosses, respectively. \textbf{(c)} VLT/SPHERE-ZIMPOL image of W~Aql in the VBB filter, observed in 2019 \cite{Montarges2023}, plotted with a logarithmic intensity scale. The measured centres of the AGB and F9 stars are indicated with the red and yellow crosses, respectively.}
\label{orb-phot}
\end{center}
\end{figure}

From the circular structures seen in Fig.~\ref{coarcs}, we estimate the orbital period to be $1082^{+89}_{-108}$ years. Based on the expansion time of the inner circle and the arc of SiN, we estimate the time since the most recent periastron to be $172\pm22$ years (Methods \ref{orbit-from-alma}). The SiN PV diagram indicates an orbital inclination of $i = 90\pm7\deg$ (Methods \ref{sinsic}). Combining these results with resolved images of W~Aql (Methods \ref{photmethods}), we found a series of numerical solutions that reproduce the observations within their uncertainties (Methods \ref{sec:solution}). All our solutions (Table \ref{tab:solutions} in the Extended Data) have high eccentricities ($e>0.9$) and small periastron distances ($r_p \leq 2\e{14}$~cm~$=13$~au), with long periods $\sim 1100$~years. A solution with $e=0.93$, $r_p = 1.5\e{14}$~cm ($10$~au) and period 1051~years is plotted in Fig.~\ref{orb-phot}, where it is superposed on resolved images to show the agreement with the positions of the stars.

\section{Discussion} 

For the first time, we have identified in observations, with the aid of astrochemistry, molecular species that formed during a periastron passage of an AGB + main sequence (F9) binary system. Through our analysis of these species, in combination with structures in the CO and resolved images of the two stars, we were able to constrain the binary orbit to a limited number of solutions, all having high eccentricities and almost edge-on inclinations. Our analysis opens up a new method for studying binary systems containing AGB stars by observing spatially resolved emission of key molecular species.

SiN was crucial to our analysis because it is distributed asymmetrically in the W~Aql CSE --- alerting us to a non-standard formation pathway --- and was detected with a sufficiently high SNR to be readily analysed. The other two molecules that we identified as being created during periastron, SiC and NS, strengthened our argument but their lower SNR in the present observations would not have allowed us to draw firm conclusions in the absence of SiN. However, we note that all three molecules have the potential to serve as diagnostic tools for identifying binary interactions in other systems, especially with targeted observations at high SNR. Based on the predictions of chemical models that consider the presence of a Sun-like companion \cite{Van-de-Sande2022}, SiN and SiC are expected to be good tracers of stellar companions to S-type and oxygen-rich AGB stars, but \edits{probably} not carbon stars (unless notably asymmetric emission is detected), because carbon-rich CSEs are expected to have higher abundances of both molecules without the presence of a companion. NS is predicted to have higher abundances around carbon stars in the presence of a white dwarf companion, but not if the companion is a main-sequence star. For S-type and oxygen-rich AGB stars, NS is expected to be a good tracer of either a white dwarf or a Sun-like companion. While there may be other molecules that are enhanced or destroyed in the presence of a companion, a comprehensive list is difficult to compile \cite{Van-de-Sande2022}. After checking all detected molecular lines for asymmetries, we do not find any additional candidates for tracers of binary-induced chemistry towards W~Aql.

The timing between our observations and the present orbital configuration of W~Aql contributed to our being able to use SiN to characterise the orbit. If the W~Aql system was instead observed $\sim 200$ years prior to the next periastron, rather than $\sim 200$ after the most recent periastron, it is unlikely that SiN would have been detected. In that case, in the $\sim 900$ years since the previous periastron, the SiN arc would have expanded with the CSE to around 4 times farther from the AGB star than what we presently observe. At that radial distance, most of the SiN would have been destroyed through photodissociation by the interstellar radiation field \cite{Van-de-Sande2022}. This is also why we do not detect SiN that was created contemporaneously with the white circle in CO (Fig.~\ref{coarcs}a) during the second most recent periastron passage, $\sim 1300$ years ago. 
\editstwo{That said, SiN has already persisted for $\sim200$ years since the periastron interaction, and may continue to be detectable for another 50 to 100 years, based on the expansion velocity and excitation conditions. This means that the imprint of the periastron interaction will be potentially detectable for around a quarter of the total orbital period, a much larger portion than if we had to rely on, for example, observing changing stellar positions or radial velocities around periastron (see Table~\ref{tab:solutions}). The high eccentricity and small periastron separation of the system also contributed to favourable conditions for the formation of SiN around W~Aql. As noted above and in Methods \ref{chemmethods}, the companion-initiated photochemistry is most impactful in the dense inner CSE, meaning that the tracers of this photochemistry --- SiN, SiC and NS --- may not be formed in sufficiently high quantities to be detected for binary systems with wider orbits, where the companion passes through regions of the CSE with lower number densities.}
Despite \editstwo{these} potential limitations, molecular tracers in the CSE generally persist for a relatively long time (hundreds of years, depending on the molecule) and allow us to probe the system on longer time scales than direct imaging or radial velocity measurements, which can only be taken on human timescales. Hence, molecular tracers are invaluable for constraining binary systems with long orbital periods.


The W~Aql system may be unusual for having such a highly eccentric orbit, but it is not unique nor is it impossible for it to have formed with such a high eccentricity. In fact, studies of eccentricity distributions that include wider binaries find a tendency for the mean eccentricity to be higher for subsamples with larger periods \cite{Tokovinin2016,Boffin1993}. 
Indeed, for long-period binaries, orbital circularisation during their formation is not expected \cite{Moe2017}.
Furthermore, a large statistical analysis of binary systems found that solar-type stars in binaries are more likely to have long periods than short periods, i.e. the companion frequency distribution for solar-type primaries peaked at periods of $\log P[\mathrm{days}] = 5.5$ \cite{Moe2017}, very close to the period we found for W~Aql ($\log P_\mathrm{W Aql}[\mathrm{days}] = 5.6$). 
Both the aforementioned studies focussed primarily on main sequence stars, but our result for W~Aql shows that wide binaries with high eccentricities can survive to the AGB phase. {Our hydrodynamic model, which takes into account the gravitational effect of the secondary star on the wind and vice versa, exhibits a very slightly increasing orbital period (owing to the mass lost by the AGB star) but negligible changes in eccentricity, and does not show precession over $\sim5000$~years.} While 5000 years may seem too short a time to make a definitive judgement, we point out that the expansion of the CSE during this time represents a larger spatial extent than the cool dust emission imaged by Herschel/PACS (at 70 and $160\um$ \cite{Mayer2013}).
Despite the high eccentricity that we find for W~Aql, none of our orbital solutions (Table \ref{tab:solutions}) have periastron separations smaller than the Roche limit, so no direct interaction between the two stars is expected and no evidence of such an interaction is seen in the ALMA observations. 
This suggests a relatively stable, if slowly evolving, system from which we could expect the eventual formation of a planetary nebula characterised by elongation to the southwest and perhaps a variety of additional arcs, analogous to what is presently seen in the AGB CSE, including at larger scales \cite{Ramstedt2011,Mayer2013,Ramstedt2017}.


Other binary systems containing AGB stars have also been found to have long periods (based mainly on spiral-like structures in CO observations) including AFGL~3068 (\mbox{$\sim800$} years \cite{Kim2017}), R~Scl (445 years \cite{Maercker2012,Maercker2016a}), and II~Lup (128 years \cite{Lykou2018}). In comparison, AGB stars that have close companions, such as $\pi^1$~Gru (current projected separation 6~au \cite{Doan2017,Homan2020}, period unknown) and V~Hya (8.5 year period \cite{Sahai2016,Sahai2022}), both of which are triple systems that also have wide companions, have less spherical and more disrupted CSEs with, for example, bipolar outflows. Unlike the former group with more spherical CSEs, these triple systems are more likely to go on to form bipolar planetary nebulae. The very high eccentricity of W~Aql precludes the presence of a stable third companion and, despite the small periastron separation, we can consider it to be a relatively undisrupted system, suggesting the eventual formation of a relatively regular planetary nebula, i.e. perhaps more closely resembling the Ring Nebula than the Butterfly Nebula.

The study we have presented here adds to the small number of AGB stars with known companions and orbital parameters. While previous studies have struggled to explain the range of eccentricities observed for e.g. post-AGB stars, most of these have focussed on shorter orbital periods, ranging up to 1000 days, rather than 1000 years, owing to observational limitations \cite{Oomen2018}. The W~Aql system provides further evidence that highly eccentric systems with long orbital periods exist during the AGB phase and that such eccentricity could be inherited by binary systems in later evolutionary phases, such as post-AGB stars and Barium stars \cite{Jorissen1998}.
The method used here --- which entails the combination of chemical tracers and hydrodynamical models --- can be used to detect the characteristic effects of main sequence binary companions in other AGB CSEs. Rather than solely searching for structures in the CSE, future studies can also check for anisotropies in molecular emission and the production of particular molecular species to confirm or rule out the presence of a stellar companion.

\section{Methods}\label{methods}

\subsection{Distance}\label{dist-notes}

Many of the measurements and calculations in the present work rely on the value of the distance to the W~Aql system and, more specifically, to the AGB component. Previous modelling of W~Aql has assumed a distance of 395~pc, calculated from a period-magnitude relation \cite{Danilovich2014}. Prior to this, a variety of distances were assumed for W~Aql, ranging from 230 to 680 pc \cite{Ramstedt2009,Knapp1985,Jura1988,Groenewegen1998}. Recently, distances have been calculated based on high-precision parallax observations from the Gaia mission \cite{Gaia-2016}. Values of  $374\pm22$~pc \cite{Danilovich2021} and $380^{+68}_{-49}$~pc \cite{Andriantsaralaza2022} were found using different methods based on the Gaia Early Data Release 3 \cite{Gaia-EDR3}. In this work, we continue to use a distance of 395~pc because this value falls within the uncertainties of both Gaia-derived values, and because it has been previously used in many radiative transfer models for the AGB star \cite{Danilovich2014,Ramstedt2017,Brunner2018,Danilovich2021} and various stellar and circumstellar parameters such as luminosity and mass-loss rate have been derived relative to this value (Table \ref{tab:parameters}).
\edits{We note that if the true distance is not exactly the adopted one, then the derivations of various parameters would be altered in the following way: mass-loss rate and relative molecular abundances would tend to increase for a larger distance, although molecular abundances may not change significantly after the mass-loss rate was updated, owning to a degeneracy between the impact of distance and density (the latter being directly related to mass-loss rate) on the line intensity. Our derived projected separations would increase linearly with distance, which would in turn result in a larger calculated orbital period.}

\subsection{Stellar masses}\label{masses}

The companion to the AGB star was identified as a main sequence star classified as F8 to G0 \cite{Danilovich2015}, implying the stellar mass is in the range $1.09-1.04\,\msol$ \cite{Habets1981}.
For the purposes of this study we have assumed the companion is an F9 star with a mass of $1.06\,\msol$.

The situation for the AGB component is more complicated. Previous studies comparing oxygen isotopic ratios with stellar evolution models have calculated an initial stellar mass for the AGB star of $1.6\pm0.2\,\msol$ \cite{De-Nutte2017,De-Beck2020}. Although the current mass-loss rate of the AGB star is relatively high at $\dot{M} = 3\e{-6}\spy$ \cite{Ramstedt2017}, stellar evolution models indicate that a significant decrease in stellar mass (i.e. $>0.1\,\msol$) is not expected to occur until the final stages of the thermally pulsing AGB phase (i.e. during and after the last one or two thermal pulses \cite{Vassiliadis1993}). Ergo, we assume $1.6\,\msol$ for the present AGB mass and hence assume a total system mass of $2.66\,\msol$. 


\subsection{Spatially resolved imaging}\label{photmethods}

W~Aql was observed with the Advanced Camera for Surveys (ACS) on the Hubble Space Telescope (HST) at 400 nm on 12 October 2004 (Fig. \ref{orb-phot}b, \cite{Ramstedt2011}).
It was observed again with VLT/SPHERE-ZIMPOL at 735.4 nm on 9 July 2019 (Fig. \ref{orb-phot}c, \cite{Montarges2023}). Both HST and SPHERE images were taken during a similar phase of the AGB pulsation, approximately halfway between maximum and minimum light.
Another HST observation was taken with the Wide Field/Planetary Camera (WFPC) at 550 nm in 1993 \cite{Mayer2013}, but this was taken before the first servicing mission, and the degraded angular resolution makes it unusable for our study.

We measured the positions of the AGB and F9 stars using the python package \texttt{lmfit}\footnote{\url{https://lmfit.github.io/lmfit-py/index.html}}. We find the separation between the two stars is \edits{$475\pm1.0$~mas} in the HST epoch, and \edits{$491\pm1.8$~mas} in the SPHERE epoch. \edits{For HST, the astrometry is well characterised and} the uncertainties were estimated based on the noise of the images. \edits{For SPHERE, the astrometric uncertainty includes the orientation with respect to north, the distortion, the plate scale stability and the statistical position uncertainty \cite{Maire2021}} The change in projected distance between the two stars is then calculated to be \edits{$16\pm0.25\pm1.79$~mas (to distinguish between the systematic and statistical uncertainties)} in 14.75 years, with the projected motion of the F9 star approximately following a straight line away from the AGB star. This motion \edits{does not contradict} a highly inclined, nearly edge-on orbit, with inclination, $i\sim 90\deg$. The 2019 SPHERE position corresponds to a projected separation of 194~au, at our adopted distance of 395~pc. 

These results indicate that the orbital period must be long, particularly as compared with the timescale of our observations.
For example, a circular orbit with a radius of 194~au gives a period of 1660~years for our assumed system mass of $2.66\,\msol$. An extremely elliptical orbit with an apastron of 194~au and a periastron of 3~au (a value chosen so the F9 star does not pass through the AGB star, since we see no evidence of such an extreme interaction) results in a period of 600~years. Note that neither of these orbits properly consider the motion seen between the HST and SPHERE epochs and are merely illustrative.
The ephemeris of such a long orbit cannot be constrained through direct photometric imaging in a reasonable timeframe, because the observations would need to be taken decades and centuries apart. Hence, we require other markers in the circumstellar environment of the AGB star to constrain the orbital parameters of the W~Aql system.

\subsection{ALMA results}\label{almadata}

High spatial resolution observations of W~Aql were obtained with the Atacama Large Millimetre/submillimetre Array (ALMA) as part of the ATOMIUM Large Programme\footnote{Programme ID: 2018.1.00659.L, PI L. Decin} \cite{Gottlieb2022}. More than 110 molecular lines were detected towards W~Aql, including CO, SiN, SiC, and HC$_3$N, which are analysed here. We detected the SiC and NS radicals for the first time towards an S-type star. Previously, SiN was detected and HC$_3$N was tentatively detected towards W~Aql with the APEX telescope \cite{De-Beck2020}. We present spatially resolved emission of SiN and \hc3n for the first time. The SiN, SiC and HC$_3$N emission show two types of asymmetric morphologies, both different to the more extensive circumstellar structures revealed by the CO observations at high spatial resolution.

\subsubsection{Data reduction}\label{datared}

W~Aql was observed with three array configurations of ALMA. This enabled us to observe small structures at high angular resolutions (down to $0.024\arcsec \times 0.021\arcsec$) while still retrieving larger structures (up to a maximum recoverable scale, or MRS, of $8.9\arcsec$) that would otherwise be resolved out \cite{Gottlieb2022}. 
\edits{While these are the extremes of resolution and MRS available in the ATOMIUM dataset, the precise properties the data we analyse can be found in Table \ref{resolutions} for each transition.}

We combined the three data sets to maximise the sensitivity of images, using the Common Astronomy Software Applications for Radio Astronomy (CASA \cite{CASA2022}). We used the combined data to make spectral image cubes for each transition in Table \ref{resolutions}, weighting the contributions of the baselines to optimise the resolution and surface brightness sensitivity. The velocity resolution is 1.1--$1.3\,\kms$ depending on frequency, and in some cases we averaged 2 or more channels to increase sensitivity. The typical rms noise is $\lessapprox2$ mJy.
All velocities are adjusted to the LSR frame. The relative astrometric accuracy of the extended configuration alone is $\sim0.002\arcsec$ and $\sim0.005\arcsec$ for the combined data at slightly lower resolution. The flux scale for the combined images is accurate to $\sim10$\%. 
\edits{The chances of interferometric noise causing artefacts $\geq 5 \sigma$ in these images is negligible. The relative position accuracy of  measurements is at least equal to the synthesised beam divided by the SNR \cite{TMS}, so for SNR = 5 this is  $\sim 40$ mas for SiN, SiC, NS, HC$_3$N, and $^{13}$CN, around 25~mas for CO, and 12~mas for SiO, SiS, CS and HCN.}

Moment zero (integrated intensity) maps were made by summing all the channels with emission above $\sim3\sigma_{\mathrm{rms}}$. Position-velocity (PV) diagrams were made by selecting a tilted rectangular slice (`slit') covering the moment zero emission (spanning a width of $3\arcsec$) at the angle shown in Fig.~\ref{sin} (though other angles were tested, see Methods \ref{sinsic}), and  measuring the flux density in this region for each channel in increments along the slice.
The peak of the continuum emission was assumed to be the position of the AGB star. In the channel maps and moment zero maps, the position of the AGB star is at (0,0). {A small secondary peak, associated with the position of the F9 star, was detected in the continuum emission and will be analysed in a future paper.}

\begin{table}[t]
\caption{Molecular lines in the ground vibrational state used in our analysis.}\label{resolutions}
\begin{center}
\resizebox{\textwidth}{!}{%
\begin{tabular}{cccccccc}
\hline
Molecule & Transition & Frequency & Ref. for & $\upsilon_\mathrm{cent}$ & Ang. res. & MRS & Recovered\\
 & & [GHz] & freq. & [$\kms$] &[$\arcsec$] & [$\arcsec$] & flux\\
\hline\hline
CO 	 & 	$J=2 \to 1$	 & 	230.538	 & 	\cite{Winnewisser1997}	 & 	\dots	 & 	$0.132 \times 0.123$	 & 	5.3	 & 	\dots	 \\
	 & 		 & 		 & 		 & 	-23.4	 & 	$0.829 \times 0.679$	 & 	8.9	 & 	33\%	 \\
SiN 	 & 	$N,J,F=6,{13}/{2}, {13}/{2} \to5,{11}/{2},{11}/{2}$	 & 	\phantom{$^\dagger$}262.156$^\dagger$	 & 	\cite{Turner1992}	 & 	-23.7	 & 	$0.222\times0.198$	 & 	4.7	 & 	100\%	 \\
SiC 	 & 	$^3\Pi_2\;\,\, J=6\to5$	 & 	236.288	 & 	\cite{Cernicharo1989,Bogey1990}	 & 	-23.8	 & 	$0.199 \times 0.184$	 & 	2.6	 & 	\dots	 \\
NS 	 & 	$^2\Pi_{1/2}\, f\;\,\,  J, F={11}/{2} , {13}/{2} \to {9}/{2} , {11}/{2}$	 & 	\phantom{$^\ddagger$}253.968$^\ddagger$	 & 	\cite{Lee1995}	 & 	\dots	 & 	$0.187\times0.171$	 & 	2.5	 & 	\dots	 \\
HC$_3$N	 & 	$J=25\to24$	 & 	227.419	 & 	\cite{Yamada1995}	 & 	-22.5	 & 	$ 0.204\times 0.181$	 & 	5.4	 & 	\dots	 \\
	 & 	$J=26\to25$	 & 	236.513	 & 	\cite{Yamada1995}	 & 	-21.8	 & 	$ 0.208 \times 0.191 $	 & 	2.6	 & 	\dots	 \\
	 & 	$J=27\to26$	 & 	245.606	 & 	\cite{Yamada1995}	 & 	-21.4	 & 	$0.213 \times 0.172$	 & 	5.0	 & 	\dots	 \\
	 & 	$J=28\to27$	 & 	254.700	 & 	\cite{Yamada1995}	 & 	-20.8	 & 	$0.190 \times 0.172$	 & 	2.5	 & 	\dots	 \\
SiO 	 & 	$J = 5\to4$	 & 	217.105	 & 	\cite{Muller2013}	 & 	-22.6	 & 	$0.063 \times 0.055$	 & 	5.7	 & 	85\%	 \\
SiS 	 & 	$J=12\to11$	 & 	217.818	 & 	\cite{Muller2007}	 & 	-21.4	 & 	$0.063 \times 0.055$	 & 	5.7	 & 	91\%	 \\
CS 	 & 	$J=5\to4$	 & 	244.936	 & 	\cite{Gottlieb2003}	 & 	-22.5	 & 	$0.078\times 0.066$	 & 	5.0	 & 	79\%	 \\
HCN 	 & 	$J=3\to2$	 & 	265.886	 & 	\cite{Ahrens2002}	 & 	-23.0	 & 	$0.061\times0.053$	 & 	2.4	 & 	72\%	 \\
H$^{13}$ CN	 & 	$J=3\to2$	 & 	259.012	 & 	\cite{Fuchs2004}	 & 	-22.2	 & 	$0.073\times 0.064$	 & 	4.8	 & 	\dots	 \\
$^{13}$CN	 & 	$N, F_1, F_2, F =2, 0, 2, 3, \to 1, 0, 1, 2$	 & 	\phantom{$^\ddagger$}217.303$^\ddagger$	 & 	\cite{Bogey1984}	 & 	\dots	 & 	\multirow{ 2}{*}{$0.213\times0.199$}	 & 	\multirow{2}{*}{5.7}	 & 	44\%	 \\
 	 & 	$N, F_1, F_2, F =2, 1, 3, 4, \to 1, 1, 2, 3$	 & 	\phantom{$^\ddagger$}217.467$^\ddagger$	 & 	\cite{Bogey1984}	 & 	\dots	 & 	 	 & 	 	 & 	48\%	 \\
\hline
\end{tabular}}
\end{center}
\footnotesize
\textbf{Notes:} All frequencies are rest frequencies and all velocities are are with respect to the local standard of rest. ($^\dagger$) Frequency and corresponding quantum numbers of central hyperfine component are given.
($^\ddagger$) Frequency and corresponding quantum numbers of the brightest hyperfine component are given.
Listed in column 4 are the primary references which provide the measured frequencies and the spectroscopic designation of
the transitions observed here.  
The Cologne Database for Molecular Spectroscopy, CDMS \cite{Muller2001,Muller2005}, provides a comprehensive list
of the best estimate of the transition frequencies, the excitation energies, and the quantum mechanical line strengths.
Column 5 gives the central velocity of the line as obtained from fitting a soft parabola (see Methods \ref{datared}).
Column 7 gives the maximum recoverable scale for the ALMA observations.
\end{table}%

To check whether our observations suffered from resolved-out flux, we compared spectra extracted from the ALMA data with previous
observations of the same lines taken with the APEX single antenna  \cite{De-Beck2020} as shown in  Fig.~\ref{lostflux}. 
For CO we found 66\% of the flux was resolved out, whereas all the flux was recovered by ALMA for SiN. We were unable to make the same comparison for SiC, which is a first detection, or HC$_3$N, which was at best only tentatively detected with APEX \cite{De-Beck2020}. Although only a third of the CO flux was recovered by ALMA, it is only smooth large-scale flux that is resolved out. 
This large scale flux is mostly associated with smoother bulk outflows, whereas our analysis in the present work focuses on
smaller structures in the wind --- i.e., the missing CO flux does not impede the present study.

Out of the other molecular lines discussed here and which have previously been observed, we found that about 28\% of the flux in H$^{12}$CN $J=3\to2$ was resolved out (Fig. \ref{otherlostflux}). Some degree of lost flux was expected because this line was not observed with the most compact configuration of ALMA. The corresponding transition in H$^{13}$CN was not observed with APEX \cite{De-Beck2020} but since it was observed with the compact configuration of ALMA and shows more extended emission than H$^{12}$CN, we can assume very little, if any, flux was resolved out for H$^{13}$CN. For SiO, SiS and CS, most of the flux was recovered, with only about 10--20\% lost, as can be seen in Fig. \ref{otherlostflux}, where we have compared the spectra of these three molecules and H$^{12}$CN observed with APEX and ALMA.

The $^{13}$CN emission in $N=2\to1$ at 217~GHz has a low SNR. Therefore, to better determine the spatial distribution of $^{13}$CN, we combined the two most intense components of the many possible fine and hyperfine structure transitions of the $N = 2 \to 1$ transition that span a 450 MHz wide range centred on 217.257 GHz.
We extracted the channels in the calibrated visibility data in the frequency ranges corresponding to $\upsilon_\mathrm{LSR}=-23\pm50$ km s$^{-1}$ around each of the rest frequencies and combined the channel selections aligned in velocity.  The combined data set was assigned a fictitious rest frequency of 217.3055 GHz  so that its central velocity corresponded to $-23~\kms$, and we then made an image cube from the stacked visibility data and analysed this following the same procedure as for the other data cubes.
Finally, we checked the two multiplets of $^{13}$CN listed in Table \ref{resolutions} individually for resolved-out flux and found that a little less than half of the flux was recovered for these lines. 

For all the spectral lines studied here, except for \ce{^13CN} and NS, we fit soft parabola profiles \cite{Olofsson1993} \edits{
\begin{equation}\label{eq:softpara}
F(\upsilon) = F_0 \left( 1 - \left[ \frac{\upsilon - \upsilon_\mathrm{cent}}{\upsilon_\infty} \right] ^2\right)^{\gamma/2}
\end{equation}
where $\upsilon_\mathrm{cent}$ is the central velocity of the line profile and $F_0$ is the flux at the centre of the line. The parameters $F_0$, $\upsilon_\mathrm{cent}$, $\upsilon_\infty$, and $\gamma$ are left as free parameters in the fit. Primarily this is done} to obtain the central line velocities, \edits{which are included in Table \ref{resolutions}}. The soft parabola profile was chosen over a Gaussian profile because the majority of the lines studied here exhibit double-peaked emission and hence significantly deviate from Gaussian line profile shapes. \ce{^13CN} was excluded from this analysis because its hyperfine structure dominates its line profile, and NS was excluded because the spectrum is too noisy to obtain a reasonable fit. The central velocities of the lines were generally in agreement with the previously measured stellar LSR velocity of $\upsilon_\mathrm{LSR} = -23~\kms$ \cite{Danilovich2014,De-Beck2020} and will be discussed in more detail in the following sections. 

\subsubsection{AGB angular diameter}\label{Udisc}

We took the calibrated data for all ALMA configurations combined, excluding channels with line emission, and fit a uniform disc (UD) to the visibilities (as in \cite{Homan2021}). This gave a diameter of 16.6~mas, containing 8.0~mJy. There was negligible ellipticity or displacement of the centroid. At mm wavelengths, a UD is expected to be a better representation of stellar brightness distribution than a Gaussian distribution. The SNR is $>100$, suggesting sub-mas precision, \edits{based on the nominal uncertainty of beam size divided by SNR}, but taking into account possible irregularities in the stellar disc, we adopt a conservative uncertainty of 3 mas.
\edits{The diameter of $16.6\pm3$~mas is the size of the the surface where electron-neutral free-free emission dominates and is optically thick (at these wavelengths \cite{Vlemmings2019}) and corresponds to a radius of $3.3\pm0.6$~au at our adopted distance.}
\edits{We note that the resolution of the continuum image from the extended array is $21\times24$~mas \cite{Gottlieb2022}, while for the combined continuum image it is $40\times33$~mas.}
The optical diameter is \edits{$11.6\pm1.8$~mas \cite{Richichi2005}, 34\% smaller than our value}.  Vlemmings et al. \cite{Vlemmings2019} found that the mm-wave diameters of a small sample of AGB stars were 15--50\% greater than the optical diameters, consistent with our finding. \edits{It has also been found that, in general, the mm-wave diameters of the ATOMIUM sample are 30--100\% larger than the optical diameters \cite{Baudry2023}.}

\subsubsection{SiN and SiC}\label{sinsic}

The SiN line we observe towards W~Aql ($N,J=6,{13/2}\to5,{11/2}$) is a blend of three closely-spaced hyperfine components separated by about 0.8 and 0.5~MHz (Fig.~\ref{lostflux}), and the frequency of the centroid is 262,155.78~MHz. 
The lower spin-rotation component ($N,J=6,{11/2} \to 5,{9/2}$) at 262.650~GHz falls just outside of the frequency range covered by our observations.
The SiC line detected towards W~Aql corresponds to the $J= 6\to5$ transition in the lowest fine structure ladder $^3\Pi_2$ \cite{Cernicharo1989}. 
The corresponding $J= 6\to5$ rotational transitions in the $^3\Pi_1$ and $^3\Pi_0$ upper fine structure ladders fell between the frequency bands covered by our observations.

\editstwo{Neither SiN nor SiC were detected for any other stars in the ATOMIUM sample, all of which are oxygen-rich aside from one other S-type AGB star ($\pi^1$~Gru).} SiC has been previously detected towards 12 carbon-rich AGB stars \cite{Massalkhi2018}, but the present work represents the first detection of SiC in the envelope of an S-type AGB star.
SiN has been previously detected towards W~Aql \cite{De-Beck2020} and only one other star: the nearby carbon-rich AGB star CW~Leo \cite{Turner1992}, which is suspected of having a companion \cite{Cernicharo2015a,Decin2015,Siebert2022}. \editstwo{Spatially resolved Submillimeter Array (SMA) observations towards CW~Leo show the SiN mainly distributed in a shell-like pattern, with some brighter, asymmetric, emission to the south-west \cite{Patel2013}.} 
\editstwo{However, a detailed analysis of these observations has not yet been published and,} consequently, we lack detailed spatial information for SiN around other stars with which to compare our W~Aql results. \edits{Spatially resolved SiC emission has also been observed towards CW~Leo, for which SiC was not detected in the innermost regions of the CSE but rather in outer shells \cite{Velilla-Prieto2019a}, \editstwo{possibly also showing some asymmetry to the south-west \cite{Patel2013}. Further} 
discussion of SiC distributions is given in Methods \ref{chem:sin}).}


The integrated intensity maps of SiN (Fig.~\ref{sin}a) and SiC (Fig.~\ref{sic}a) show emission primarily north and east of the AGB star. The SiN emission has a higher SNR and is hence more readily analysed. Therefore, we have focussed our analysis on SiN, but note that the SiC observations agree with the conclusions drawn from SiN.

We produced a series of position-velocity diagrams of SiN using a wide slit (total width $3\arcsec$) to encompass all the emission seen in the zeroth moment map (Fig.~\ref{sin}a). \edits{Using a narrower slit (such as $0.3\arcsec$) resulted in a lower SNR in the PV diagram, making an analysis more troublesome.} We tested all possible slit angles passing through the position of the AGB star in increments of $5\deg$ and then $1\deg$ around the angles producing the most distinct PV diagrams. The final slit position of $33\deg$ east of north was chosen on the basis of the clarity and intensity of the associated PV diagram. Even though the slit angle was determined independently, we find that it passes through the present position of the F9 star (Fig.~\ref{sin}a). As shown in Fig.~\ref{sin}b, the PV diagram of SiN exhibits an arc-like structure in position-velocity space, tracing a little more than half an ellipse centred on the AGB star. 
We fit a parabola to the points in the PV diagram with intensities $\geq 3 \sigma$ above the noise, weighted by the flux at those points. The peak of the parabola, plotted in white in Fig.~\ref{sin}b, is at $-24.1~\kms$, which is in agreement with the central velocity we find for the spectral line (Table \ref{resolutions}). The emission distribution in the PV diagram does not precisely follow the shape of the parabola, \editstwo{particularly at the negative offset and extreme velocity edges of the emission,} so we also plot a partial ellipse based on the position of the parabola (\editstwo{using the centre and peak of the parabola and} with the half-width along the velocity axis set to $14~\kms$), which better follows the shape of the emission at the most extreme velocities.
We followed a similar procedure for SiC to produce a PV-diagram and fit a parabola to the arc of emission (Fig.~\ref{sic}b). For SiC, the peak of the parabola is at $-23.5~\kms$. We similarly plot a partial ellipse based on the parabola fit (velocity half-width $13~\kms$), which also follows the emission at the most extreme velocities more closely.

In concert, the zeroth moment map and the PV diagram show that the SiN emission forms an arc to one side of the system, which is close to edge-on \edits{or perpendicular to} the plane of the sky.
We also plot the summed blue and red channels of SiN separately in Fig. \ref{sinbluered}. Owing to the noisy edges of the contours, we could not conclusively determine whether there is an offset between them along the axis connecting the present positions of the AGB and F9 stars. Consequently, we take the orbital inclination of the system to be $i=90\pm7\deg$, where the uncertainty is derived from the beam size.
The lack of spherical symmetry in the SiN emission suggests a spatial dependence for the formation of SiN, as discussed in Methods~\ref{chemmethods} and depicted in Fig.~\ref{SiNsketch}. Despite the lower SNR of the SiC emission, the similar structure seen in the PV-diagram for SiC indicates a similar formation history for both SiN and SiC.

\subsubsection{NS}\label{NSdet}

Two rotational transitions of NS were covered by the ATOMIUM observations --- { the $J=11/2 \to 9/2$ hyperfine split multiplets in the $^2\Pi_{1/2}$ 
and $^2\Pi_{3/2}$ spin-orbit fine structure components}. 
Neither rotational transition was detected in spectra centred on the AGB star.
However guided by predictions from chemical models (Methods \ref{nschem} and \cite{Van-de-Sande2022}), 
we conducted a more careful search for NS. 
The transition in the ground $^2\Pi_{1/2}$ component lies very close to the edge of our frequency band and is difficult to discern in the spectra,
but we successfully detected it in the zeroth moment map (Fig.~\ref{nsplot}a), { which constitutes the first detection of NS towards an S-type AGB star.}
The corresponding rotational transition in the upper ($^2\Pi_{3/2}$) spin-orbit component at 255.597~GHz \cite{Lee1995} 
{ {lies about 322~K above the ground state and is estimated to be about three times less intense.
We found an upper limit for the $^2\Pi_{3/2}$ component of 3$\sigma = 0.047$~Jy~beam$^{-1}$~km~s$^{-1}$ in a zeroth moment map  
that covers the same velocity extent as that observed for the ground $^2\Pi_{1/2}$ component.}}
%

Prior to this, NS had been detected towards just one AGB star, the oxygen-rich IK~Tau \cite{Velilla-Prieto2017,Decin2018} (and notably has not been detected towards the nearby carbon star CW~Leo). An enhanced abundance of NS is expected to be a good tracer of binarity for S-type or oxygen-rich AGB stars with main sequence or white dwarf companions \cite{Van-de-Sande2022}. 
\editstwo{We checked the ATOMIUM data for NS detections towards other sources. While we could rule out NS detections in several sources, for a selection of others (the AGB stars IRC\,+10011 and IRC\,$-$10529, and the red supergiants VX~Sgr and AH~Sco) we could not conclusively confirm or rule out the presence of NS for three reasons. First, the $^2\Pi_{1/2}$ component at 253.968~GHz lies close to the edge of an observed band in frequency space, meaning that the line may be partially truncated, as it is for W~Aql. Second, that line lies close to the SO$_2$ ($J_{K_a,K_c} = 15_{6,10}\to 16_{5,11}$) line at 253.957~GHz and, for the oxygen-rich sources mentioned above, we cannot easily disentangle which emission comes from SO$_2$ and which might come from NS. (This is not a problem for W~Aql, towards which no SO$_2$ lines are detected, including more intrinsically intense lines covered by our observations.) Disentangling NS and \ce{SO2} emission is made more difficult because both lines are truncated by the edge of the observed band. Finally, we also checked for emission from the $^2\Pi_{3/2}$ component at 255.597~GHz but could not confirm the detection of this line of NS. For the AGB stars mentioned above, we did not detect emission above the noise of our observations. However, if we take the expected intensity of the $^2\Pi_{3/2}$ component to be a third that of the truncated and possibly blended line around 253.968~GHz, we determine that the expected intensity is below the noise of our observations. For the two red supergiants, the potential NS line is blended with a high-energy SO$_2$ line, ($J_{K_a,K_c} = 51_{7,45}\to 50_{8,42}$) at 255.595~GHz. Therefore, to determine whether NS is present in these or other ATOMIUM stars, observations of other NS transitions that do not overlap with SO$_2$ or other molecular lines are required.}

In addition to the zeroth moment map, we also constructed a PV diagram of NS (Fig.~\ref{nsplot}b). The only significant region of emission that is $3\sigma$ above the noise in the PV diagram is located on the red side of the PV diagram and not notably offset from the position of the AGB star. This is close to some of the most intense regions seen in the SiN and SiC PV diagrams. We note that because the NS line is on the edge of the observed band, some redder emission might not have been recovered by our observations. To emphasise that this is a true detection of NS rather than a misidentified line, we plot the spectrum of the NS line with the spectra of the SiN and SiC lines in Fig.~\ref{nsplot}c. All lines were extracted from circular apertures with radii $0.25\arcsec$, centred on the continuum peak, which was chosen to best show the NS line. All three lines have a double-peaked profile, with SiN and NS having a brighter red peak than blue peak. Although the NS spectrum is truncated at $-9~\kms$, it can be seen rising in a profile similar to the SiN and SiC red peaks. Deeper observations targeting NS would confirm this behaviour.


\subsubsection{HC$_3$N}\label{sec:hc3n}

Four successive rotational lines of \hc3n were detected towards W~Aql as part of the ATOMIUM project (Table~\ref{resolutions}). 
Prior to this, the three lowest transitions in this group were tentatively detected towards W~Aql with APEX \cite{De-Beck2020}.
A comparison of the lines tentatively detected with APEX and our ALMA data suggests that the ALMA data does not suffer from resolved-out flux. 
It should also be noted  that the $J=25\to24$ and $J=27\to26$ lines were observed with all three ALMA configurations  (including the compact 
configuration), while the $J=26\to25$ and $J=28\to27$ lines were observed with only the extended and medium configurations. 
All four lines have similar intensities when the spectra are extracted from our combined data cubes,  
as expected for lines with similar energies (the lower level energies span 131--165~K) and Einstein $A$ coefficients.
Taken together, our observations confirm that there is no flux resolved out for the observations with the medium configuration.
Most of the \hc3n flux is located south and west of the present location of the AGB star (Figs. \ref{fig:cn} and \ref{hc3n}),
in direct contrast with the observed flux of SiN and SiC  (Figs.~\ref{sin} and \ref{sic}). 




\subsubsection{CO}

The CO $J=2\to1$ line has the most extended emission distribution of all the spectral lines observed towards W~Aql as part of the ATOMIUM Programme. Although only one third of the flux was recovered by ALMA (Methods \ref{datared}), our analysis focuses on smaller structures in the wind, which are not affected by resolved-out flux.

Many complex structures are seen in the CO emission, making a definitive analysis difficult. We first examined the inner wind region, where an overdensity thought to be (part of) a spiral arm was reported \cite{Brunner2018}. In this region, we found an approximately circular structure that corresponds very well to the location of the overdensity and to the radius of the observed SiN emission. In Fig. \ref{coarcs}b, we plot the CO emission close to the AGB star using a logarithmic colour scale and overplot the contours of SiN (as seen in Fig. \ref{sin}a) and a black circle to guide the eye to the roughly circular structure. The radius of this circle is $1.35\arcsec$ and its centre is $0.1\arcsec$ to the north of the AGB star.

Additional circular structures in the CO emission were more difficult to concretely identify, so we plotted the radial intensity against anticlockwise angle to help find such structures (Fig.~\ref{angrad}). Circular structures centred on the AGB star would appear as horizontal lines in such a plot, whereas off-centre circular structures appear as sinusoids. Using the angle-radius plot, we found off-centre circles corresponding to: (red) the edge of the bright central region with a radius of $4\arcsec$, (pink) a circular structure surrounding this region, with a radius of $5.5\arcsec$, and (white) another circle with radius $10.75\arcsec$ which falls close to the edge of the ALMA field of view. The white circle is offset in the same direction (north) as the black circle. Note that the sinusoid corresponding to the black circle identified above can be seen more clearly in the angle-radius plot when it is zoomed in on the structures closer to the AGB star \edits{(bottom panel of Fig.~\ref{angrad})}. \editstwo{In Fig.~\ref{angrad-no-lines} we show the same angle-radius plots as in Fig.~\ref{angrad}, but exclude the coloured lines highlighting the aforementioned structures.}

We plot all these circular structures in Fig.~\ref{coarcs}a over the averaged central three channels of the CO emission. From our analysis with the hydrodynamic model (Methods \ref{hydro}) we come to the conclusion that the black and white circles were formed during the periastron passage of the two stars, in which case they are expected to be offset to the opposite side of the AGB star from the F9 star. The periastron origin of the black circle is also supported by its co-location with the SiN arc. The red and pink circles, \edits{and other irregular structures,} are not directly reproduced by the hydrodynamic model, but this is likely because of limitations in the model including missing physics around pulsations and the wind launching mechanism (see discussion in Methods \ref{hydro}). \edits{Significantly, the wind is launched at $13~\kms$ in the hydrodynamic model, whereas previous studies assume a much lower initial velocity of $3~\kms$, close to the sound speed. This discrepancy prevents a dense inner region forming in the hydrodynamic model, such as the region encircled in red in Fig.~\ref{coarcs}a.}
\edits{We also note that the formation timescales of the red and pink circles and other neighbouring features are $~\lesssim300$~years (taking $\beta$-law wind acceleration into account) and do not match the longer timescale of the binary orbit inferred from resolved imaging (Methods \ref{photmethods}).} 

\edits{When comparing these circular structures with the lower-resolution ($0.47\times0.41$) ALMA observations of CO ($3\to2$) around W Aql \cite{Ramstedt2017}, in which several circular arcs were identified, we find that our red, pink and white circles correspond to the locations of those arcs. In particular, the outermost arc in the earlier data corresponds well with our white circle, and the innermost two arcs (north and south-west) match the position of our red circle. The circular region of higher flux that we have indicated in red in Fig.~\ref{coarcs}a for CO ($2\to1$) also corresponds to the region of higher flux seen in CO ($3\to2$). The remaining arcs identified by \cite{Ramstedt2017} match our pink circle and a few other structures seen in our data which do not form full circles. Note that our black circle is too small to be well resolved in the earlier data.}

\edits{The shell-like structures seen around W Aql have some similarity to previously reported shells around the carbon star CW~Leo, which are also not perfectly centred on the AGB star \cite{Cernicharo2015a, Decin2011, Guelin2018}. Many more shells are seen for CW Leo than W Aql, likely in part because CW~Leo is closer, making emission easier to detect. Studies of the CW~Leo shells have concluded that they could be caused by an eccentric binary orbit seen perpendicular to the line of sight, and assumed some periods of enhanced mass loss \cite{Cernicharo2015a,Guelin2018}. Our hydrodynamic models do not assume a variable mass-loss rate (see Methods \ref{hydro}) but still form shell-like structures when viewed perpendicular to the orbital plane. This does not mean that the mass-loss rate of W~Aql cannot be variable --- indeed variable or anisotropic mass-loss might account for some of the other structures seen in the CO emission.} \editstwo{The possible effects of variable and anisotropic mass loss are discussed in more detail in the Supplementary Materials \ref{sec:anisotropies}.}

We also analysed the higher and lower velocity channels of the W~Aql CO emission, particularly in comparison to the hydrodynamic model. A long-standing unexplained phenomenon is excess emission in the blue wings of the line profiles of CO and other molecules towards W~Aql \cite{Danilovich2014}. 
In our ALMA observations of CO (Fig. \ref{cochan}), it is clear that the blue- and red-shifted channel maps are not symmetric around the LSR velocity. The blue channels ($-37$ to $-30~\kms$) show slightly asymmetric emission, with an elongation in the south-west direction, while the red channels ($-14$ to $-8~\kms$) show emission with more circular symmetry. These differences in shape account for the excess emission in the blue wing of the line profiles. We also compared these different emission distributions with the equivalent distributions produced by the hydrodynamic model after processing by the radiative transfer code MCFOST (Methods \ref{hydro}). In Fig. \ref{otherCOchans} we plot two CO channels equidistant from the LSR velocity and the equivalent model channels. The model also shows the elongated CO emission for the blue channel and the more circular emission for the red channel, reinforcing that the asymmetry arises from the companion's interactions with the AGB wind.

\subsubsection{Other molecular species}\label{sec:othermolecules}

The species SiO, SiS, HCN, and CS are commonly observed in the envelopes of many AGB stars of all chemical types \cite{Ramstedt2009,Schoier2013,Danilovich2018,Massalkhi2019}. All four molecules were observed previously towards W~Aql at a lower spatial resolution of $0.55\arcsec \times 0.48\arcsec$ \cite{Brunner2018} and were analysed using radiative transfer models under the assumption of spherical symmetry. Our new observations were obtained at a much higher angular resolution and the emitting regions for all four molecules are very well resolved (Table \ref{resolutions} and Fig. \ref{othermol}). The increased angular resolution allows us to observe asymmetries in the emission. The emission from all four molecules is more extended to the north-east than to the south-west. This is a qualitatively similar anisotropy to that seen in SiN, but unlike SiN, the more common species exhibit roughly spherically symmetric emission across a much wider fan-like region, running clockwise from east to north west (Fig. \ref{othermol}). In the context of an eccentric binary companion, we interpret this not as enhanced production of SiO, SiS, HCN, and CS triggered during the periastron passage (as we conclude in the cases of SiN and SiC), but as enhanced destruction through photodissociation of SiO, SiS, HCN, and CS by the F9 companion, during the large portion of the orbital period it spends to the southwest of the AGB star. If this were not the case, we should see significantly less emission to the northwest and southeast (i.e. the other regions where we do not see SiN), but the contours in Fig. \ref{othermol} have similar extents from the southeast to northeast to northwest. \edits{This is especially apparent in plots of the central channels of SiS and CS, shown in Fig.~\ref{cssissinglechans}, which show significantly reduced emission near the F9 star as opposed to on the opposite side of the AGB star.}
\editstwo{For CS, the $3\sigma$ contour centred on the AGB stars extends out to $0.33\arcsec$ ($\sim2\e{15}$~cm) from the AGB star in the direction of the F9 star, compared with $0.71\arcsec$ ($\sim4\e{15}$~cm) in the opposite direction. For SiS, the $3\sigma$ contour centred on the AGB star extends out to $0.09\arcsec$ ($\sim5\e{14}$~cm) in the direction of the F9 star and out to $0.23\arcsec$ ($\sim1\e{15}$~cm) in the opposite direction.}

Furthermore, the PV diagrams of CS, SiO and \ce{H^13CN}, taken along the same axis as we used for SiN and plotted in Fig.~\ref{CSPV}, show the brightest emission spatially close to the AGB star, not in an arc as for SiN or SiC. They also show that the emission is consistently less extended and less intense on the side of the AGB star where the F9 star is located. Notably, this is not the case for CO, also plotted in Fig.~\ref{CSPV}, which does not show evidence of photodissociation by the F9 star, as expected given its stronger bond energy \edits{and self-shielding \cite{Morris1983}}. 
The reduced emission seen in the spectra around the F9 star (Methods \ref{sec:F9spec}) is further evidence of most molecules being destroyed by the F9 flux.

Another molecular species that displays highly asymmetric emission around W~Aql is $^{13}$CN. Although the main isotopic species, $^{12}$CN, was not covered in the ATOMIUM observations, it has previously been observed towards W~Aql with the IRAM 30m telescope \cite{Bachiller1997}. We find that, unlike the common molecular species discussed above, the $^{13}$CN emission is mainly seen on the opposite side of the AGB star. As can be seen from the zeroth moment maps of H$^{13}$CN and $^{13}$CN in Fig.~\ref{fig:cn}, the $^{13}$CN emission is mainly observed where the H$^{13}$CN emission is absent, which is consistent with the generally accepted notion that CN is a photodissociation product of HCN \cite{Huggins1984}. This is discussed in more detail in Methods \ref{chemmethods}.





\subsubsection{Molecular emission around F9 star}\label{sec:F9spec}

We extracted spectra in circular apertures of radii 100~mas (corresponding to a projected radius of 40~au) centred on the F9 star to check for anomalous emission. Very few lines were detected above the noise in these spectra, with lines originating only from CO, SiO, CS, and HCN. We compared the line profiles extracted from the region around the line-of-sight position of the F9 star with profiles of the same sized aperture centred on the AGB star and plot comparisons for CS, HCN and SiO in Fig. \ref{F9spec}. Notably, the F9-centred line profiles exhibit relatively more flux in the blue channels and less flux in the red channels than the corresponding AGB-centred profiles. The F9-centred profiles also tend to have relatively less emission in the channels close to the LSR velocity. 
\edits{From this, we can estimate that the F9 star is located, spatially, in the region that corresponds to gas with velocities close to the AGB stellar LSR velocity, i.e.~gas with motions approximately in the plane of the sky. This estimate is possible because, in an expanding circumstellar envelope, the velocity axis has a correspondence to the line of sight spatial axis (see, for example, \cite{Guelin2018}). Although this does not say anything about the present velocity of the F9 star (it need not be moving at the same velocity as the AGB circumstellar gas that it is moving through), it is consistent with}
the stars being in a highly eccentric orbit, as the present relative motion of the F9 star would be predominantly in the plane of the sky rather than into or out of the plane of the sky, and would, in any case, have a low absolute total velocity of $\sim2~\kms$.


We also checked the shape of the line profiles extracted for an equivalent 100~mas aperture on the opposite side of the AGB star from the F9 star (at the same projected separation) and found that those line profiles were more similar to the AGB-centred line profiles than those centred on the F9 star (Fig. \ref{F9spec}).
Finally, we note that the phenomenon of the blue peaks being brighter than the red peaks for the F9-centred profiles is the opposite of what we see for the line profiles of SiN and NS (Fig. \ref{nsplot}c) centred on the AGB star. This is easily explained by the different formation/destruction times of the two groups of molecules: SiN and NS formed during the periastron passage, whereas CS, HCN and SiO are presently being (partly) photodissociated by the UV flux of the F9 star.

\edits{The intensity of the UV flux from the F9 star is proportional to the inverse square of the distance from the star. This means that the apparent UV flux close to the AGB star, taking the projected separation of 194~au, is 24 times less than the flux 40~au from the F9 star, and the flux on the opposite side of the AGB star (at a distance of 388~au) is 94 times weaker. At a distance of 10~au from the F9 star, close to the distance between the two stars during periastron, the UV flux would be 380 times higher than the flux on the same region at the present stellar separation. Note that these values are rough estimates and do not include, for example, UV extinction by dust, which would further reduce the UV flux for larger distances when there is more dust between the F9 star and the region of interest.}


%
%

\subsection{Radiative transfer modelling}\label{rtmod}

Radiative transfer calculations were undertaken to approximate the abundance of SiN in the arc seen in Fig.~\ref{sin}. To achieve this, we extracted the SiN spectra from round apertures with radii of $0.25\arcsec$, evenly spaced with centres separated by $0.3\arcsec$ starting from the continuum peak and moving outwards along the north $33\deg$ east line passing through the emission. The set-up is shown in Fig. \ref{sin-modelling}a, where the regions are labelled from A to H. The aperture size was chosen so as to not lie outside of the detected SiN emission. Furthermore, these regions are centred along the same axis for which we found the best PV diagram (Fig.~\ref{sin}b and Methods \ref{almadata}), so they are unlikely to overlap with the edges of the SiN emission. Therefore, by considering only spectra from the regions plotted in Fig.~\ref{sin-modelling}c, we can use a 1D (spherically symmetric) radiative transfer model to compare equivalent synthetic spectra and determine the SiN abundance distribution in the arc, which can also be approximated by a wedge of a spherical shell. Our approach is viable because the SiN emission is expected to be optically thin (and indeed we find a peak tangential optical depth of $\tau \lesssim 0.2$ in the model) and emission in other parts of the spherically symmetric model (at different velocities) is not expected to interact with emission in the regions of interest.

We used the accelerated lambda iteration method (ALI \cite{Rybicki1991}), which has been previously used to determine the abundances of various other molecules in the CSE of W~Aql \cite{Brunner2018,Danilovich2021}. We use previously determined circumstellar parameters for W~Aql \cite{Danilovich2014}, including a radial temperature profile, the mass-loss rate of $3\e{-6}\spy$ \cite{Ramstedt2017} and a velocity profile described by \cite{Danilovich2014}
\begin{equation}\label{betalaw}
\upsilon(r) = \upsilon_0 + (\upsilon_\infty - \upsilon_0) \left( 1 - \frac{R_\mathrm{in}}{r} \right)^\beta
\end{equation}
with $\upsilon_0 = 3~\kms$ the velocity at the dust condensation radius, $R_\mathrm{in} = 2\e{14}$~cm, $\upsilon_\infty = 16.5~\kms$ the terminal expansion velocity and $\beta = 2$. The key stellar and circumstellar parameters are summarised in Table \ref{tab:parameters}. We also included a previously implemented overdensity \cite{Brunner2018}, which was found to improve the radiative transfer model fit for ALMA observations of CS and H$^{13}$CN at lower resolutions \cite{Brunner2018}. The overdensity relates to an increase in the \h2 number density by a factor of five between the radii of $8\e{15}$~cm and $1.5\e{16}$~cm (Fig.~\ref{sin-modelling}), and is in good agreement with the location of a region of increased CO emission (a good tracer of density) traced by the black circle in Fig.~\ref{coarcs}b. (Previously the overdensity was thought to be part of an unresolved spiral arm \cite{Brunner2018}.)

We include SiN energy levels up to $N=20$ in the ground vibrational state and the 59 radiative transitions connecting those levels. The energy levels and Einstein A coefficients were calculated using \texttt{CALPGM} \cite{Pickett1991} and take fine structure into account but neglect the closely spaced hyperfine structure, which is not resolved in our observations.
There are no calculated or measured collisional (de)excitation rates for SiN, so instead we use the rates calculated for CN-He \cite{Lique2010}, scaled by 1.37 to account for the different reduced mass of the SiN-\h2 system.

On the basis that the different extraction apertures shown in Fig \ref{sin-modelling}a probe different regions of the SiN distribution, we  tried various shapes for the radial distribution of SiN abundance in an attempt to reproduce the observed distribution of SiN. These included a constant abundance, step functions of different constant SiN abundances, and Gaussian shells of various widths and positions. We also left the inner and outer radii of the SiN emitting region as free parameters. We found that while the two apertures farthest from the continuum peak, G and H, were sensitive to the outer radius and outer abundance of the SiN distribution, as expected, the inner apertures, A to D, were also sensitive to these properties, which affected the heights of the emission peaks in their double-peaked profiles. Conversely, the choice of inner radius and the innermost abundance of SiN mainly affected the heights of the line centres for apertures A to C. These dependencies were expected given the observed wedge of SiN emission.

Our best-fitting model has a constant outer SiN abundance relative to \h2 of $1.5\e{-7}$, from $6\e{15}$~cm to $2\e{16}$~cm, and a power-law distribution in the inner part, starting from an inner radius of $1.5\e{15}$~cm. This distribution is plotted in Fig.~\ref{sin-modelling}b, where we also show the \h2 number density over the same region, including the aforementioned overdensity. As can be seen from Fig.~\ref{sin-modelling}b, the extended peak of the SiN abundance spans the region of the \h2 overdensity. This further supports the idea that both phenomena have a common cause, which we postulate is the periastron passage of the AGB and F9 stars. 
The line profiles generated by the best fitting models are plotted with the spectra in Fig.~\ref{sin-modelling}c.


\subsection{Chemical modelling}\label{chemmethods}
\edits{The recent results of Van de Sande and Millar \cite{Van-de-Sande2022} focus on the effect of close companions on the circumstellar chemistry. In Fig.~\ref{fig:chemsin} we reproduce their results for stars with similar wind density to W~Aql [Model: $\dot{M}/\upsilon_\infty = 2\e{-6}~\spy/(\kms)$ compared with W~Aql: $\dot{M}/\upsilon_\infty = 1.8\e{-6}~\spy/(\kms)$], showing the predicted abundances of SiN, SiC and NS for both oxygen- and carbon-rich outflows, with and without an F9-like companion. The companion is approximated by a black body at 6000~K and does not explicitly include chromospheric UV photons. However, observations of W~Aql with GALEX in 2006 reveal a detection in the near UV (22.16~mag, 1771--2831~\AA) but not in the far UV ($>22.5$~mag, 1344--1786 \AA) \cite{Montez2017}, the latter being more important for breaking molecular bonds. If additional chromospheric UV flux is generated around periastron, as has been suggested for other types of stars in close binary systems \cite{Schrijver1987,Galvez2002}, then this would mainly serve to increase the products of UV photochemistry, such as \ce{Si+}, which are discussed below. An excessively large UV excess during periastron could possibly destroy a larger variety of molecular species than predicted, but this would occur over a relatively short timescale (see Table~\ref{tab:solutions} and Methods \ref{sec:solution}) and would not preclude further chemical interactions, including many of the formation channels discussed below, once the stars moved further apart. }

\subsubsection{SiN and SiC}\label{chem:sin}

\edits{The chemical models \cite{Van-de-Sande2022} show that,}
in the absence of a companion,
the SiN radical is expected to form in a shell-like distribution, with a peak abundance at a radius of around $10^{16}$~cm from the AGB star \edits{(Fig.~\ref{fig:chemsin})}.  The main formation pathway of SiN is via the measured reaction
\begin{equation}\label{sinform1}
\ce{NH3 + Si+ \to SiNH2+ + H}
\end{equation}
where NH$_3$ is assumed to be a parent species that is formed close to the AGB star and, \editstwo{through observations, has been} found to have a peak abundance of $\sim2\e{-5}$ relative to \h2 \cite{Danilovich2014}.
This is followed by dissociative recombination
\begin{equation}\label{sinform2}
\ce{SiNH2+ + e- \to SiN + H2}
\end{equation}
The main source of Si$^+$ is the photodissociation of SiS, i.e.
\begin{equation}\label{sisphot}
\begin{aligned}
\ce{SiS + h\nu &\to Si + S}\\
\ce{Si + h\nu &\to Si+ + e-}
\end{aligned}
\end{equation}
which occurs very readily in the presence of the F9 companion \edits{(see Fig.~\ref{fig:chemnsi})} and the UV photons it emits \cite{Van-de-Sande2022}; and is confirmed in our observations (Fig.~\ref{othermol}), because SiS is noticeably depleted to the southwest at the present position of the F9 star.
\edits{We also note that there are minor formation pathways for SiN forming from HNSi and SiC, but both pathways also depend on \ce{NH3} and \ce{Si+} and hence are also affected by UV photons driving the formation of \ce{Si+}.}

In the chemical models (\cite{Van-de-Sande2022} and \edits{Fig.~\ref{fig:chemsin})}, the main difference in the SiN abundance distributions between oxygen- and carbon-rich stars with the same wind density and no companion, is that the peak relative abundance of SiN is predicted to be $\sim10^{-8}$ for the oxygen-rich star and $\sim10^{-7}$ for the carbon-rich star. W~Aql is an S-type star whose chemistry is presumed to be intermediate between the typical carbon-rich and oxygen-rich stars \cite{Danilovich2014}, 
and that is what has been found for the abundances of HCN in S-type stars \cite{Schoier2013}. However we find that the peak abundance of SiN in W~Aql ($1.5\e{-7}$ relative to \h2, see Methods \ref{rtmod}) is in good agreement with that predicted for a carbon-rich star, although the asymmetric distribution of SiN implies that the formation process is anisotropic.

{{Van de Sande and Millar's study \cite{Van-de-Sande2022} focused on}} the impact of UV photons from stellar companions  
on the circumstellar chemistry of AGB stars. 
They include a set of models with a main sequence companion with a stellar effective temperature of 6000~K that is comparable to
the temperature of an F9 dwarf \cite{Gray2009} \edits{(reproduced in Fig.~\ref{fig:chemsin})}. 
The radial abundance profile of SiN is significantly altered by the companion
--- i.e.,  the peak abundance of SiN in both the carbon- and oxygen-rich winds is higher, and  the abundance of SiN in the inner part 
of the wind is also higher. 
For the oxygen-rich outflow, the inner abundance of SiN is higher at $\sim10^{-7}$, and it remains relatively constant until it begins 
to decrease at around $10^{16}$~cm; SiN does not exhibit a shell-like distribution, as it would in the absence of a companion, 
but rather a parent-like distribution with a high inner abundance followed by a Gaussian decline caused by photodissociation driven by the interstellar radiation field. 
For the carbon-rich outflow, a shell-like distribution is seen in the presence of the companion, but the peak abundance is higher ($\sim10^{-6}$) 
and the inner abundance of SiN is several orders of magnitude higher ($\sim2\times 10^{-9}$ relative to \h2), than it would be if the companion were not present. 
An underlying assumption in these models is that the companion star is always close to the AGB star \cite{Van-de-Sande2022}.
However, this is not the case for W~Aql, as the projected distance between the F9 and AGB stars is presently 194~au or $2.9\e{15}$~cm, rather than $2-5R_\star$ ($4-10\e{13}$~cm) as assumed in the chemical models \cite{Van-de-Sande2022}. 
A highly elliptical orbit, during which the F9 star passes within a few stellar radii of the AGB star, could result in the asymmetric 
emission by SiN that we see in Fig. \ref{sin}, if the F9 star only drove the production of SiN while it was sufficiently close to the AGB star. 
In this instance, the temporary close proximity of the two stars is relevant, because the wind region close to the AGB star is the densest 
and the chemical reactions will occur more readily. 
For example, at 5~au from the AGB star, the \h2 number density is $3\e{9}$~cm$^{-3}$, whereas at the current projected distance 
of the F9 star, the number density is four orders of magnitude smaller, at $3\e{5}$~cm$^{-3}$. 
Because the rates of chemical reactions generally depend on (the square of the) number density, a lower number density 
results in a corresponding decrease in reaction rates, and hence much lower SiN production. \edits{Even very fast periastron interactions (Table \ref{tab:solutions}) are still long enough to produce SiN, particularly as, for example, the photoionisation of Si to \ce{Si+} (Eq. \ref{sisphot}) proceeds very quickly in the presence of the companion.}

\edits{Once formed, we expect SiN to persist in the expanding circumstellar envelope until it is photodissociated by the interstellar radiation field, based on chemical modelling \cite{Van-de-Sande2022} and because it is not expected to participate in the formation of dust or other molecular species. In general, the photodestruction timescale of a molecule being dissociated by the interstellar radiation field depends on the photodissociation rate for that molecule and on the extinction, with higher extinctions meaning that fewer photons will penetrate to that region. This is taken into account in the chemical models and accounts for the drop off in abundance in the outer regions of the CSE (Fig.~\ref{fig:chemsin}), which, for SiN, agrees with the location of the drop off we found from radiative transfer modelling (Methods \ref{rtmod}). The additional UV photons originating from the F9 star only have a relatively local effect on the chemistry of the CSE; as discussed in Methods \ref{sec:othermolecules} and \ref{sec:F9spec} and shown in Figures \ref{othermol} and \ref{F9spec}, the F9 star contributes to photodissociation of molecules in its vicinity, but not on the opposite side of the CSE.}




SiC behaves in a broadly similar way to SiN in the chemical models, with and without the inclusion of a main sequence companion
(\cite{Van-de-Sande2022} \edits{and the middle panel of Fig.~\ref{fig:chemsin}}). 
For both carbon- and oxygen-rich CSEs without a companion, SiC is expected to be distributed in a shell around the star, 
albeit with a more shallow gradient on either side of the peak than for SiN. 
For the carbon-rich star with a density similar to W~Aql, the peak abundance of SiC is located at $\sim10^{16}$~cm from the AGB star 
and is found to be $\sim10^{-6}$ relative to \h2, while for the oxygen-rich CSE, the peak of $\sim 5\e{-9}$ is found slightly farther 
from the star at $\sim 3\e{16}$~cm. 
The presence of an F9-like companion alters the SiC distribution in a similar way as for SiN, changing it from a shell-like distribution to a more centralised distribution. The abundance in the inner part 
of the distribution (i.e., in the region from the inner edge of the model to $\sim10^{16}$~cm) increases up to $\sim 2\e{-5}$ 
for carbon-rich CSE; 
and $\sim 5\times 10^{-9}$ for the oxygen-rich CSE, where there is previously negligible SiC in this region \edits{without a companion (Fig.~\ref{fig:chemsin})}.

Analogous with SiN (Eqs. \ref{sinform1} and \ref{sinform2}), SiC mainly forms via 
\begin{equation}
\begin{aligned}
\ce{CH3 + Si+ &\to SiCH2+ + H}\\
\ce{SiCH2+ + e- &\to SiC + H2}
\end{aligned}
\end{equation}
with the same source of \ce{Si+} as explained in Eq. \ref{sisphot}.
\ce{CH3} is formed either via photodissociation of \ce{CH4}, or through the successive hydrogenation of C. The former pathway is dominant for carbon-rich CSEs, while the latter is more likely in oxygen-rich CSEs. For an S-type star such as W~Aql, both pathways may contribute to \ce{CH3} formation.


\edits{The formation of both SiN and SiC is driven by \ce{Si+}, which forms through the photoionisation of Si (Eq.~\ref{sisphot}). In Fig.~\ref{fig:chemnsi} we plot the predicted abundances of \ce{Si+} with and without the presence of the F9 companion \cite{Van-de-Sande2022}. While the abundance of \ce{Si+} naturally rises in the outer part of the envelope (beyond $\sim 10^{16}$~cm), owing to the interstellar radiation field, the inner abundance rises significantly in the presence of an F9-like companion.} \editstwo{We note that although the abundance of \ce{Si+} rises to $10^{-9}$ to $10^{-7}$, for oxygen- and carbon-rich CSEs, respectively, this is still significantly less than the total abundance of Si ($6.5\e{-5}$ relative to \h2, assuming solar elemental abundances \cite{Asplund2021}), meaning that the photoionisation process driven by the F9 star is not expected to ionise all the Si.}

\subsubsection{NS}\label{nschem}

In the absence of a companion, NS is expected to form in shell with a peak at about $10^{16}$~cm \cite{Van-de-Sande2022}. 
For a carbon-rich CSE, the addition of an F9 companion does not cause a significant difference in the NS distribution.
For an oxygen-rich CSE, however, the chemical model with an F9 like companion predicts a higher abundance of NS by 
almost an order of magnitude and significantly changes the shape of the distribution, resulting in a high abundance of NS  
in the inner wind ($\sim10^{-6}$ which decreases from around $5\e{15}$~cm).

NS is formed via the photodissociation of \ce{N2} \cite{Van-de-Sande2022}
\begin{equation}\label{nsform}
\begin{aligned}
\ce{N2 + h\nu &\to N + N}\\
\ce{N + HS &\to NS + H}
\end{aligned}
\end{equation}
Even though the rate of photodissociation of \ce{N2} is relatively low because of the strong bond, the abundance of \ce{N2} is thought to be high ($4\e{-5}$ relative to \h2 \cite{Agundez2020}). Therefore, even if only $\leq 1\%$ of \ce{N2} is destroyed, enough N is liberated to form NS \cite{Van-de-Sande2022}.  \edits{The predicted abundance distribution of N, taking into account the presence of an F9-like companion, is plotted in Fig.~\ref{fig:chemnsi}.} The detection of NS is tentative (Fig.~\ref{nsplot}), but its co-location with the brightest region of SiN (especially in the PV diagram) and the predictions of chemical models that include an F9-like companion (\cite{Van-de-Sande2022} \edits{and Fig.~\ref{fig:chemsin})}, suggest that NS was likely formed during the periastron passage of the W~Aql system, when the F9 star irradiated part of the inner wind.

\subsubsection{HCN, CN and HC$_3$N}\label{HCNchem}

HCN, CN and \ce{HC3N} are closely linked species which have a wide astronomical literature in the context of the cyanopolyyne 
(H--(C$\equiv$C)$_n$--C$\equiv$N) family of molecules.
HCN is a parent species formed close to the star \cite{Agundez2020}, and CN has long been known to be a photodissociation 
product of HCN \cite{Huggins1984}. At low temperatures  (below 800~K \cite{Agundez2017}), where \ce{HC3N} is seen towards W~Aql, the main formation pathway for \ce{HC3N} is from the two parent species HCN and \ce{C2H2} \cite{Agundez2017,Cordiner2009}:
\begin{equation}
\begin{aligned}
\ce{HCN +h\nu &\to CN + H}\\
\ce{CN + C2H2 &\to HC3N + H}
\end{aligned}
\end{equation}

For most molecular species, chemical fractionation of isotopologues is expected to be negligible around AGB stars. Hence, we can use the observations of \ce{H^13CN} and \ce{^13CN} to understand the formation of \ce{H^12C3N}. For the rest of this section, we omit the isotope labels. As can be seen in Fig.~\ref{fig:cn}, CN is preferentially detected on the side of the CSE where the F9 star is presently located, coinciding with a region of HCN depletion. We refer to this phenomenon as depletion because it aligns well with the location of the F9 star and of CN, and because the extent of HCN to the north east agrees well with the predicted extent of HCN in the chemical models, in the absence of a companion \cite{Van-de-Sande2022} (see also discussion in \ref{otherchem} below). Although the F9 star passes close to the AGB star during the eccentric orbit, the amount of time the stars spend close together is relatively short, $\lesssim 2\%$ of the orbital period (Table \ref{tab:solutions} and Methods \ref{sec:solution}), compared with the amount of time the F9 star spends to the south west of the AGB star, providing a relatively consistent source of UV radiation. A similar pattern of molecular depletion is seen for SiO, SiS and CS (Fig.~\ref{fig:cn}) for the same reason. 

HC$_3$N is present on the same side of the CSE as CN \editstwo{(Figs.~\ref{fig:cn} and \ref{hc3n})}, from which we can infer that the presence of CN preferentially drives 
the formation of HC$_3$N to the south and west of the AGB star.
Although HC$_3$N has long been known to be present around carbon stars, W~Aql is the first S-type AGB star towards 
which HC$_3$N has been observed. \edits{Although HC$_3$N and other carbon-bearing molecules such as C$_2$H and SiC$_2$ seem to indicate a carbon-rich circumstellar chemistry for W~Aql \cite{De-Beck2020}, the spectroscopic classification of W~Aql marks it as an S-type star \cite{Danilovich2015}. It is possible such carbon-bearing species are common around (some subset of) S-type stars more generally, but, to date, W~Aql has been studied in the most detail.}

\ce{HC3N} has been most widely studied around the nearby carbon star CW~Leo, where it is located mainly in a spherical shell centred on the star, well-resolved in ALMA observations and as predicted by chemical models \cite{Agundez2017,Van-de-Sande2022}, with some enhancement in the inner regions which is thought to be driven by a companion \cite{Siebert2022}. 
We do not see a symmetric shell-like distribution of HC$_3$N around W~Aql (Fig.~\ref{hc3n}), however we interpret the \ce{HC3N} 
that we observe as part of a broken shell that is formed where CN is more abundant.

Although we expect that some CN --- and subsequently \hc3n --- would have formed during, or as a result of the periastron passage 
of the W~Aql binary, these two molecules will have expanded with the CSE (as SiN has), to a radius that is comparable to the black circle in CO 
($1.35\arcsec$, Fig.~\ref{coarcs}). 
At this distance from the AGB star, some CN might remain but is not \editstwo{easily} detectable above the noise in our observations. 
\editstwo{Some traces of \ce{^13CN} are seen north of the AGB star in Fig.~\ref{fig:cn}, but the SNR of the \ce{^13CN} observation is relatively low, partly because more than half of the flux was resolved out (Table \ref{resolutions}). We also note that \ce{^12CN}, expected to be around 10 to 30 times more abundant \cite{Danilovich2014}, was not covered by our observations. Hence, we cannot conclusively determine how much CN is present to the north east of the AGB star, relative to the apparently higher abundance of CN to the south west, closer to the F9 star. More sensitive observations, ideally covering \ce{^12CN} and not subject to resolved out flux, would be required to fully understand the distribution of CN around W~Aql.}
\editstwo{We note the CN we expect to be co-located with SiN, which should have formed during the periastron interaction, is harder to detect than SiN is, for several reasons relating to the molecular physics and energy level distributions of the two molecules. Although SiN is also subject to hyperfine splitting, the three most intense hyperfine components are only separated by $\sim$1.4~MHz, a tiny separation compared with the $30~\kms$ (22~MHz) width of the SiN line, as can be seen in Fig.~\ref{lostflux}(b). In comparison, the spectrally resolved hyperfine splitting of CN results in especially wide lines which have lower peak intensities than they would in the absence of hyperfine splitting. This makes them harder to detect above the noise. Furthermore, the dipole moment of SiN is around 1.8 times larger than for \ce{^13CN} \cite{Kerkines2005,Thomson1968}, resulting in intrinsically brighter lines for SiN.}

The excitation conditions of the observed lines of \hc3n are such that we do not expect to see these same lines of \hc3n lines farther 
out in the wind than we do in Fig.~\ref{fig:cn} ($< 0.5\arcsec$). 
Therefore, if any \hc3n is present at a radius of $1.35\arcsec$ from the AGB star, we would not have detected it in the present observations. \editstwo{We predict that \ce{HC3N} in this region could be detected in more sensitive observations covering lower-energy transition lines.}

\subsubsection{SiO, SiS and CS}\label{otherchem}

The emission seen from SiO, SiS and CS (Fig. \ref{othermol}) --- like that of HCN --- indicates photodissociation driven by the F9 star 
during its time to the southwest of the AGB star, 
unlike SiN, SiC, and NS whose formation is driven by the brief but intense addition of UV photons from the F9 star to the inner CSE 
during the periastron passage. 
This process works because it is the products of photodissociation and photoionisation that go on to form the observed SiN, SiC and NS.
However, this is not the case for SiO, SiS, CS and HCN, which are considered to be parent species in most chemical models 
\cite{Van-de-Sande2022,Agundez2020}. 
Accordingly, the effect of a stellar companion is generally not to increase the abundances of these molecules in the inner 
CSE, \editstwo{but may potentially deplete them} \cite{Van-de-Sande2022}. 
Taking the case of a main sequence companion in the inner wind, the predictions are as follows:
(i) for SiO in an oxygen-rich CSE, a minimal decrease of the inner abundance is predicted, compared with a decrease of almost 
an order of magnitude for the carbon-rich CSE;
(ii) for SiS, the models predict a significant decrease of several orders of magnitude (4--6 dex) for both chemical types; however, 
this dramatic change could be the result of an uncertain photodissociation rate for this molecule;
(iii) for CS and HCN, the change in abundance for both chemical types in the presence of a stellar companion is negligible.
Hence we can conclude that the asymmetric distributions seen for SiO, SiS, CS and HCN (Fig.~\ref{othermol} \editstwo{and \ref{cssissinglechans}}) are caused by photodissociation from the F9 companion, rather than enhanced formation during periastron.


\subsection{Hydrodynamic simulations}\label{hydro}

To better understand the structure in the CO emission, we performed high-resolution 3D smoothed particle hydrodynamic (SPH, \cite{Monaghan2005,Price2012}) simulations of highly eccentric systems with parameters similar to the W~Aql system. 
These simulations were performed with the SPH code Phantom \cite{Price2018,Siess2022,Maes2021,Malfait2021}. The AGB star and companion are represented by gravity-only sink particles, and the wind consists of $\sim 7\e{6}$ SPH gas particles that are gradually launched from boundary shells around the AGB star, with a velocity of $13~\kms$, mimicking a free wind \edits{and a constant mass-loss rate \cite{Siess2022}}. Cooling within the wind is regulated by the equation of state for an ideal gas with polytropic index $\gamma =1.2$, and the pulsations and rotation of the AGB star are not taken into account. 
It is important to note that these hydrodynamic simulations are necessarily simplified compared to observations, as they mainly account for the gravitational impact of the companion on the wind, and neglect the impact of additional effects such as radiation, \edits{radiation pressure,} pulsations, realistic cooling, and \editstwo{variable or anisotropic mass loss}. We also note that the free wind approach does not reproduce velocities lower than $13~\kms$, even though lower velocities are expected in the inner wind region (within $\sim80$~au of the AGB star). All of these contribute to the differences between the model and observations. Hence we aim for a qualitative understanding rather than a direct fit to the data.

We present results for a model with orbital parameters close to the W~Aql system, with eccentricity $e = 0.92$ and semi-major axis $a = 125$~au, and taking the masses of the W~Aql system (Methods \ref{dist-notes}).
The Phantom model was evolved for around 5000 years and the snapshot that we plot in various figures was selected from a time step a little earlier than this to better match the current positions of the two stars. From a detailed analysis of the Phantom model we found that the orbital period increased slightly with time, owing to the mass being lost by the AGB star. This corresponded to a small increase in the semimajor axis but no change in the eccentricity over the time of the simulation.
In Fig.~\ref{coarcs}c, we show the density distribution in a slice perpendicular to the orbital plane of this model.
Plots of the same model showing the inner regions and a slice through the orbital plane are given in Fig.~\ref{faceonSPH}.
In general, we expect the companion to generate a spiral-like structure in the wind \cite{Maes2021,Malfait2021,Mastrodemos1998,Mastrodemos1999}. However, owing to the high eccentricity of this system, concentric near-spherical density structures are created in the wind, visible as the near-circular structures in the edge-on density distribution in Figs.~\ref{coarcs}c and \ref{faceonSPH}b. The circular structures are not quite centred on either of the central stars but rather offset to the opposite side from the F9 position at apastron, similar to the offsets we find in the ALMA CO observations.
These structures are remarkably similar to the circular structures traced out by the black and white circles in Fig.~\ref{coarcs}a. 
The offset centres of the circles, particularly the outer circle, agree well with the observed ALMA data (white circle in Fig.~\ref{coarcs}a).
Similar structures at a $90\deg$ edge-on inclination were seen for other highly eccentric SPH simulations we performed, and are also seen in previous studies with $e=0.8$ \edits{and mass ratio $M_1/M_2= 2.75$, compared with 1.5 for the W Aql system} \cite{Kim2019}.

From a close study of our hydrodynamic simulations, we determine that the concentric circles are formed during the relatively quick periastron passage of the F9 star. During the periastron passage, the stars reach their maximal orbital velocity \edits{($\sim 17~\kms$ for our chosen orbital parameters) and move hypersonically through the wind (which has a sound speed of $\sim 3~\kms$ at 10~au), resulting in near-spherical shocks. The funnel-like structure (see Fig.~\ref{coarcs}c) is formed through gravitational interactions between the companion and the wind. More concretely, when the companion moves towards the AGB star shortly before the periastron passage, its gravitational force results in a high-density wake behind the companion (see first and second columns in Fig.~\ref{fig:peri}). Because there is a velocity dispersion within this wake, it is delimited by a radially faster outer edge and a denser inner edge. As the companion and the AGB star pass each other quickly during periastron passage, the inner edge is shaped as a circular high-density shock, that travels radially outwards and expands as the left side ($x<0$) of the 3D sphere-like structure. Because the wind-companion interaction around periastron passage is strong, the outer edge of the wake becomes a bow shock after periastron passage (second and third columns of Fig.~\ref{fig:peri}, \cite{Malfait2021}).
The formation of the spherical high-density shock is enhanced, and is completed on the right side ($x>0$), by the fast wobble of the AGB star. The orbital velocity of the stars reaches a maximum absolute value during this close encounter, however, the direction of the orbital velocity vectors changes by almost 180 degrees due to the elliptical nature of the orbit. The wobble of the AGB creates a strong gradient in the radial wind velocity (mainly of the material on the $x>0$ side of the AGB, where the wind is not disturbed by the companion shock wake).  The transition from faster outflowing material to slower wind particles results in a low-density region (around $x=40$--80 au in the right column of Fig.~\ref{fig:peri}). The edge between this low-density region and the inner denser material completes the spherical high-density shock (see the bottom row of Fig.~\ref{fig:peri}, showing the orbit with an inclination of $90\deg$).
The spherical structures are slightly offset because of the movement of the stars.}
From this, \edits{and the similar results of \cite{Kim2019} and \cite{Cernicharo2015a},} we can conclude that such circular structures are typical of highly elliptical systems, including when those systems are seen edge on.

\edits{We emphasise that the circular structures are a consequence of binary interaction and do not, in our model, represent a period of enhanced mass loss. This is in contrast with the simplified model of CW Leo \cite{Cernicharo2015a} where the increase in density was caused by an assumed increase in mass-loss rate during periastron, in addition to the wobble of the AGB star. \editstwo{Some discussion of the impact of anisotropic mass loss is given in the Supplementary Materials \ref{sec:anisotropies}.}
To illustrate the effect of our constant mass-loss rate, we extracted the number density of our model along the $x$-axis with $z=y=0$ and compared this with the 1D smooth model with an overdensity described in Methods \ref{rtmod} and \cite{Brunner2018}. In Fig.~\ref{fig:nh2}a we show the number densities from the hydrodynamic model along the positive and negative $x$ directions. Because the orbital parameters of our main hydrodynamic model do not precisely match the orbital parameters that we derive in this work, we performed an additional hydrodynamic model using the orbital solution discussed in Methods \ref{sec:solution} ($e=0.93, r_p=10$~au). To reduce the required computational resources, we set a large accretion radius for the F9 star (1~au compared with 0.05~au in our main model), which reduces the more complex (and computationally expensive) close gravitational interactions between the companion star and the wind particles. This eliminates the funnel-like structure seen on the right of Fig.~\ref{coarcs}c but retains the sphere-like structures resulting from the motions of the two stars. For this model, the same number density plot, Fig.~\ref{fig:nh2}b, reveals density peaks at radii in good agreement with the overdensity found by \cite{Brunner2018}. Note that, overall, the number density of the hydrodynamic models can be averaged to equal the number density of a smooth 1D model (without any overdensity). However, we also note that our main model, which better allows for the close gravitational interactions between the wind and the F9 star, results in a less symmetric distribution of over- and under-dense regions (as shown for the $x$-axis in Fig.~\ref{fig:nh2}a and seen in the funnel-like structure in Fig.~\ref{coarcs}c) and contributes to the large-scale asymmetries discussed below.}

Based on the circular structures formed during periastron, we can estimate the time of the most recent periastron from the expansion time of the black circle in Fig. \ref{coarcs}a and the orbital period from the difference in expansion times between the black and white circles. These calculations are outlined in Methods \ref{orbit-from-alma}. The fact that the black circle overlaps with the edge of the SiN emission (Fig. \ref{coarcs}b) also supports our hypothesis that the SiN was created during the most recent periastron passage.

To enable a better comparison of the SPH model to the observations, we processed the Phantom model with the radiative transfer code MCFOST \cite{Pinte2006,Pinte2009}, using the effective stellar temperatures of both the AGB (2300~K) and F9 (6000~K) stars and silicate dust from \cite{Weingartner2001}. The computation was sped up by only considering the lowest 6 CO levels since this was sufficient for the task at hand. MCFOST includes a routine for determining the photodissociation of CO by the interstellar UV field \cite{Pinte2018}, which we used to determine the drop off in CO distribution (set to $6\e{-4}$ relative to \h2 at the centre of the model), based on our 3D structures. This resulted in the near-complete photodissociation of CO in the outermost density structures and left only (parts of) the innermost four circular structures visible in CO. The resultant central velocity channel is plotted in Fig.~\ref{coarcs}d, rotated to match the orientation of the W~Aql system on the sky. Although the model is not a perfect reproduction of the observed CO emission (expected in light of the missing physics mentioned above), there are many qualitative similarities. We also extracted an angle-radius plot from the central channel of the MCFOST output (Fig.~\ref{mcfostangrad}), in which we see similar sinusoidal structures as those found in the observations (Figs.~\ref{angrad} \editstwo{and \ref{angrad-no-lines}}). 
The structures outlined by the pink and red circles identified in Fig.~\ref{coarcs}a are not apparent in the MCFOST output, although they do qualitatively resemble the structures formed at periastron. The main distinguishing feature is that the pink and red circles are offset in the opposite direction (south rather than north). If we were to ignore the offset and assume that one or both of these circles have the same origin as the black and white circles, we find that the period calculated between all the identified circles would be too short to agree with the HST and SPHERE observations of the stellar separations. Therefore, the red and pink circles cannot have formed during periastron.
Noting that the Phantom model overestimates the wind velocity in the inner regions, we suggest that the difference between the observed and modelled structures partially arises from this as well as the other missing physics mentioned at the start of this section.


We also examined the channel maps generated by MCFOST at high and low velocities and compared these with equivalent channels from the ALMA observations in Fig.~\ref{otherCOchans}. The observations are taken from channels $\pm 13~\kms$ from the LSR velocity of $\upsilon_\mathrm{LSR} = -23~\kms$. The blue channel exhibits CO emission elongated to the southwest, approximately along the companion axis, while the red channel has a more circular CO emitting region. These differences are qualitatively reproduced in their respective MCFOST channels. This asymmetry in velocity space is also responsible for the enhanced blue emission seen in the wings of several line profiles observed towards W~Aql \cite{Danilovich2014}. The asymmetry arises from the orientation of the orbital plane such that the observations are reproduced if the Phantom model is orientated so that motion of the F9 star at periastron is into the plane of the sky. 

Finally, we comment on the large-scale asymmetry to the southwest, revealed by past observations, in the CO \cite{Ramstedt2017} and dust \cite{Ramstedt2011,Mayer2013} emission on scales of $10\arcsec$ and $60\arcsec$. Although this more extended emission is in the same direction as the F9 star, the emission extent is much larger than the current or maximal separation between AGB and F9 stars ($\sim0.5$ to $0.8\arcsec$, Fig. \ref{orb-phot}). The luminosity of the F9 star is insufficient for its radiation to drive the dust outwards, as the AGB star does (Supplementary Materials \ref{dustdrive}); instead, it contributes to the large-scale shaping of the wind through its gravitational pull. We do not detect any accretion disc around the F9 star, either in the ALMA continuum or in any molecular lines, and an accretion disc is not predicted for the W~Aql system by the SPH model. However, the F9 star does gravitationally attract some circumstellar material, which is then pushed outwards by the radiation pressure from the AGB star, and results in the large scale asymmetry seen in the dust and more extended gas \cite{Ramstedt2011,Mayer2013,Ramstedt2017}, and reproduced in our hydrodynamic model. The enhanced emission in this direction can be seen in the full extent of the central CO channel output by MCFOST (Fig. \ref{faceonSPH}c), where the CO extends out farther to the southwest.

\subsection{Orbital parameters from ALMA observations}\label{orbit-from-alma}

Here we constrain some orbital parameters from the ALMA observations. First we make an estimate of the period based on the round structures seen in the CO observations. As determined in Methods \ref{hydro}, the black and white circles shown in Fig.~\ref{coarcs} were created during periastron interactions between the AGB and F9 stars. Assuming the velocity profile from Eq.~\ref{betalaw}, we find the expansion time between the two circles, and hence the orbital period, through the integral:
\begin{equation}\label{exptime}
T = \int_{R_\mathrm{black}}^{R_\mathrm{white}} \frac{dr}{\upsilon(r)} = \int_{R_\mathrm{black}}^{R_\mathrm{white}} \frac{dr}{\upsilon_0 + (\upsilon_\infty - \upsilon_0) \left(1 - \frac{R_\mathrm{in}}{r}\right)^\beta} 
\end{equation}
where $R_\mathrm{black}$ and $R_\mathrm{white}$ are the radii of the black and white circles, and $R_\mathrm{in} = 2\e{14}$~cm is the dust condensation radius, with $\upsilon_0 = 3~\kms$ the velocity for $r<R_\mathrm{in}$, taken to be the sound speed \cite{Danilovich2014}. The period is found to be $1082^{+89}_{-108}$~years. The uncertainty is based on the width of the circles as fit from the angle-radius plot (Fig.~\ref{angrad}). There we found the uncertainties in the radii of the circles to be $10.75\pm0.75\arcsec$ for the white circle and $1.35\pm0.10\arcsec$ for the black circle.

Another crucial parameter needed to constrain the orbital solution of the W~Aql system is the time since periastron. As previously discussed, the most recent periastron passage generated the black circle seen in CO (Fig.~\ref{coarcs}) and the arc of SiN (Fig.~\ref{sin}). We can estimate the time of periastron by calculating the expansion time of these two structures. Since we are now considering expansion in the inner part of the envelope, 
we need to also consider the velocity inside the dust condensation radius, which we assume to be close to the sound speed at $\upsilon=3~\kms$. Equation \ref{exptime} can then be rewritten:
\begin{equation}\label{timesinceperi}
\Delta t  = \int_{R_\mathrm{in}}^{R_\mathrm{black}} \frac{dr}{\upsilon_0 + (\upsilon_\infty - \upsilon_0) \left(1 - \frac{R_\mathrm{in}}{r}\right)^\beta} + \int_{R_\mathrm{form}}^{R_\mathrm{in}} \frac{dr}{\upsilon_0} 
\end{equation}
where $R_\mathrm{black}$ is the radius of the black circle and the radial extent of the SiN arc, and $R_\mathrm{form}$ is the radial distance at which these two structures formed.

The value of $R_\mathrm{form}$ is uncertain so we take it to be the periastron distance between stars. The smallest periastron distance we obtain is $\sim3$~au, while the largest is equal to the dust condensation radius. Using these values as a guide and assuming a constant velocity of $\upsilon_0 = 3~\kms$ for $r<R_{in}=2\e{14}$~cm, we estimate the time since the most recent periastron as $172\pm22$~years ago.
These derived values are listed with other orbital parameters in Table \ref{tab:parameters}.

%

Finally, we can determine the direction of the orbit from the PV diagrams of the species formed at periastron, namely SiN, SiC, and NS. Taking into account that 1) the redder emission is brighter for all three of these molecules (and indeed only red emission is seen above the noise in the NS PV diagram, Fig. \ref{nsplot}) and 2) the line profiles of SiN and SiC are slightly blue-shifted relative to the stellar LSR velocity, suggests that these species formed first on the blue side of the envelope and then more recently on the red side. Hence, there has been slightly more time for the blue emission to expand, shifting the line profiles and PV diagrams bluewards. From this we conclude that the direction of the periastron passage was, for the F9 star, into the plane of the sky. This agrees with the evidence from the SPH model discussed above.

\subsection{Orbital solutions}\label{sec:solution}

The orbit of the W~Aql system cannot be solved analytically, so instead we solve it numerically by calculating a series of possible orbits and checking which agree with the parameters derived from observations (i.e. the parameters listed in Table \ref{tab:parameters}).
We adjust our basic orbital solution by leaving as free parameters the eccentricity, $e$, and the periastron distance, $r_p$. All other primary orbital parameters are either input from prior results or calculated from $e$ and $r_p$ as follows.

The apastron, $r_a$, is defined by
\begin{equation}
r_a = r_p\left(\frac{e+1}{1-e}\right)
\end{equation}
and the semimajor axis, $a$, is then
\begin{equation}
a = \frac{r_p+r_a}{2}\,.
\end{equation}
Working in the reference frame of the AGB star, we define the focus of the ellipse traced by the F9 star as the location of the AGB star, defined here as (0,0,0) in our cartesian co-ordinate scheme.


From the system mass ($M+m = 2.66~\msol$) and the semimajor axis, we can then calculate the orbital period, $T$
\begin{equation}
T = 2\pi \sqrt{\left(\frac{a^3}{G(M+m)}\right)} \,.
\end{equation}

This is enough information to plot a top-down view of the orbit, as shown in Fig. \ref{schematic}. However, we know from observations that the orbit is inclined and rotated in the plane of the sky (relative to north). From the observations of SiN, we estimate the inclination angle of the orbit to be close to edge-on, $i=90\pm7\deg$. We plot $i=85\deg$ to better illustrate the orbit in the plane of the sky, but note that a completely edge-on system ($i=90\deg$) satisfies the observations and does not significantly change our results.
From the photometry of the two stars, we rotate the orbit in the plane of the sky by $\omega=120\deg$ to fit the SPHERE observation (Fig.~\ref{orb-phot}). We note that the uncertainty in $\omega$ comes mainly from the precise values of the inclination and eccentricity, but the selection of $\omega=120\deg$ is a good fit given the rest of our results. The sky projection of a selected orbit and the locations of the stars are plotted in Fig. \ref{orb-phot}. We assume no rotation out of the plane of the sky along the third orthogonal axis because the relative symmetry of the SiN PV-diagram (Fig. \ref{sin}b) suggests this value is small ($<5\deg$).

For a possible orbital solution, we must calculate the time since periastron and the time between the SPHERE and HST observations. For this we must consider the angle $\theta$ made between the periastron, the AGB star and the F9 star, as well as the eccentric anomaly, $E$. Both of these angles are shown in Fig. \ref{schematic} and are mathematically related by
\begin{eqnarray}
\tan{\left(\frac{\theta}{2}\right)} &=& \tan{\left(\frac{E}{2}\right)}\sqrt{\frac{1+e}{1-e}}\label{eqn:theta-E}\\
E &=& 2\tan^{-1}\left(\tan\left(\frac{\theta}{2}\right) \sqrt{\frac{1-e}{1+e}}\right) \quad .\label{eqn:E}
\end{eqnarray}
The time since periastron, $\Delta t$, can then be calculated
\begin{equation}
\Delta t = \frac{T}{2\pi} \left(E- e\sin(E)\right).\label{eqn:deltat}
\end{equation}
We also check the possible solution against the known time between the HST and SPHERE observations by comparing $\Delta t_\mathrm{SPHERE} - \Delta t_\mathrm{HST}$ against the time difference between those observations. 

To find the best solutions, we modify the input parameters ($r_p$ and $e$) until we find a suitable orbit which agrees with the values we found for the period, time since periastron and time between HST and SPHERE observations. Because of the uncertainties, we find a group of compatible solutions rather than one single definition of the orbit. 
From a grid with steps of $\Delta e = 0.01 \in [0.70, 0.99]$  and $\Delta r_p = 0.1\e{14}~\mathrm{cm} \in [0.4\e{14}~\mathrm{cm}, 5\e{14}~\mathrm{cm}]$, we found a set of compatible solutions, all of which are given in Table \ref{tab:solutions}. For the highest eccentricities $e>0.95$ we additionally tested a finer grid for $r_p$, with $\Delta r_p = 0.5\e{13}~\mathrm{cm} \in [5\e{13}~\mathrm{cm}, 1\e{14}~\mathrm{cm}]$, because the orbital timing becomes sensitive to small variations in $r_p$ at these high eccentricities.
The compatible solutions range from the extremes of $e=0.98,\,r_p = 4.5\e{13}$~cm to $e=0.91,\,r_p = 2.0\e{14}$~cm. We plot one of these solutions ($e=0.93,\,r_p = 1.5\e{14}$~cm) in Fig.~\ref{orb-phot}, where we also show the orbit superposed on the HST and SPHERE photometric observations. Note that although some of our solutions have very small periastron distances, none are smaller than the Roche limit, so direct accretion of the AGB star onto the F9 star is not expected.

In Table \ref{tab:solutions}, we also include $t_\mathrm{close}$ which we define as the time the AGB and F9 stars spend ``close'' to each other. More precisely, in the AGB frame, this is the time the F9 star takes to pass through the $-90\deg \leq \theta \leq 90\deg$ region of the orbit (see Fig.~\ref{schematic}) and can be derived from equations \ref{eqn:E} and \ref{eqn:deltat}. As noted in Table \ref{tab:solutions}, $t_\mathrm{close}$ ranges from 2 years at the highest eccentricity to 18 years at $e=0.91$. This corresponds to $\sim 0.1$ to $2\%$ of the total orbital period.




%
%

\section*{Data Availability}

The observational data used here is openly available through the data archives for ALMA (\url{https://almascience.nrao.edu/aq/}), ESO for the APEX and SPHERE data (\url{http://archive.eso.org}), and HST (\url{https://hla.stsci.edu}). Custom ALMA data products are available from TD or AMSR upon reasonable request.

\section*{Code Availability}

Phantom is open source under the GPLv3 license and can be downloaded via \url{https://github.com/danieljprice/phantom}.
MCFOST is open source under the GPLv3 license and can be downloaded via \url{https://mcfost.readthedocs.io/en/latest/overview.html}.
ALI, the 1D radiative transfer code, is available from TD upon reasonable request.

\bibliographystyle{naturemag} 
\bibliography{../master} 

\section*{}
Correspondence and requests for materials should be addressed to Ta\"issa Danilovich.

\section*{Acknowledgements}

We would like to thank Se-Hyung Cho of the Korean VLBI Network for KVN observations of W Aql to confirm consistency with our ALMA results.
TD is supported in part by the Australian Research Council through a Discovery Early Career Researcher Award (DE230100183). TD, FDC and SHJW acknowledge support from the Research Foundation Flanders (FWO) through grants 12N9920N, 1253223N and 1285221N, respectively. 
JM and SM acknowledge support from the Research Foundation Flanders (FWO) grant G099720N.
MVdS acknowledges support from European Union's Horizon 2020 research and innovation programme under the Marie Sk\l odowska-Curie grant agreement No 882991. 
MM acknowledges funding form the Programme Paris Region fellowship supported by the R\'egion Ile-de-France.
PK acknowledges funding from the European Research Council (ERC) under the European Union's Horizon 2020 research and innovation program (synergy grant project UniverScale, grant agreement 951549).
TJM is grateful to the STFC for support through grant ST/P000312/1 and thanks the Leverhulme Trust for the award of an Emeritus Fellowship. JMCP was supported by STFC grant number ST/T000287/1. LD, JMCP, SHJW, SM and DG acknowledge support from ERC consolidator grant 646758 AEROSOL. EDB acknowledges support from the Swedish National Space Agency. DG was funded by the project grant `The Origin and Fate of Dust in Our Universe' from the Knut and Alice Wallenberg Foundation.
KTW acknowledges support from the European Research Council (ERC) under the European Union's Horizon 2020 research and innovation programme (Grant agreement no. 883867, project EXWINGS).
FH, AB and LM acknowledge funding from the French National Research Agency (ANR) project PEPPER (ANR-20-CE31- 0002).
HSPM acknowledges support by the Deutsche Forschungsgemeinschaft through the collaborative research grant SFB 956 (project ID 184018867).
RS's contribution to the research described here was carried out at the Jet Propulsion Laboratory, California Institute of Technology, under a contract with NASA, and funded in part by NASA via ADAP awards, and multiple HST GO awards from the Space Telescope Science Institute. 
AZ is funded by  STFC/UKRI through grant ST/T000414/1.
This research was supported in part by the Australian Research Council Centre of Excellence for All Sky Astrophysics in 3 Dimensions (ASTRO 3D), through project number CE170100013.
This project has received funding from the Framework Program for Research and Innovation ``Horizon 2020" under the convention Marie Sk\l odowska-Curie No 945298.
	Computational resources and services used in this work were provided by the VSC (Flemish Supercomputer Center), funded by the Research Foundation Flanders (FWO) and the Flemish Government -- department EWI.
	This research was undertaken with the assistance of resources and services from the National Computational Infrastructure (NCI), which is supported by the Australian Government.
      This paper makes use of the following ALMA data: ADS/JAO.ALMA\#2018.1.00659.L. ALMA is a partnership of ESO (representing its member states), NSF (USA) and NINS (Japan), together with NRC (Canada), MOST and ASIAA (Taiwan), and KASI (Republic of Korea), in cooperation with the Republic of Chile. The Joint ALMA Observatory is operated by ESO, AUI/NRAO and NAOJ.
      Based on observations collected at the European Organisation for Astronomical Research in the Southern Hemisphere under ESO programme 0103.D-0772(A).
      We acknowledge excellent support from the UK ALMA Regional Centre (UK ARC), which is hosted by the Jodrell Bank Centre for Astrophysics (JBCA) at the University of Manchester. The UK ARC Node is supported by STFC Grant ST/P000827/1.

\section*{Author contributions}

TD conceived of and led this publication, analysed and interpreted data, performed the radiative transfer models, wrote the manuscript, and created most of the figures. JM performed and interpreted the hydrodynamics model and made figures \ref{coarcs}c, \ref{fig:peri}, and \ref{faceonSPH}a\&b. MVdS led the chemical interpretation and made figures \ref{fig:chemsin} and \ref{fig:chemnsi}. MM and PK contributed the analysis of the resolved imagining. AMSR performed the ALMA data reduction.
FDC and AC contributed to the 3D interpretation of the data. TJM and JMCP contributed to the chemical interpretation. CAG contributed to the line identifications and interpretation. CP assisted in the MCFOST modelling. DJP assisted in the Phantom modelling and interpretation. EDB contributed the fully-reduced APEX data.
The ALMA proposal was led by LD and CAG, with contributions from MM, TD, AdK, KMM, RS, AMSR, JMCP, HSPM, EDB, PK, AB, KTW, MVdS, EL, DG, JY and DJP.
All authors commented on the manuscript and analysis.

\appendix

\renewcommand{\thefigure}{\Alph{section}.\arabic{figure}}
\renewcommand{\thetable}{\Alph{section}.\arabic{table}}
\newpage

\section{Extended Data}\label{ed}

\begin{figure}[ht]
\begin{center}
\includegraphics[width=0.49\textwidth]{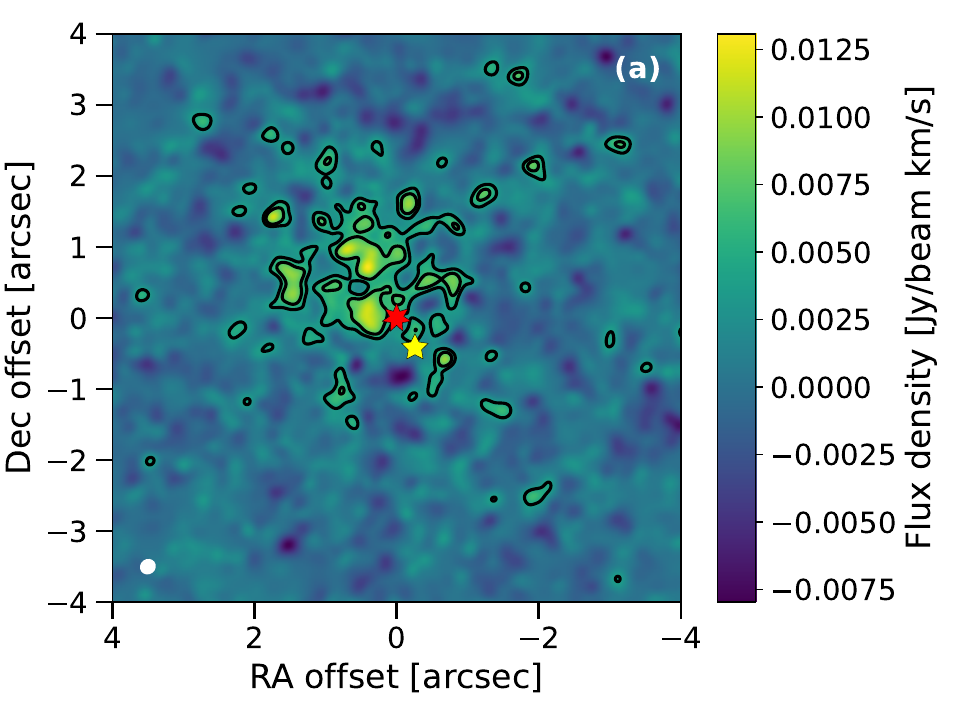}
\includegraphics[width=0.49\textwidth]{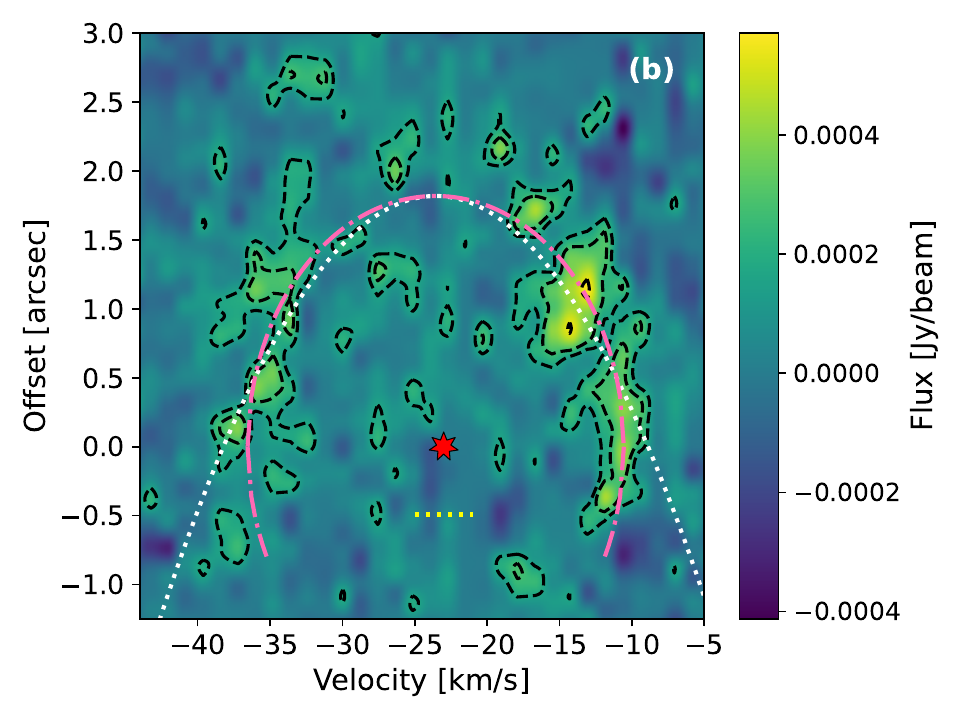}
\caption{\textbf{(a)} Zeroth moment map of SiC towards W~Aql with contours at levels of 3 and $5\sigma$. Transition details are given in Table \ref{resolutions}. North is up and east is to the left. The position of the AGB star is indicated by the red star at (0,0) and the location of the F9 companion is indicated by the yellow star to the south-west. North is up and east is left. The white ellipse in the bottom left corner indicates the size of the synthesised beam. \textbf{(b)} Position-velocity diagram of SiC towards W Aql, taken with the same wide slit as used for SiN (Fig. \ref{sin}), with a position angle of north $33\deg$ east. Dashed black contours are at levels of 3 and $5\sigma$, a dotted white parabola is fit to the data (see Methods \ref{sinsic}), and a dash-dotted pink ellipse is plotted to emphasise the shape of the emission in the PV diagram. The position and LSR velocity of the AGB star is indicated by the red star and the horizontal yellow dotted line indicates the present offset of the F9 star.}
\label{sic}
\end{center}
\end{figure}

\begin{figure}[t]
\begin{center}
\includegraphics[width=0.49\textwidth]{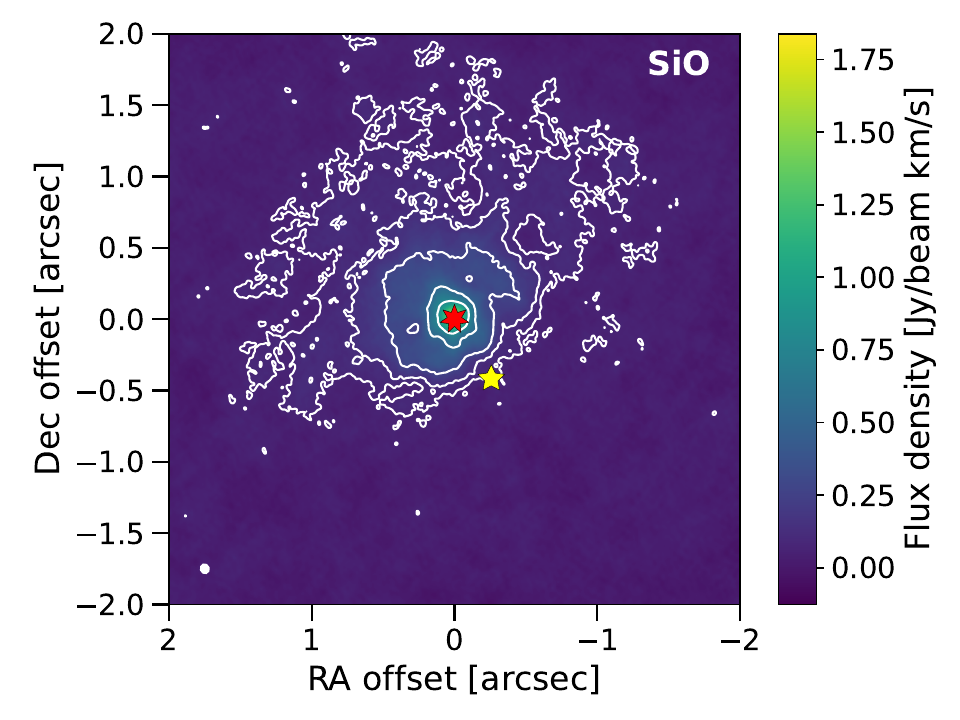}
\includegraphics[width=0.49\textwidth]{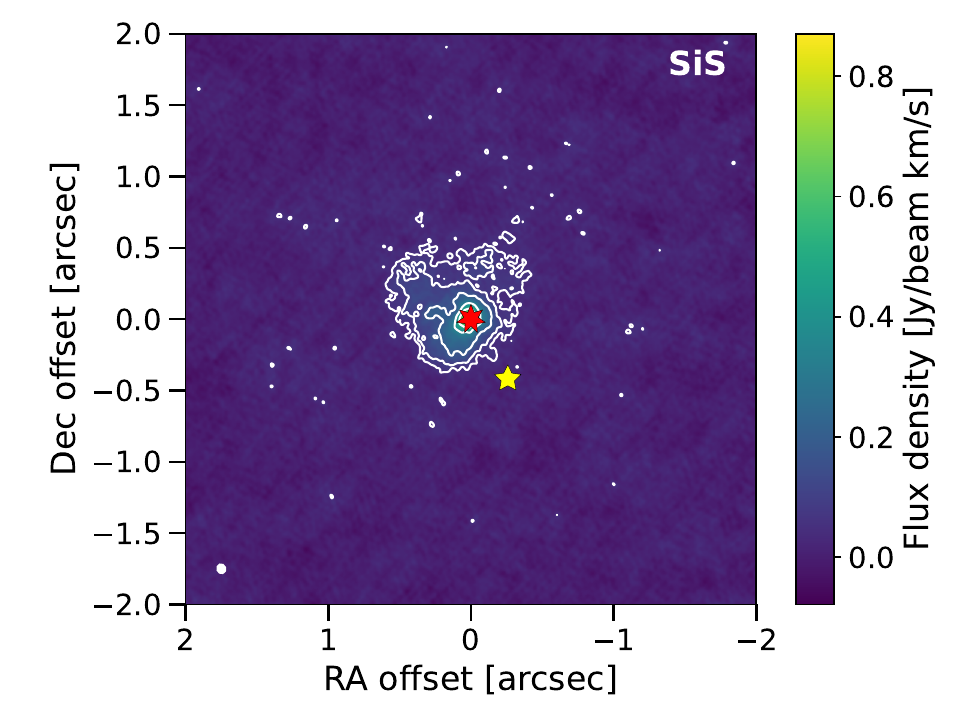}
\includegraphics[width=0.49\textwidth]{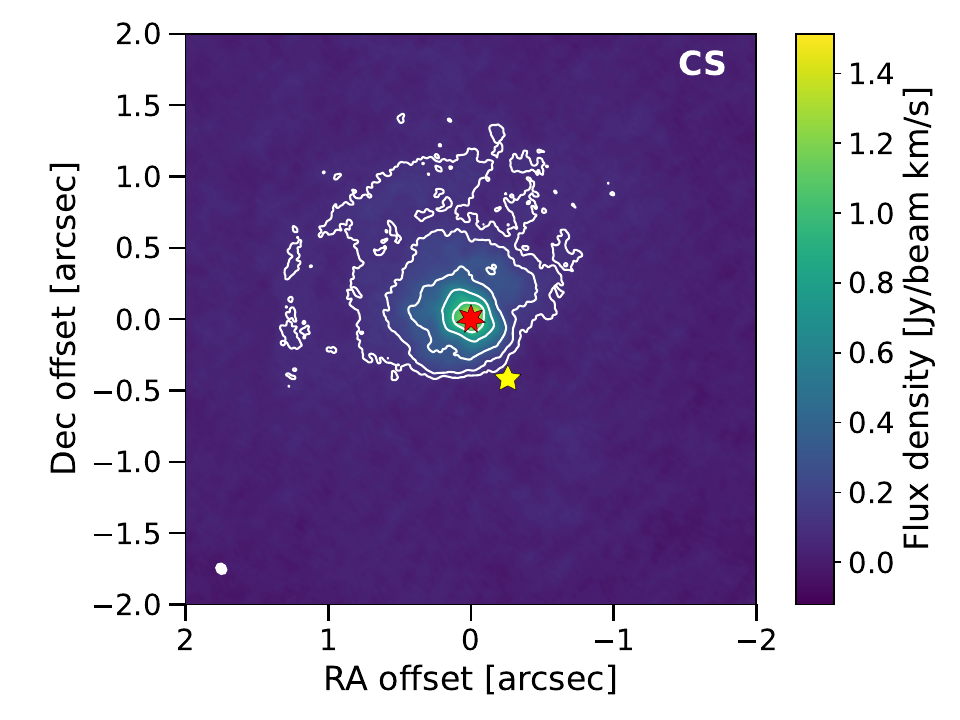}
\includegraphics[width=0.49\textwidth]{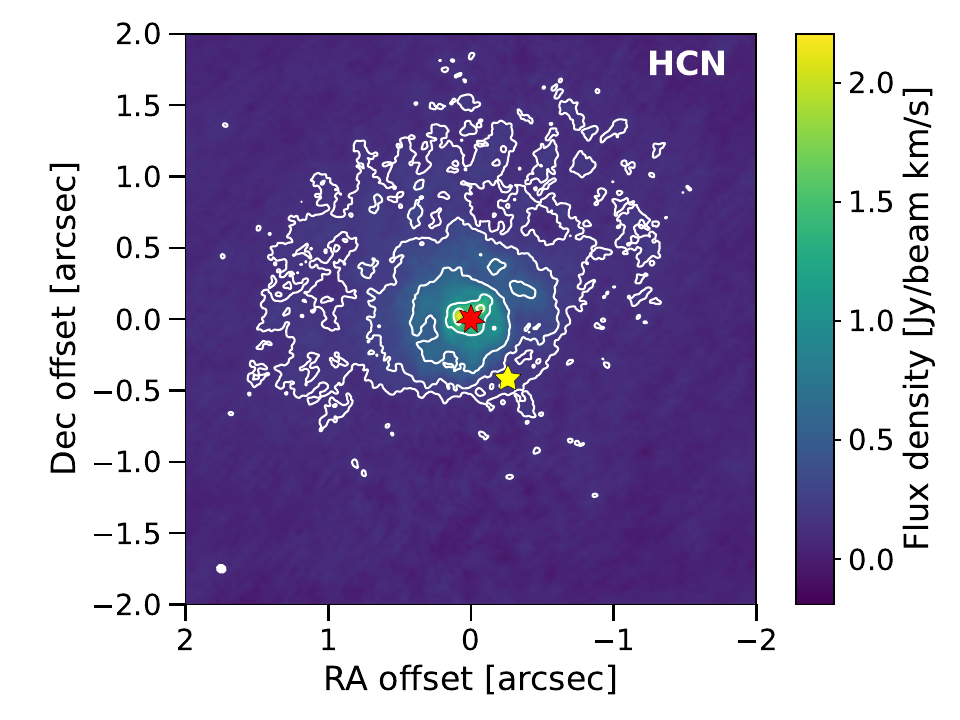}
\caption{Zeroth moment maps of SiO, SiS, CS and HCN towards W Aql (transitions give in  Table \ref{resolutions}). White contours are at levels of 3, 5, 10, 20, and 30$\sigma$. The position of the AGB star is indicated by the red star at (0,0) and the location of the F9 companion is indicated by the yellow star to the south-west. North is up and east is left. The white ellipses in the bottom left corners indicate the sizes of the synthesised beams.}
\label{othermol}
\end{center}
\end{figure}

\begin{figure}[t]
\begin{center}
\includegraphics[width=\textwidth]{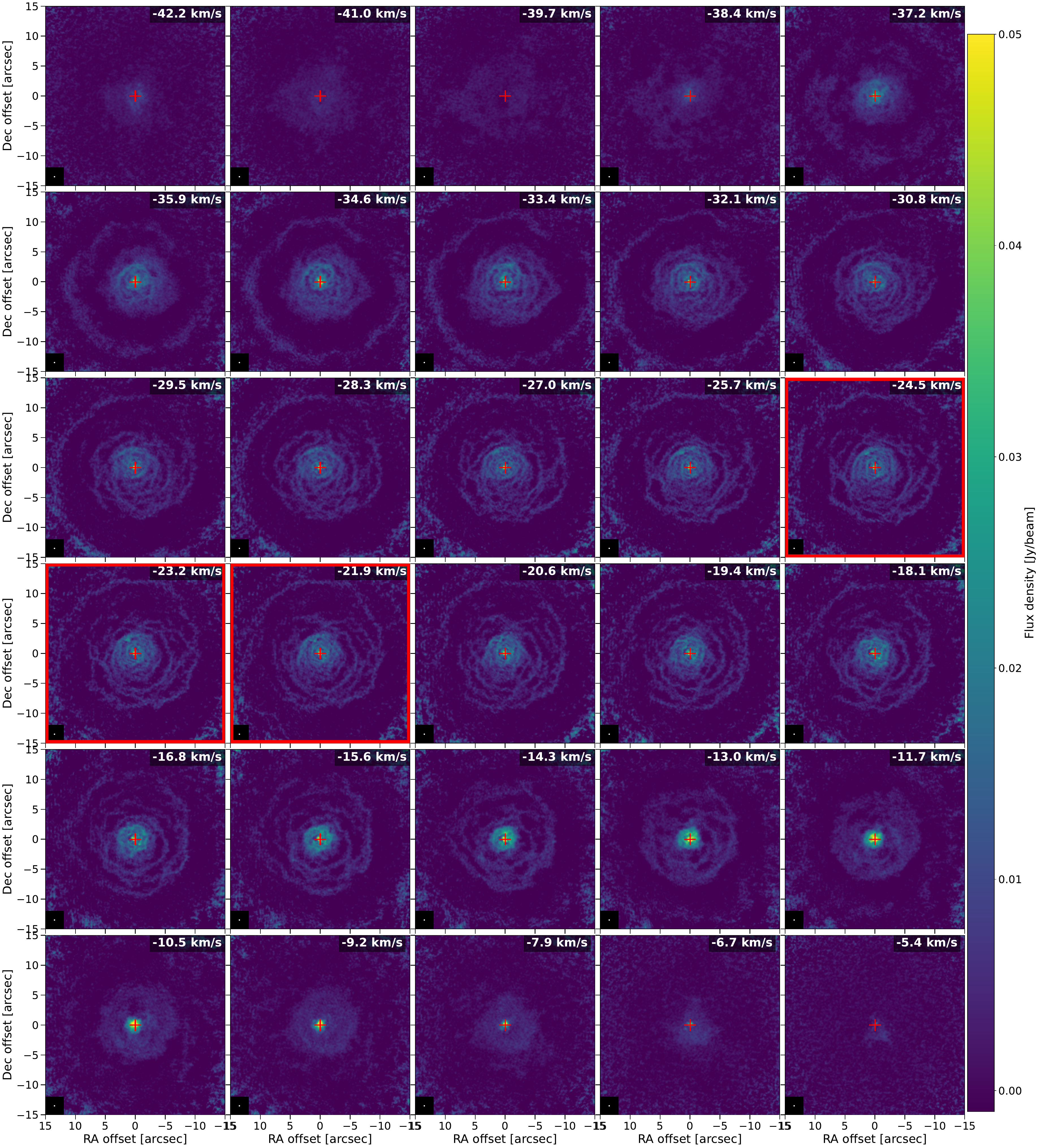}
\caption{Channel maps of CO ($J=2\to1$) towards W~Aql, obtained by combining observations from three configurations of ALMA. The AGB star is located at (0,0) and is marked by a red cross. The LSR velocity of each channel is given in the top right hand corner and the three channels closest to the W~Aql $\upsilon_\mathrm{LSR} = -23~\kms$ are highlighted with red borders and summed for Fig. \ref{coarcs}. The synthetic beam is given by the white ellipse in the bottom left corner of each channel. North is up and east is left.}
\label{cochan}
\end{center}
\end{figure}

\begin{figure}[t]
\begin{center}
\includegraphics[width=\textwidth]{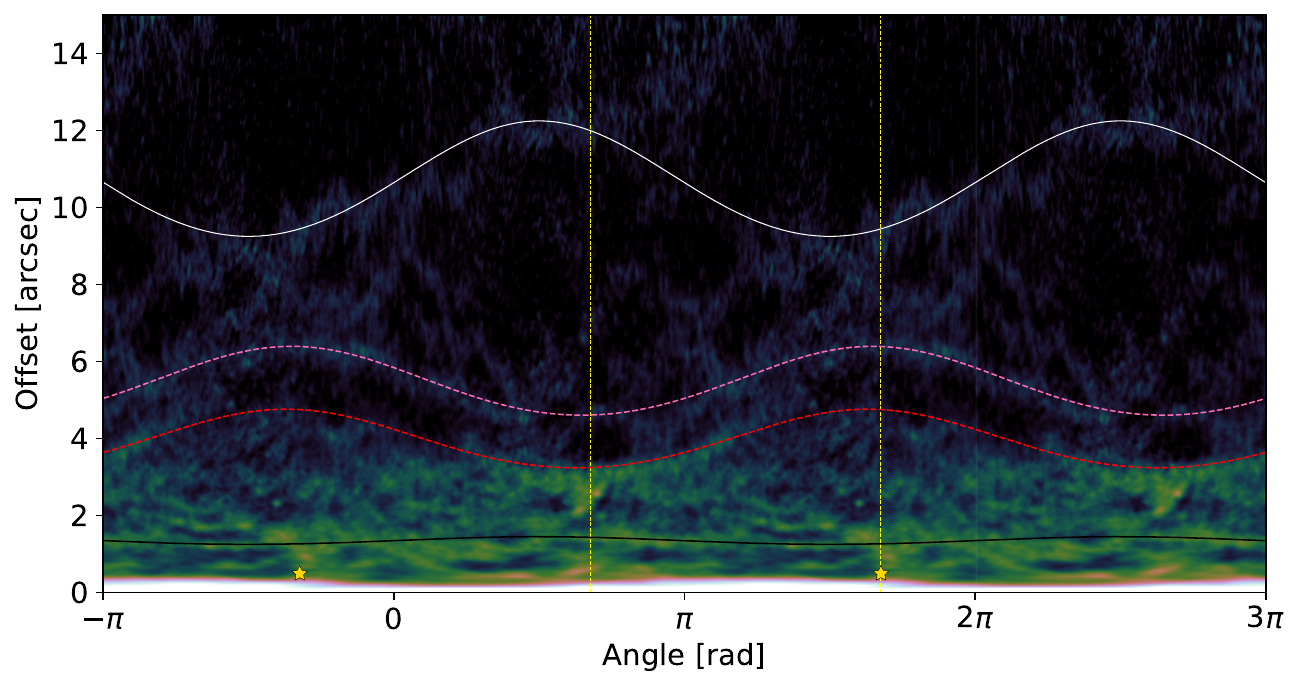}
\includegraphics[width=\textwidth]{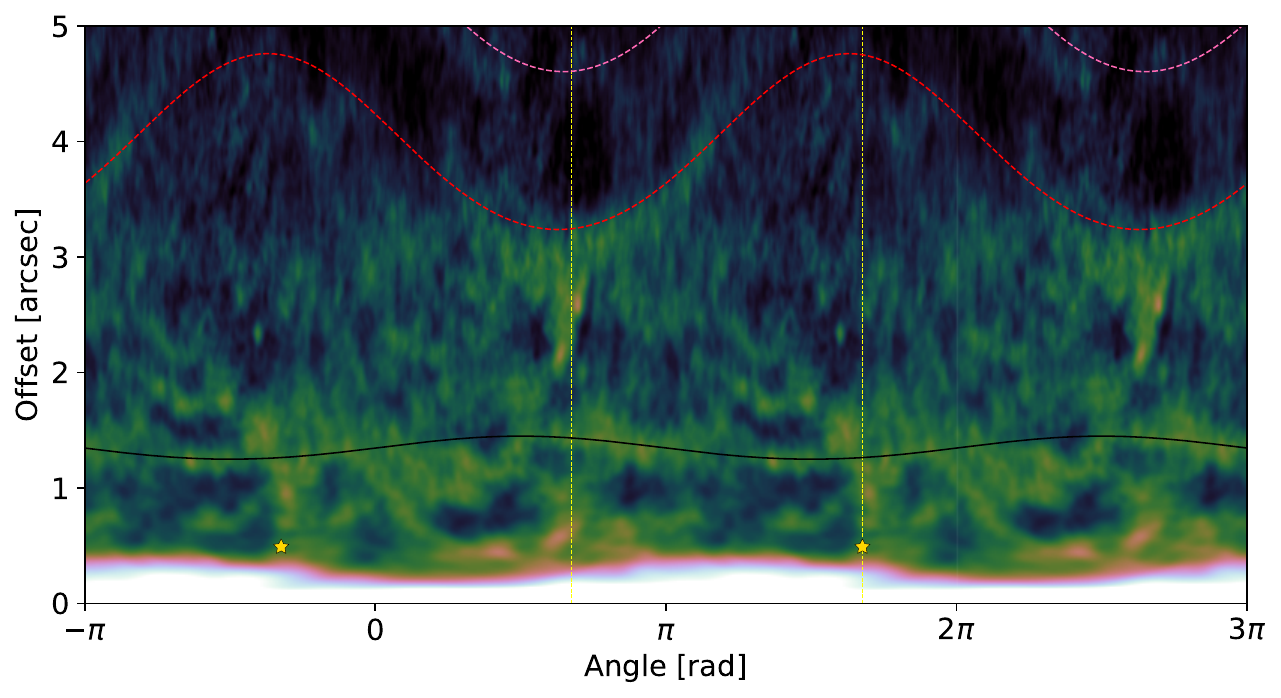}
\caption{Plots showing the radial emission distribution against angle for the summed central three channels of CO (Fig. \ref{coarcs}) with a full revolution \editstwo{shown in the centre (0 to $2\pi$) and half a revolution is shown on either side ($-\pi$ to 0 and $2\pi$ to $3\pi$) to show how the structures extend onwards.}
The location of the F9 star is indicated by the yellow star and a yellow dotted line which passes through both stars and is plotted in the central winding to guide the eye. The black, red and white curves correspond to the same features highlighted in Fig. \ref{coarcs}. The top plot shows the full observed extent of the CO emission (out to $15\arcsec$) and the bottom plot focuses on the regions out to $5\arcsec$ from the AGB star. \editstwo{These plots are reproduced in the Supplementary Materials Fig.~\ref{angrad-no-lines} without the additional curves.}}
\label{angrad}
\end{center}
\end{figure}

\begin{figure}[t]
\begin{center}
\includegraphics[width=0.49\textwidth]{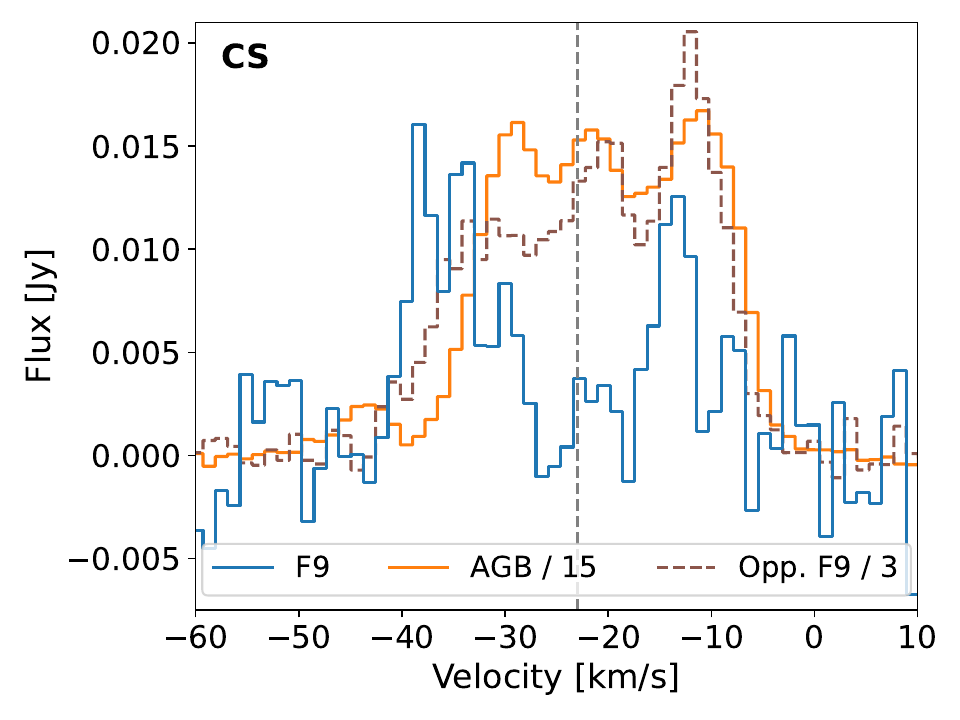}
\includegraphics[width=0.49\textwidth]{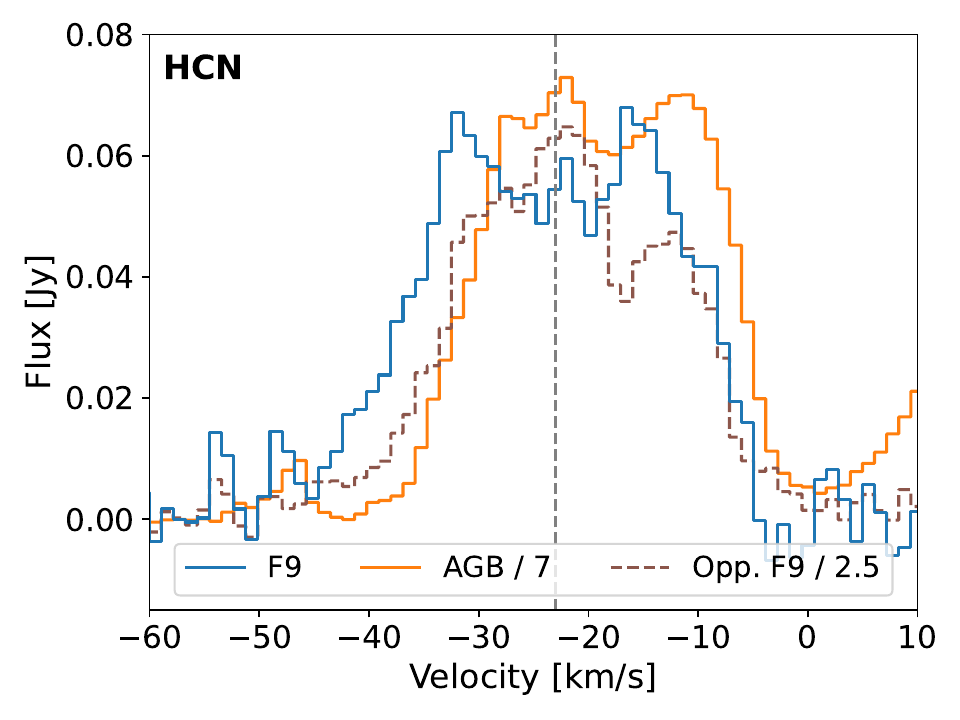}
\includegraphics[width=0.49\textwidth]{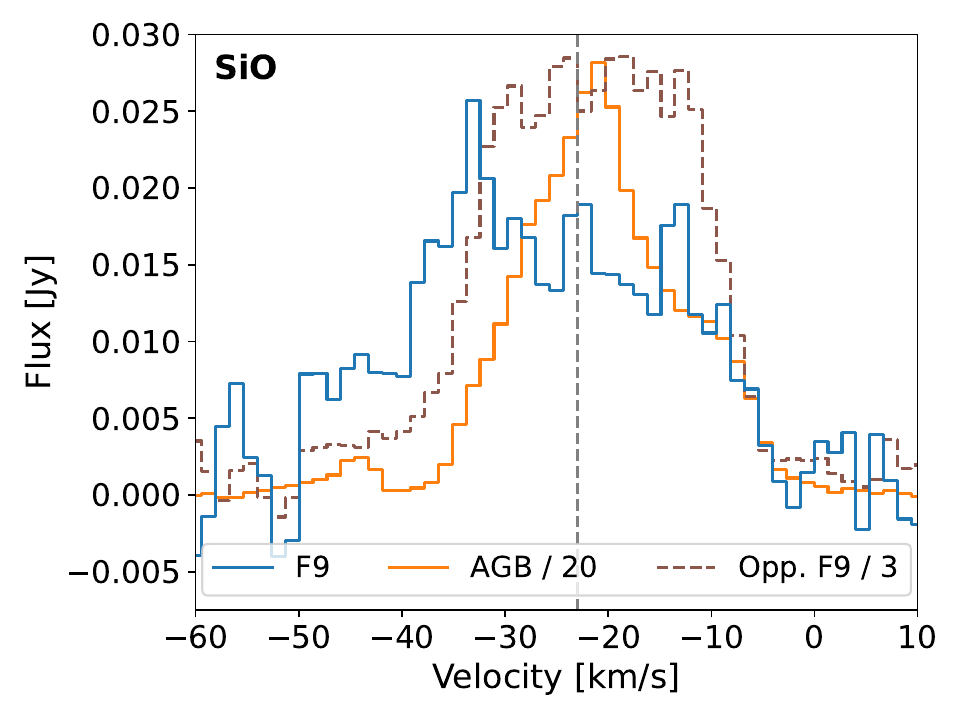}
\includegraphics[width=0.49\textwidth]{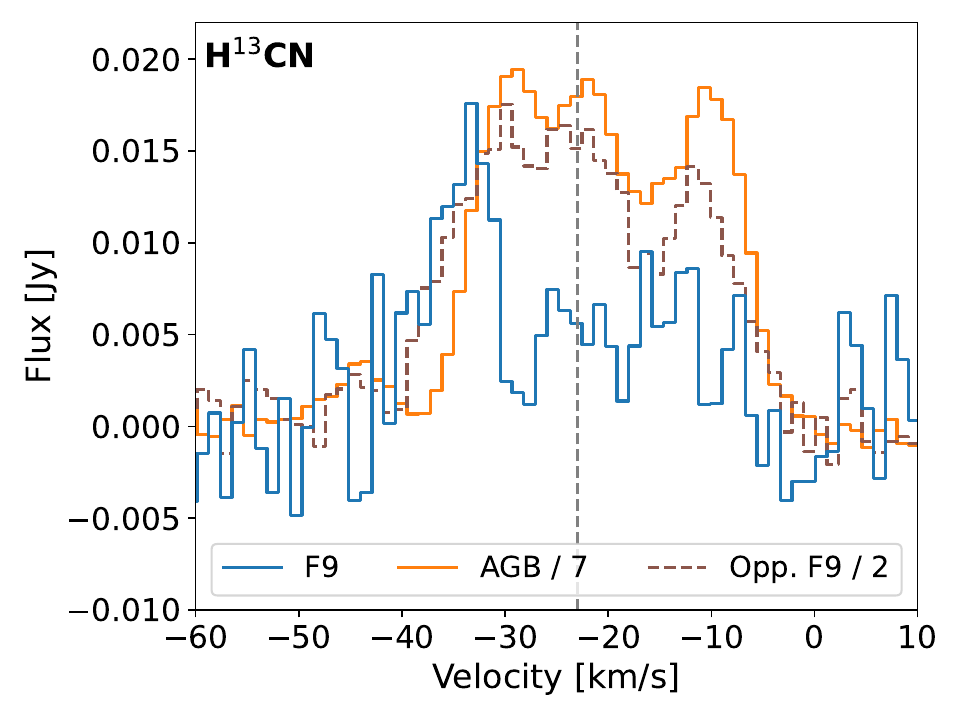}
\caption{Plots of CS, HCN, SiO and H$^{13}$CN emission extracted from circular apertures with 100~mas radii centred on the F9 star (blue), on the AGB star (orange) and at the same separation as the F9 star but on the opposite side of the AGB (Opp. F9, brown, dashed).
(See Table \ref{resolutions} for line frequencies.) The AGB and Opp. F9 line profiles are scaled by the factor given in the legend to facilitate comparison with the F9 line profiles. The vertical grey line indicates $\upsilon_\mathrm{LSR}=-23~\kms$.}
\label{F9spec}
\end{center}
\end{figure}

\begin{figure}[t]
\begin{center}
\includegraphics[width=\textwidth]{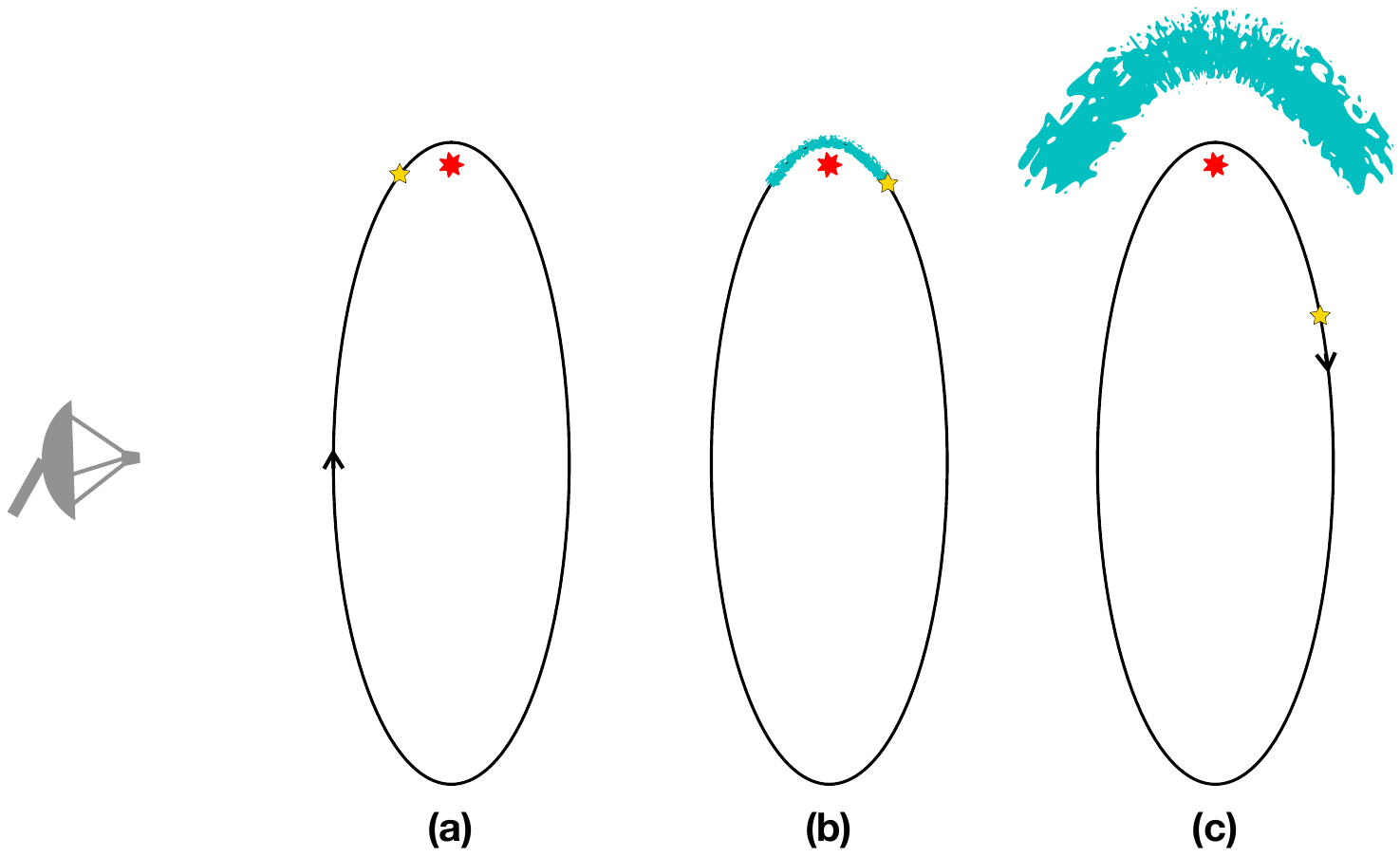}
\caption{A series of sketches illustrating the formation of SiN (or, similarly, SiC or NS) during the periastron passage of the W~Aql system. The orbit (black line) is shown face on in the frame of the AGB star and the F9 star is assumed to be moving clockwise. Relative to our observations, the observer is located to the left, represented by the radio dish. \textbf{(a)} The F9 star (yellow) approaches the AGB star (red) and enters the dense inner wind region ($n_{\mathrm{H}_2} \sim 10^{8}$ to $10^{10}$~cm$^{-3}$). \textbf{(b)} The rapid periastron passage is completed and SiN has formed in the wake of the F9 star (cyan region), with formation initiated by the F9 UV flux (see Methods \ref{chemmethods}). \textbf{(c)} As the F9 star continues on its orbit, the arc of SiN expands away from the AGB star, along with the stellar wind in which it is embedded. The present-day configuration of SiN can be seen in Fig. \ref{sin}, where the PV diagram is a good approximation of the final arc shape that would be seen around the AGB star were the orbit viewed face-on.}
\label{SiNsketch}
\end{center}
\end{figure}

\begin{figure}[t!]
\begin{center}
\includegraphics[width=0.505\textwidth]{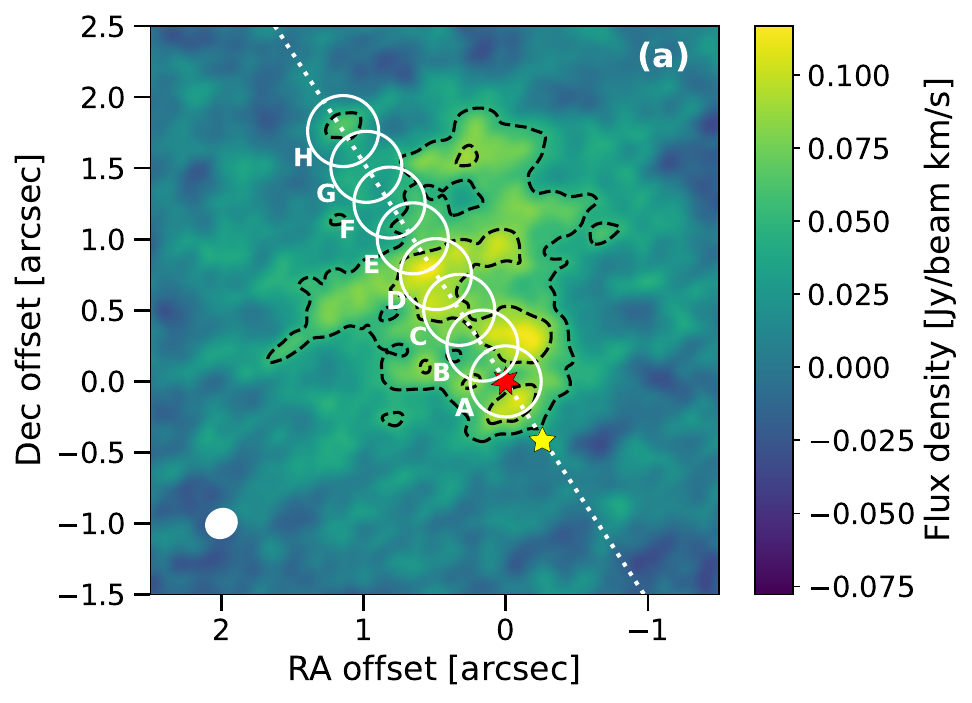}
\hspace{0.1cm}
\includegraphics[width=0.475\textwidth]{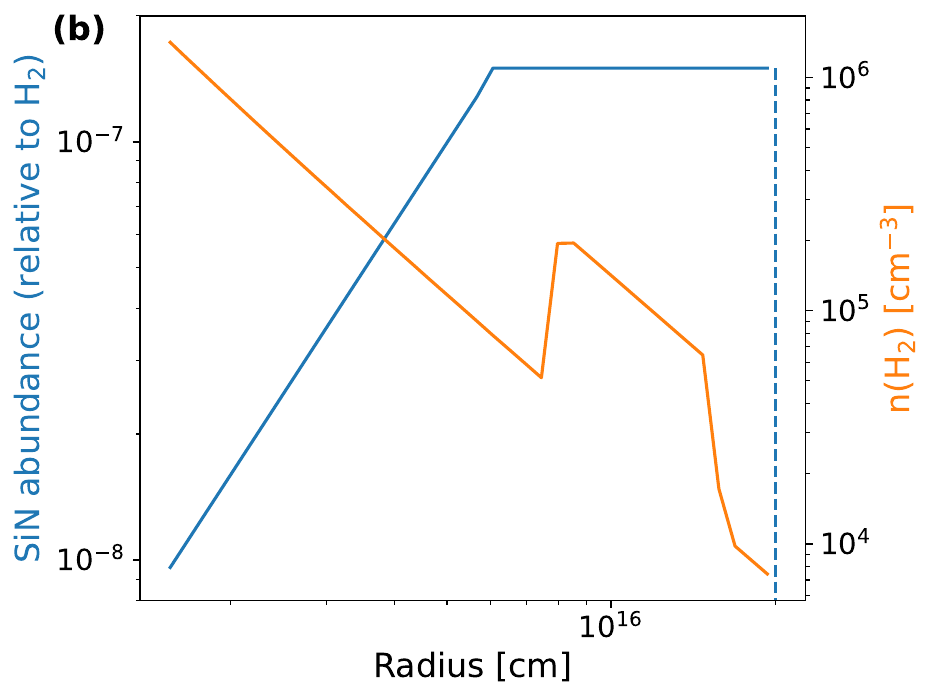}
\fbox{\includegraphics[width=0.99\textwidth]{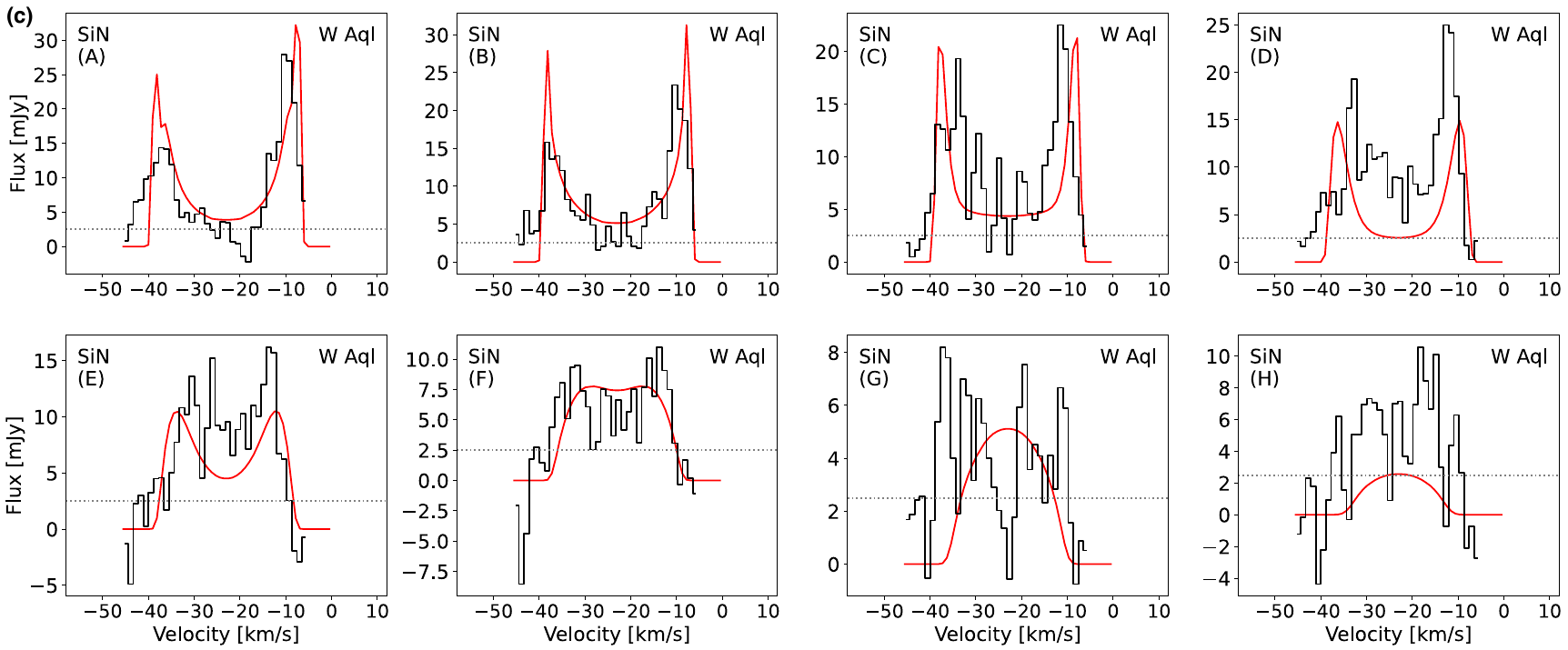}}
\caption{\textbf{(a)} SiN zeroth moment map, as shown in Fig. \ref{sin}, with the circular extraction apertures, labelled A to H, used to obtain spectra for radiative transfer modelling. The white dotted line lies at an angle of north $33\deg$ east, passing through the continuum peak. \textbf{(b)} SiN abundance (blue) and \h2 number density (orange) for the region of the CSE for which we model SiN. The dashed blue line represents the edge of the model, beyond which we do not include any SiN. \textbf{(c)} SiN spectra (black histograms) extracted for the regions (A to H) defined in (a) plotted with the results of the radiative transfer model (red curves). For these spectra rms = 2.5 mJy and is indicated by the dotted grey lines.}
\label{sin-modelling}
\end{center}
\end{figure}

\begin{figure}[t]
\begin{center}
\includegraphics[width=0.49\textwidth]{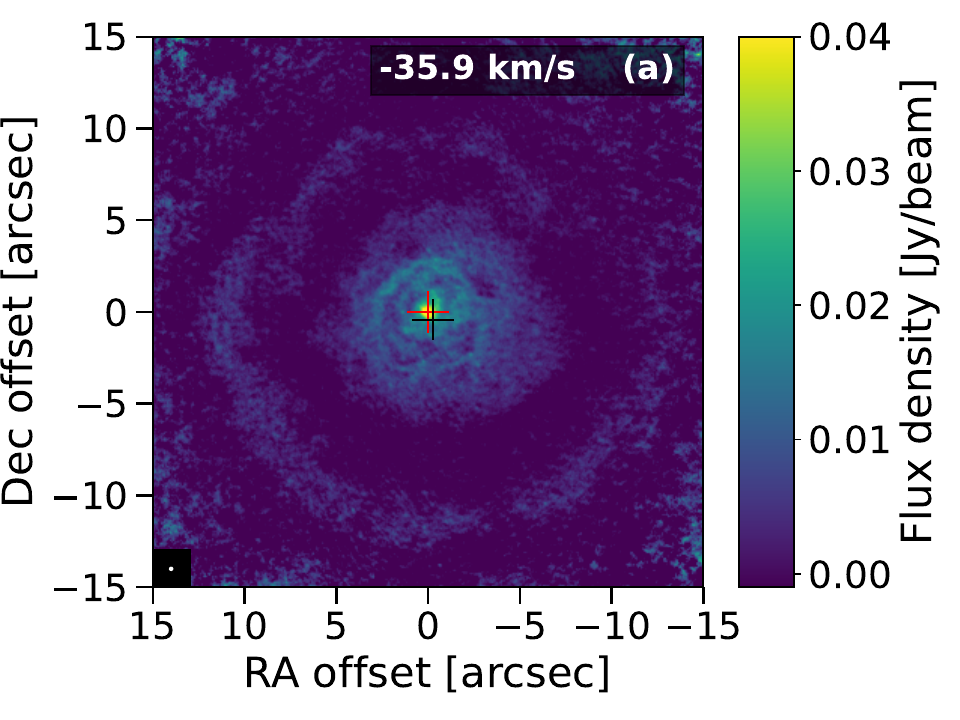}
\includegraphics[width=0.49\textwidth]{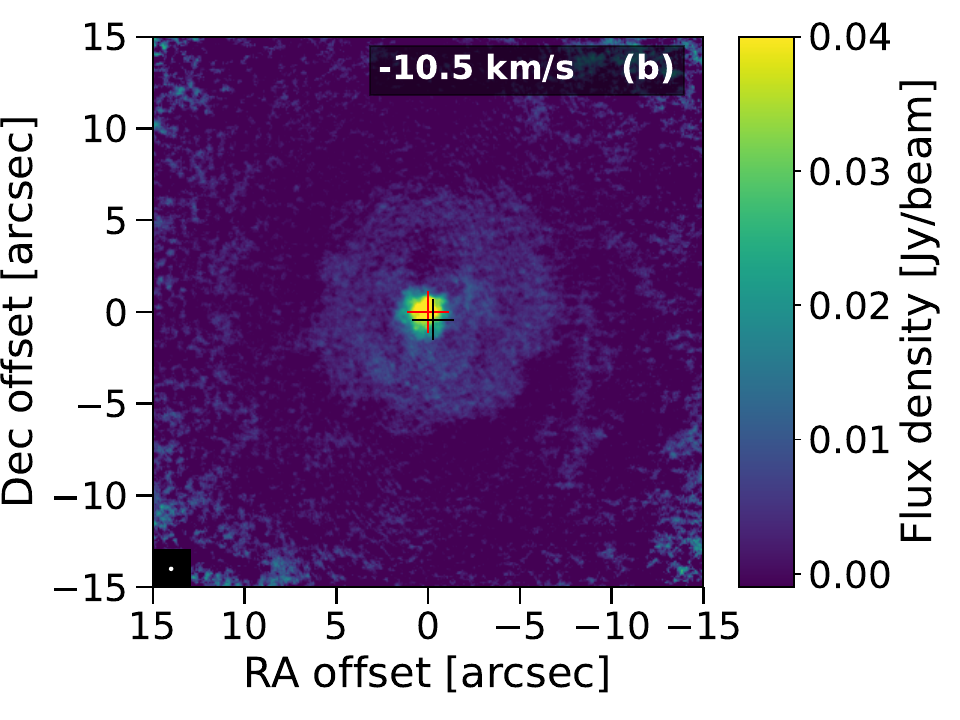}
\includegraphics[width=0.49\textwidth]{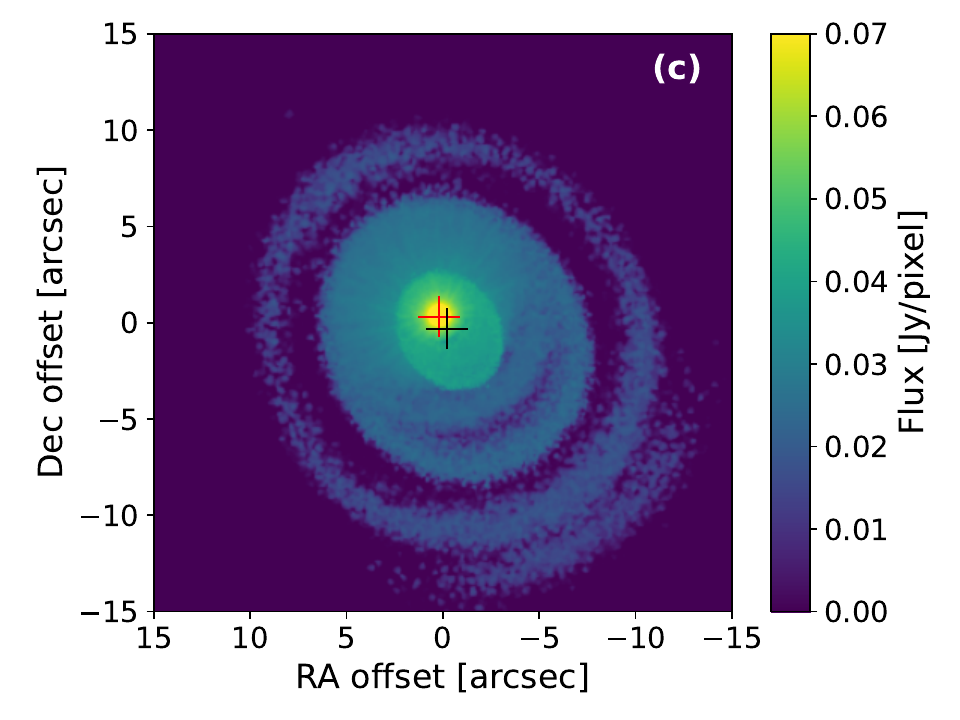}
\includegraphics[width=0.49\textwidth]{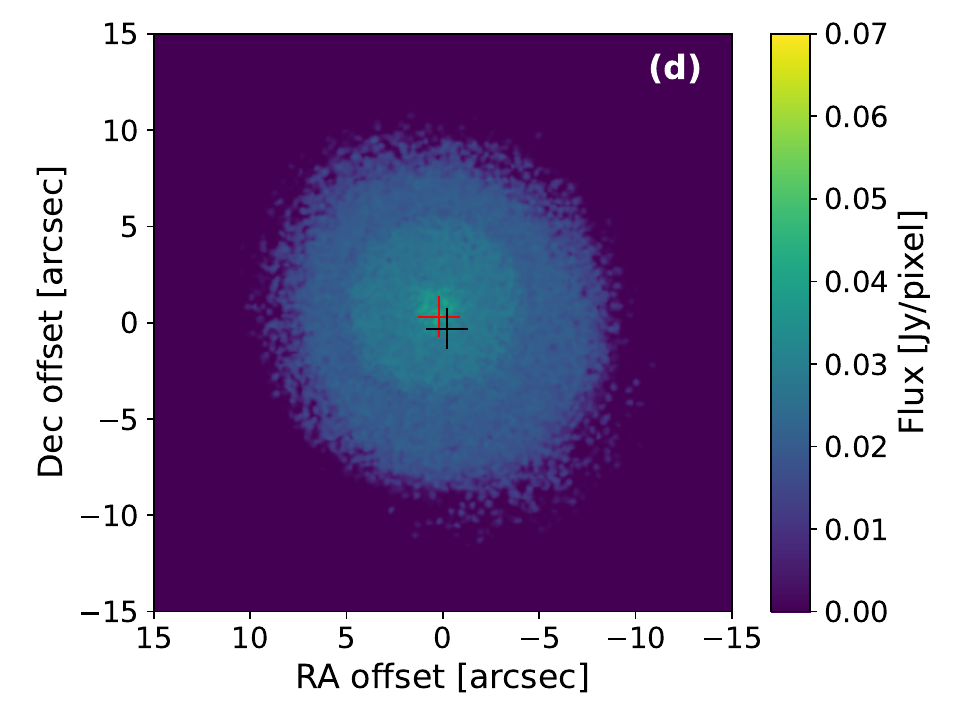}
\caption{Plots showing that blue (a and c) and red (b and d) channels equidistant from the stellar LSR velocity ($\upsilon_\mathrm{LSR} = -23~\kms$) in velocity space do not exhibit identical CO emission patterns. The ALMA observations (a and b) show an elongated emission region on the blue side (a) and an approximately round emission region on the red side (b). The same pattern is mimicked in the red (c) and blue (d) channels of the hydrodynamic model processed with MCFOST. The red and black crosses correspond to the locations of the AGB and F9 stars. Note that the modelled and observed positions do not exactly correspond. Details are given in Methods \ref{hydro}.
}
\label{otherCOchans}
\end{center}
\end{figure}

\begin{table}[hb]
\caption{Physical parameters of the W Aql system}
\begin{center}
\begin{tabular}{lr}
\hline\hline
\multicolumn{2}{c}{\textit{AGB and circumstellar parameters}}\\
LSR velocity, $\upsilon_\mathrm{LSR}$ & $-23~\kms$ \\
Mass-loss rate, $\dot{M}$ & $3\e{-6}\spy$\\
Stellar effective temperature, $T_\mathrm{eff}$ & 2300~K\\
Luminosity, $L_\star$ & $7500~\lsol$\\
Stellar radius, $R_\star$ & 8.3~mas\\
\hline
\multicolumn{2}{c}{\textit{System parameters}}\\
Distance, $D$ & 395~pc\\
AGB mass, $M_\mathrm{AGB}$ & $1.6~\msol$\\
F9 mass, $M_\mathrm{F9}$ & $1.06~\msol$\\
\hline
\multicolumn{2}{c}{\textit{Orbital parameters from ALMA observations}}\\
Orbital period, $T$ & $1082^{+89}_{-108}$~years\\
Time since periastron, $\Delta t$ & $172\pm22$~years\\
Rotation in the plane of the sky, $\omega$ & $120\pm5\deg$\\
Inclination, $i$ & $90\pm7\deg$\\
\hline\hline
\end{tabular}
\end{center}
\label{tab:parameters}
\end{table}%

\begin{table}[b!]
\caption{Possible orbital solutions for the W~Aql system.}
\begin{center}
\begin{tabular}{ccccccc}
\hline\hline
$e$ & $r_p$ [cm] & $r_p$ [au] & $a$ [au] & $T$ [years] & $\Delta t$ [years] & $t_\mathrm{close}$ [years]\\
\hline
0.98	&	$	4.5\e{13}	$	&	3.0	&	150	&	1131	&	157 	&	1.9 \\
0.97	&	$	6.5\e{13}	$	&	4.3	&	145	&	1069	&	163 	&	3.3 \\
0.96	&	$	8.5\e{13}	$	&	5.7	&	142	&	1038	&	167 	&	5.0 \\
0.96	&	$	9.0\e{13}	$	&	6.0	&	150	&	1131	&	165 	&	5.4 \\
0.95	&	$	1.1\e{14}	$	&	7.4	&	147	&	1093	&	170 	&	7.3 \\
0.94	&	$	1.3\e{14}	$	&	8.7	&	145	&	1069	&	174 	&	9.3 \\
0.93	&	$	1.5\e{14}	$	&	10	&	143	&	1051	&	179 	&	12 	\\
0.93	&	$	1.6\e{14}	$	&	11	&	153	&	1158	&	177 	&	13 	\\
0.92	&	$	1.7\e{14}	$	&	11	&	142	&	1038	&	183 	&	14 	\\
0.92	&	$	1.8\e{14}	$	&	12	&	150	&	1131	&	181 	&	15 	\\
0.91	&	$	1.9\e{14}	$	&	13	&	141	&	1028	&	187 	&	16 	\\
0.91	&	$	2.0\e{14}	$	&	13	&	149	&	1110	&	186 	&	18 	\\
\hline
\end{tabular}
\end{center}
Notes: $e$ is the eccentricity, $r_p$ is the periastron, $a$ is the semimajor axis, $T$ is the orbital period, $\Delta t$ is the time since the most recent periastron, and $t_\mathrm{close}$ is the amount of time the two stars spend close together (see Methods \ref{sec:solution}).
\label{tab:solutions}
\end{table}%

\clearpage
\newpage
\section{Supplementary Materials}\label{sup}

\subsection{Radiation pressure on dust}\label{dustdrive}

Here we compare the contribution to the radiation pressure on dust from the AGB and F9 stars.
The ratio of the radiation pressure force on dust grains, ${F}_{P_r} = |\vec{F}_{P_r}|$, over the gravitational attraction, ${F}_\mathrm{grav}=|\vec{F}_\mathrm{grav}|$, is defined as
\begin{equation}\label{Gamma}
\Gamma = \frac{F_{P_r}}{F_\mathrm{grav}} \simeq \frac{\sigma_d \bar{Q} \Psi}{4 \pi c m_\mathrm{dust} GM_\star} L_\star = \left( \frac{\bar{Q} \Psi}{3 \pi c a \rho_d G}\right) \frac{L_\star}{M_\star}
\end{equation}
where $\sigma_d = \pi a^2$ is the cross-section of the assumed spherical grain, with $a$ the radius, $\bar{Q} = 2\e{-2}$ is the mean radiation pressure efficiency of the grains \cite{Tielens1983}, $\Psi = 2\e{-3}$ is the dust to gas ratio \cite{Danilovich2014}, $c$ is the speed of light, $m_\mathrm{dust} = \frac{4}{3}\pi a^3 \rho_d $ is the mass of a dust grain (derived from volume, assuming a sphere, and a specific dust density of $\rho_d = 3.3$~g~cm$^{-3}$), and $M_\star$ and $L_\star$ are the stellar mass and luminosity. A dust driven wind is achieved for $\Gamma > 1$. 

When comparing the ability of the AGB and F9 stars to drive the wind through radiation pressure, the properties in brackets on the right-hand side of equation \ref{Gamma} do not change, so the dust driving potential comes mainly from the luminosity of the star. For the AGB star, the luminosity is $7500~\lsol$ \cite{Danilovich2014}, while for the F9 star it is $\sim 1.5~\lsol$. We use the system mass of $2.66~\msol$ as this is the maximum possible gravitational force that must be overcome by the radiation pressure. 
For relatively small grains with $a=0.03~\mu$m, we find $\Gamma_\mathrm{AGB} = 1.1$ and $\Gamma_\mathrm{F9} = 2.3\e{-4}$, indicating that the F9 star's contribution to driving the wind is negligible.

\subsection{\editstwo{Anisotropic mass loss}}\label{sec:anisotropies}

\editstwo{Recent observational studies of AGB stellar discs and inner winds at near-infrared wavelengths have shown asymmetric and clumpy surface brightnesses \cite{Ohnaka2016,Khouri2016,Wittkowski2017,Paladini2018,Khouri2020}. These broadly agree with 3D hydrodynamical simulations of AGB atmospheres, which predict the formation of large convective shells in the low-gravity environment of the AGB star's extended atmosphere, resulting in a clumpy and non-spherical atmospheric structure and asymmetric dust formation \cite{Freytag2008,Freytag2017,Freytag2023}. Similar asymmetric features have also been observed in the millimetre range with ALMA, in both the continuum emission and for molecular lines that originate in, or close to, the stellar atmosphere \cite{Vlemmings2017,Takigawa2017,Khouri2019,Velilla-Prieto2023}. A recent study of the nearby carbon star CW~Leo determined that the asymmetries in the stellar atmosphere and inner wind are unlikely to have been formed as a result of binary interactions, but rather as a result of varying temperature and density conditions caused by convection cells \cite{Velilla-Prieto2023}. This is despite the larger-scale shells observed around this star being thought to have formed as a result of binary interactions \cite{Cernicharo2015a}.
In light of these observational and modelling results, we analysed whether the various anisotropies reported in the molecular emission around W~Aql could be related to random convection cells rather than formed through binary interactions.}

\begin{figure}[t]
\begin{center}
\includegraphics[width=0.49\textwidth]{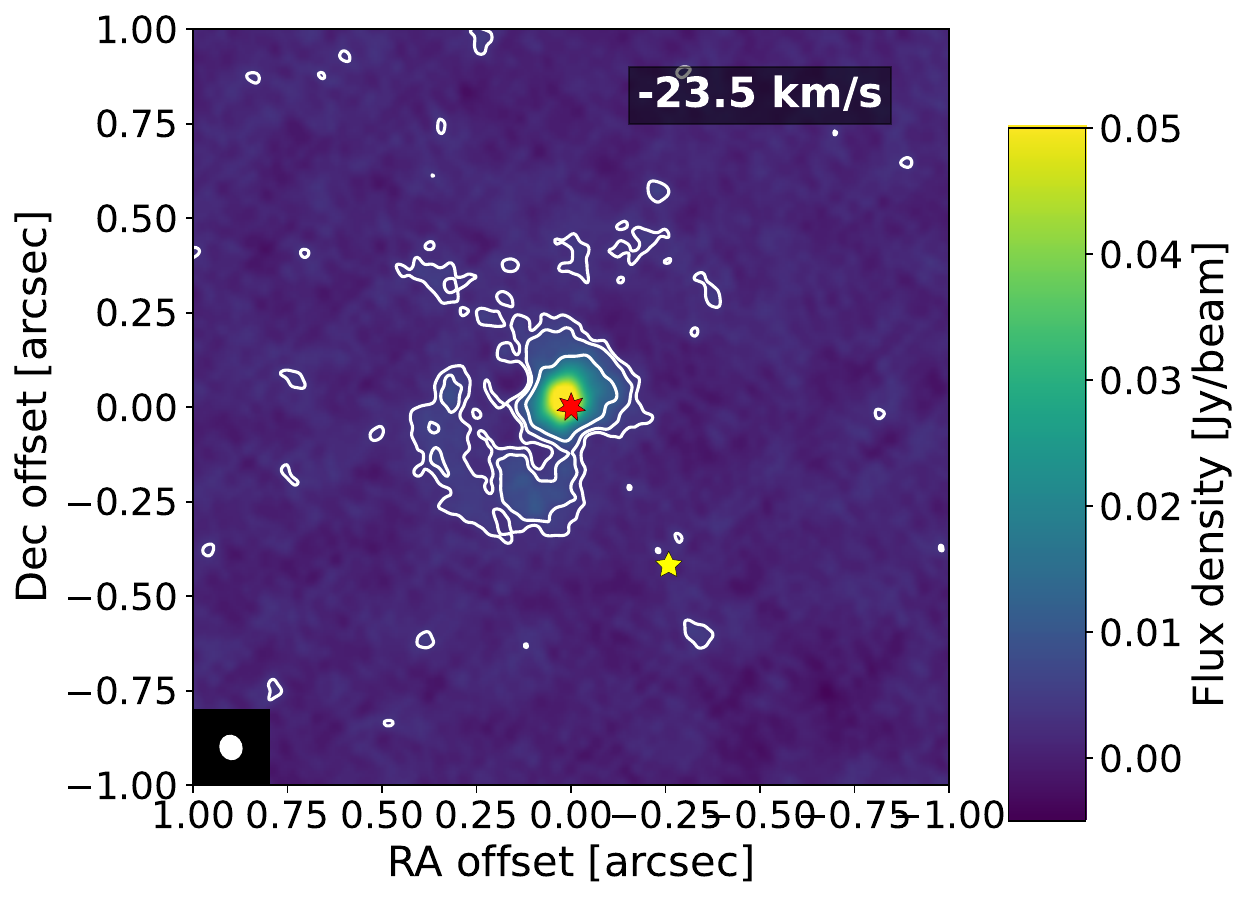}
\includegraphics[width=0.49\textwidth]{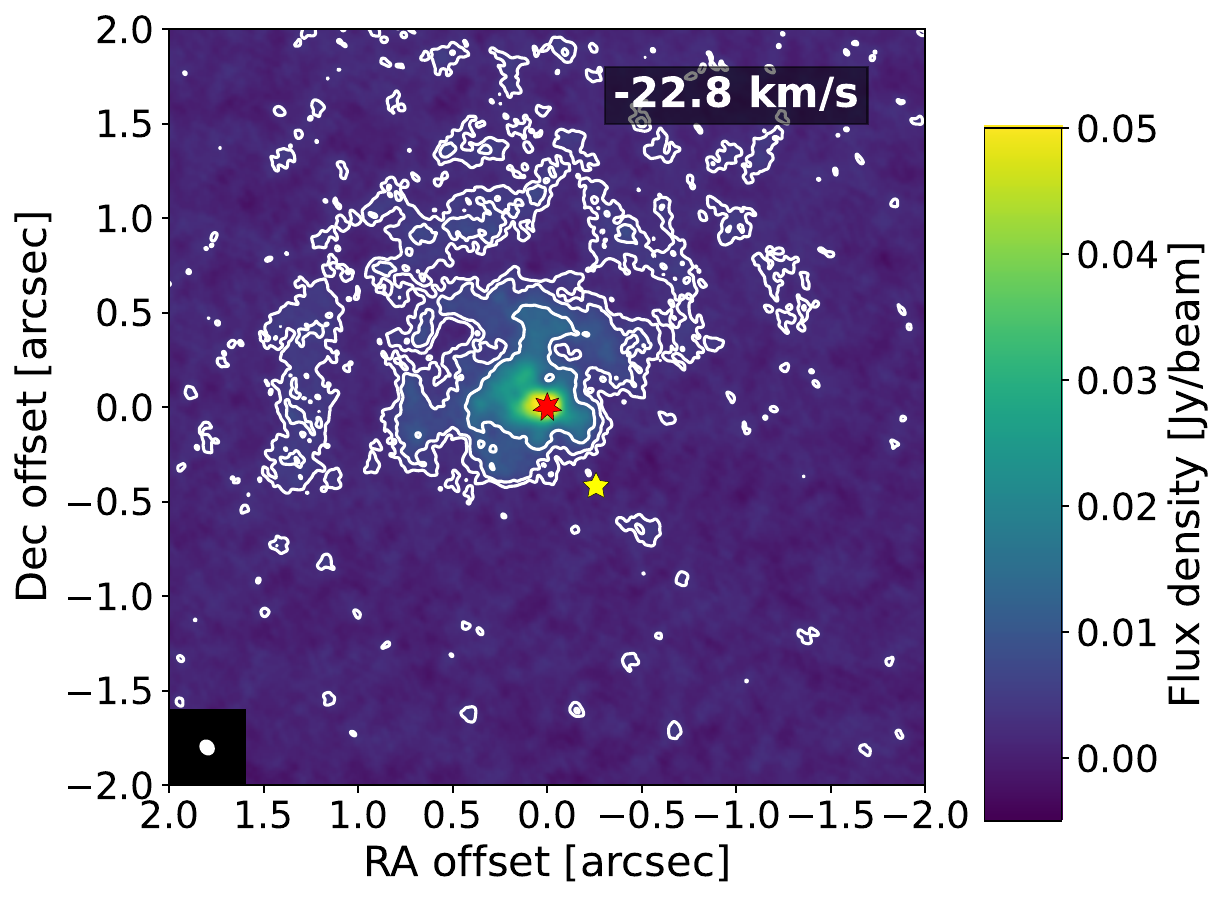}
\caption{\edits{Plots of the central channels of SiS (left) and CS (right), showing the asymmetric distribution of these molecules caused by the flux from the F9 star. The positions of the AGB and F9 stars are indicated by the red and yellow stars, respectively. The channel velocities are given in the top right corners and the beam is shown in the bottom left corner. Contours are plotted for levels of 3, 5, and 10$\sigma$. North is up and east is left.}}
\label{cssissinglechans}
\end{center}
\end{figure}

\editstwo{The asymmetries we see on the largest scales in the molecular emission around W~Aql are those that we associate with the photodissociation of common species (SiO, SiS, CS, HCN) by the F9 companion, as discussed in the Results and in Methods \ref{sec:othermolecules} and \ref{otherchem}. While we do see some smaller-scale asymmetries in these molecular lines, which may have originated as a result of the chaotic distributions of convective cells before expanding in the wind (e.g. see the non-uniform distributions of SiS and CS in their central channels, shown in Fig.~\ref{cssissinglechans}, or the smaller-scale arcs and clumps seen in the CO emission in Figs.~\ref{coarcs} and \ref{cochan}), these are unlikely to account for the overall asymmetry on a larger scale.}

\editstwo{The asymmetric emission detected for SiN, SiC and NS is generally seen on smaller scales than the asymmetries in the common species discussed above and has higher degrees of asymmetry. Such an arc-like distribution is unlikely be formed as a result of localised (and necessarily very specific, based on the observations) fluctuations of temperature and density caused by convective cells. If these asymmetries were formed as a result of random fluctuations, we would expect them to be formed in several directions (e.g. see \cite{Freytag2023}), not just in an arc on one side of the AGB star, and would expect additional similar fluctuations to have occurred in the $\sim170$ years since the formation of the observed SiN arc. For example, the model by Freytag et al. \cite{Freytag2023} that most closely resembles W~Aql has events of elevated dust production on a time scale of a few to tens of years.
We also note that the formation of both SiN and SiC is driven by \ce{Si+} (Methods \ref{chem:sin}), which has an ionisation energy of 8.2~eV \cite{Martin1998} and hence is most easily ionised by UV photons. 
Ergo, as discussed in Results and Methods \ref{sinsic} and \ref{chem:sin}, the formation of SiN in an arc to one side of the AGB star suggests formation during a close periastron interaction between the AGB and F9 stars.}

\begin{figure}[t!]
\begin{center}
\includegraphics[width=0.3\textwidth]{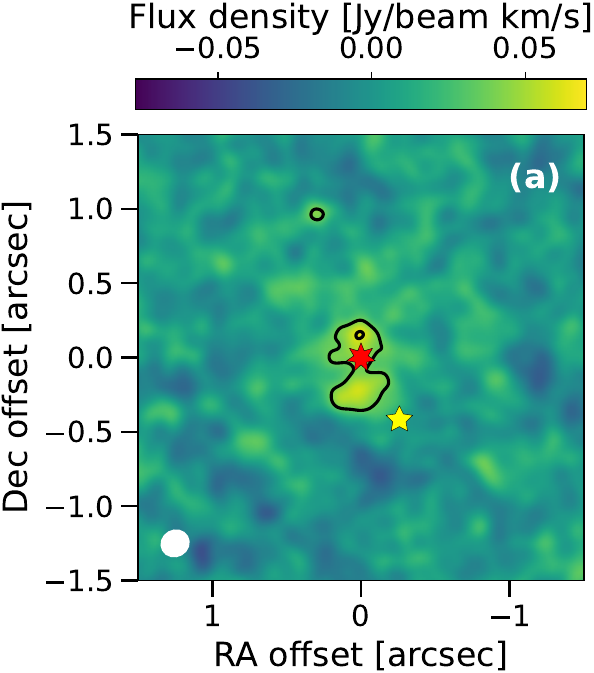}
\includegraphics[width=0.33\textwidth]{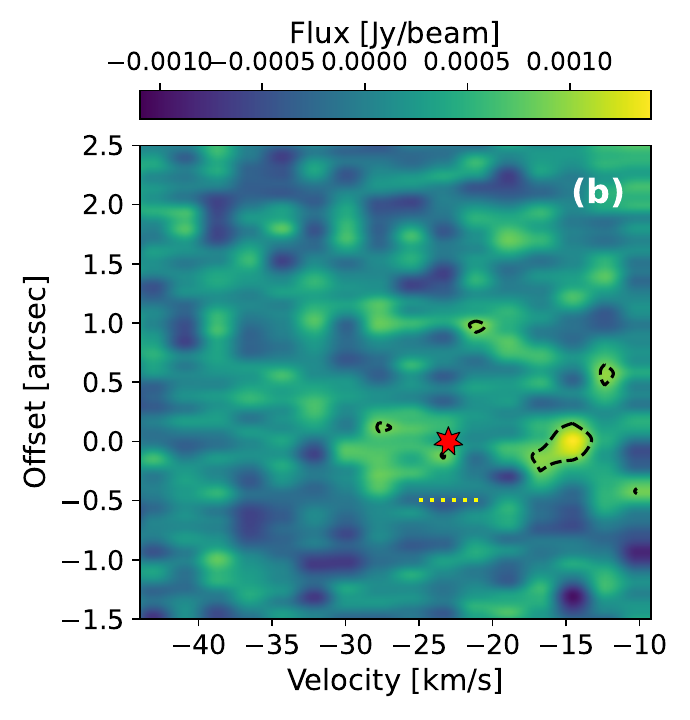}
\includegraphics[width=0.33\textwidth]{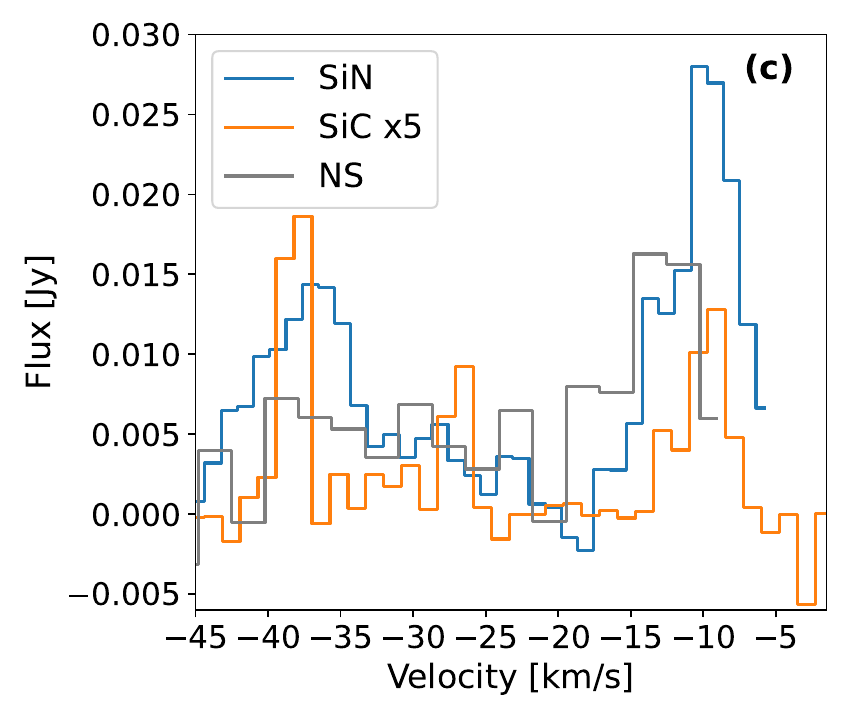}
\caption{\textbf{(a)} Zeroth moment map of NS towards W~Aql with contours at levels of 3 and $5\sigma$. North is up and east is to the left. The position of the AGB star is indicated by the red star at (0,0) and the location of the F9 companion is indicated by the yellow star to the south-west. The white ellipse in the bottom left corner indicates the size of the synthesised beam. \textbf{(b)} Position-velocity diagram of NS taken with the same wide slit as used for SiN (Fig. \ref{sin}). The position and LSR velocity of the AGB star is indicated by the red star and the horizontal dotted yellow line indicates the present offset of the F9 star. \textbf{(c)} Spectra of the NS, SiN and SiC lines given in Table \ref{resolutions}. All lines were extracted for circular apertures with radii $0.25\arcsec$, centred on the continuum peak. The flux of the SiC spectrum is multiplied by 5 to allow for a more direct comparison to SiN and NS.}
\label{nsplot}
\end{center}
\end{figure}

\editstwo{However, the distribution of SiN emission seen in the zeroth moment map and PV diagram in Fig.~\ref{sin} does not reveal a perfectly uniform structure. For example, there are regions of brighter flux in the zeroth moment map, enclosed in the $5\sigma$ contours, which are not symmetric along the axis connecting the AGB and F9 stars. Similarly, the PV diagram is not perfectly symmetric across the LSR velocity axis and shows clumps of brighter emission. These clumpy asymmetries are more likely to be caused by variations in density and temperature driven by chaotic motions of convection cells, similar to the clumpy emission seen in the inner wind of CW~Leo \cite{Velilla-Prieto2023}. It is also possible that some of the enhancements were caused by the interaction between the shock created by the companion's passage and the pulsation of the AGB star, as modelled in the simulations of Aydi et al \cite{Aydi2022}. Such varying conditions could explain why, for example, there is only one bright clump of NS in the PV diagram (Fig.~\ref{nsplot}) but several bright clumps of SiN and SiC.}

\editstwo{We conclude that the large arc-like structure of the molecular emission is more consistent with enhanced formation during the periastron interaction of the AGB and F9 stars. However, anisotropic mass loss processes may also have contributed to the precise small-scale structure of the SiN, SiC and NS emission. }


\subsection{Additional figures}

\begin{figure}[h]
\begin{center}
\includegraphics[width=0.49\textwidth]{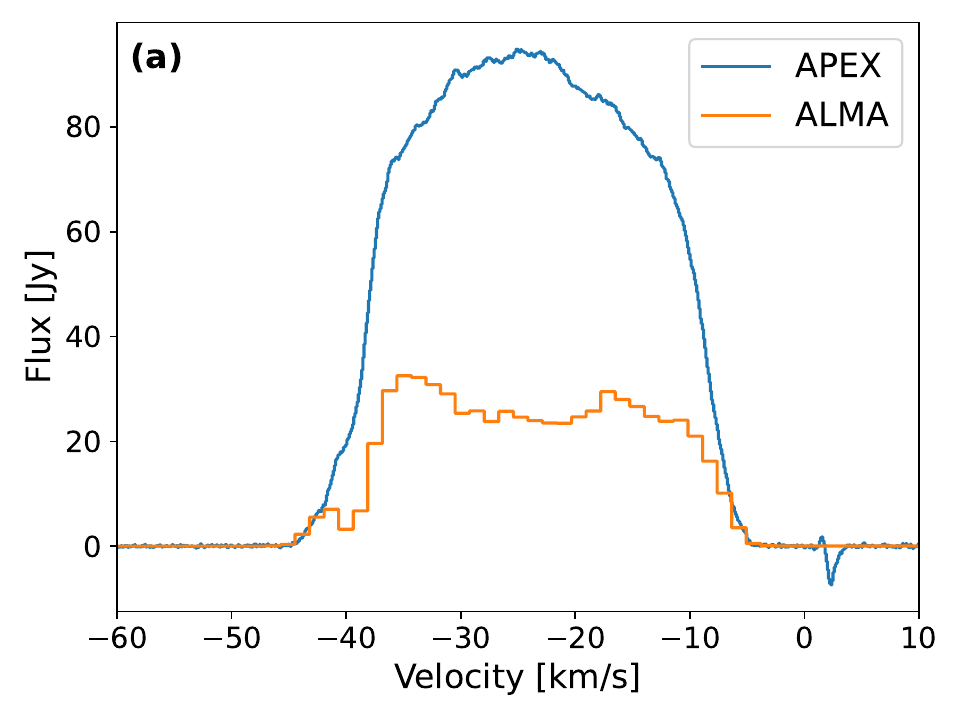}
\includegraphics[width=0.49\textwidth]{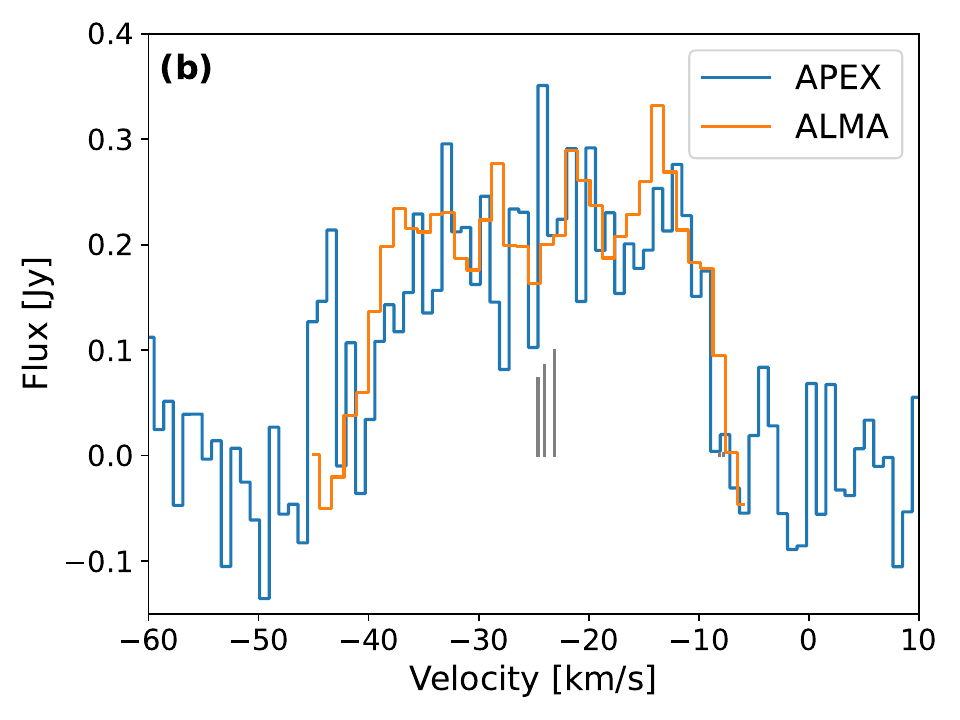}
\caption{A comparison between ALMA (orange) and APEX (blue, \cite{De-Beck2020}) observations of the same molecular lines. \textbf{(a)} Spectra of CO ($J=2\to1$), showing that around 66\% of the CO flux was not recovered with ALMA for a spectrum extracted from an aperture with radius $5.4\arcsec$ from the low-resolution ALMA data.  \textbf{(b)} Spectra of SiN ($N,J=6,{13/2}\to5,{11/2}$) extracted from an aperture with radius $2.5\arcsec$, showing that all the SiN flux has been recovered by ALMA. The vertical grey lines indicate the relative velocities and intensities of the hyperfine components of the SiN, assuming an LSR velocity of $-24~\kms$.}
\label{lostflux}
\end{center}
\end{figure}

\begin{SCfigure}
\includegraphics[width=0.5\textwidth]{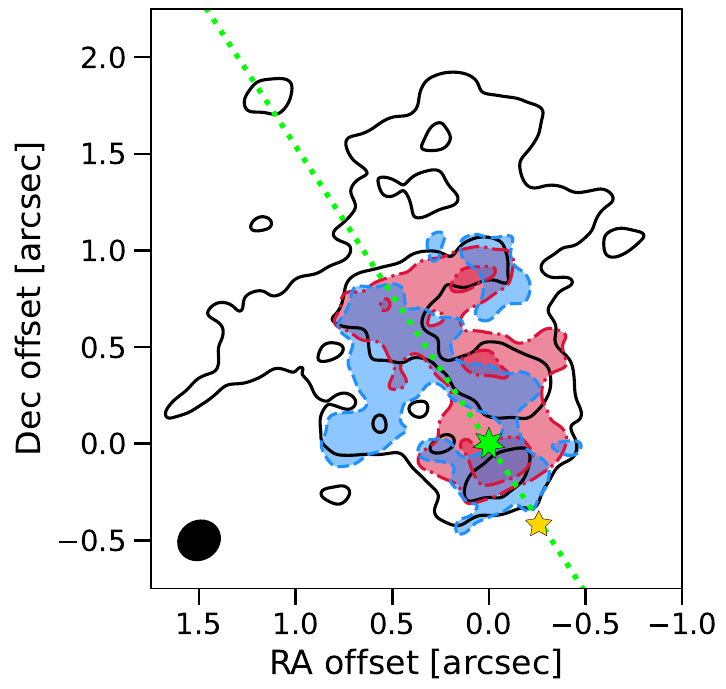}\caption{The blue- and red-shifted components of the SiN emission overplotted with the full zeroth moment map contours from Fig. \ref{sin}a. In all cases, contours are at levels of 3 and $5\sigma$. The synthetic beam size is given by the black ellipse in the bottom left corner and the locations of the AGB and F9 stars are indicated by the green and yellow stars, respectively. The dotted green line is at north $33\deg$ east.}
\label{sinbluered}
\end{SCfigure}

\begin{figure}[t!h]
\begin{center}
\includegraphics[height=5.8cm]{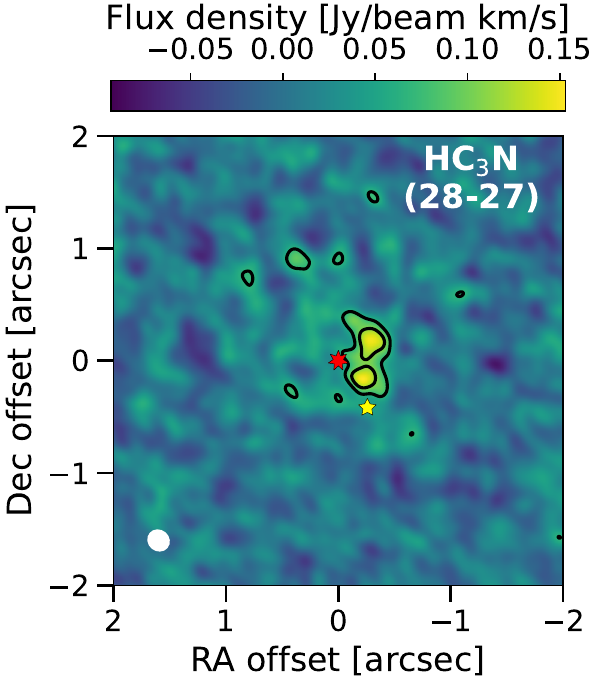}
\includegraphics[height=5.8cm]{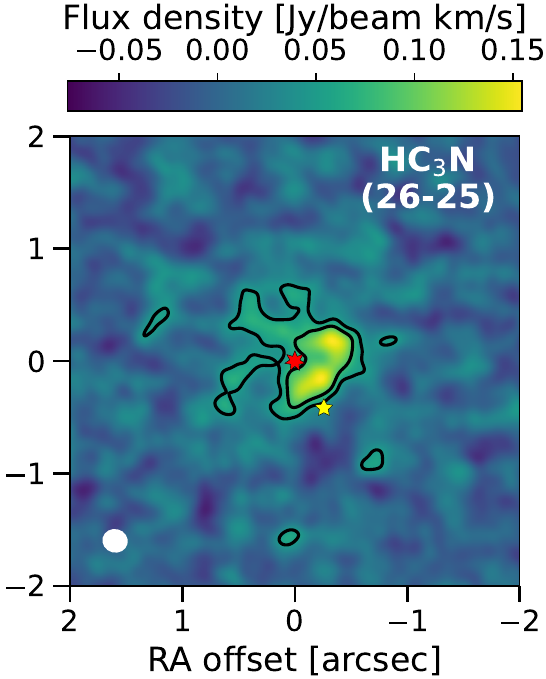}
\includegraphics[height=5.8cm]{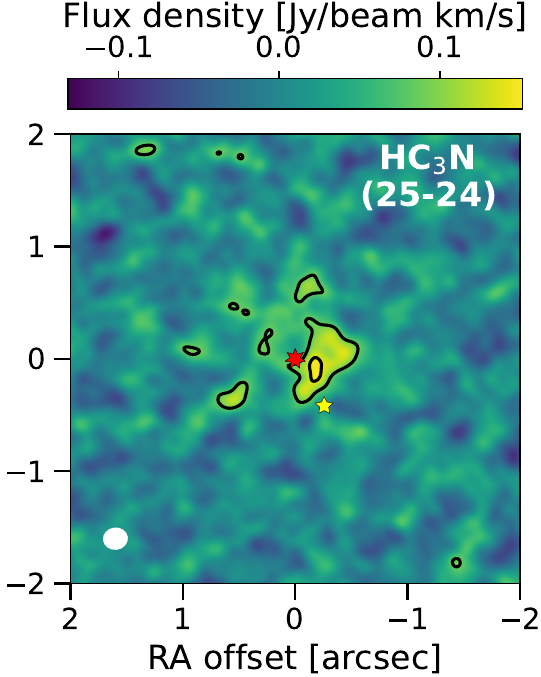}
\caption{Zeroth moment maps of HC$_3$N towards W~Aql with contours at levels of 3 and $5\sigma$. The transition is given in the top right of each map. North is up and east is to the left. The position of the AGB star is indicated by the red star at (0,0), also corresponding to the continuum peak, and the location of the F9 companion is indicated by the yellow star to the south-west. The white ellipse in the bottom left corner indicates the size of the synthesised beam.}
\label{hc3n} 
\end{center}
\end{figure}


\begin{figure}[t]
\begin{center}
\includegraphics[width=\textwidth]{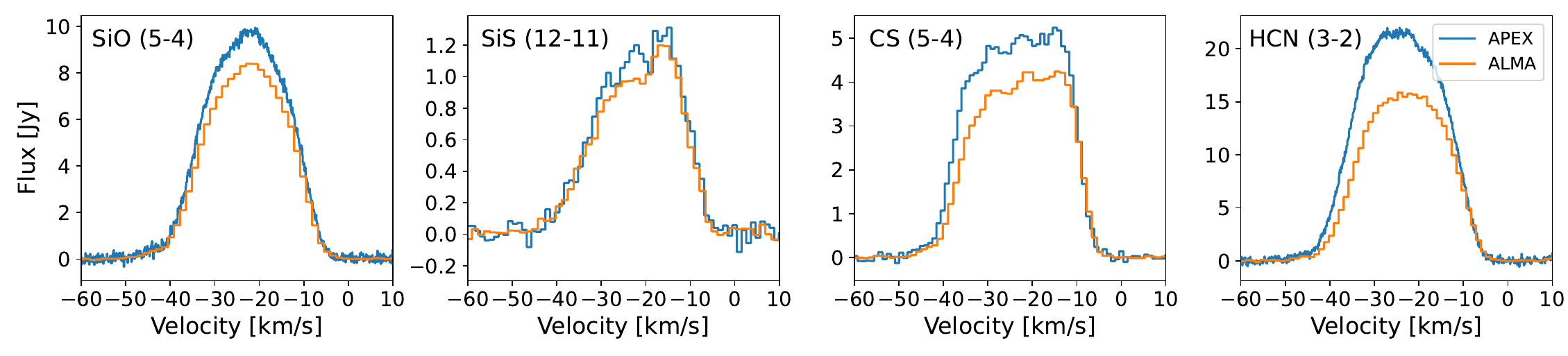}
\caption{Comparisons of ALMA and APEX data for SiO, SiS, CS and HCN, showing relatively low levels of resolved out flux (10--30\%) for the ALMA observations. All ALMA spectra were extracted from apertures with radii of $5.4\arcsec$.}
\label{otherlostflux}
\end{center}
\end{figure}

\begin{figure}[t]
\begin{center}
\includegraphics[width=0.49\textwidth]{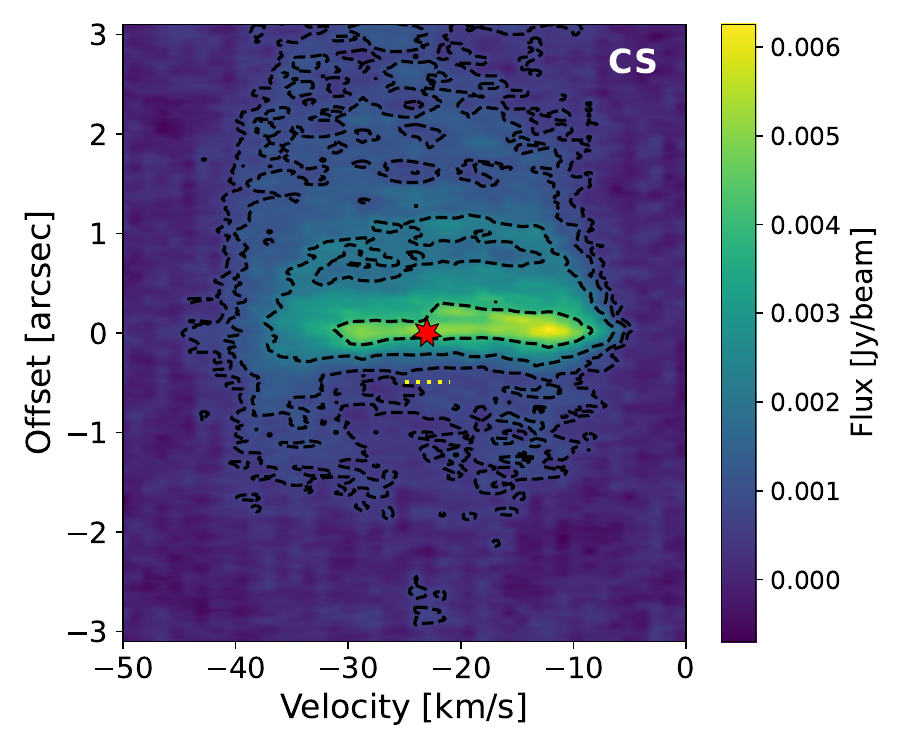}
\includegraphics[width=0.49\textwidth]{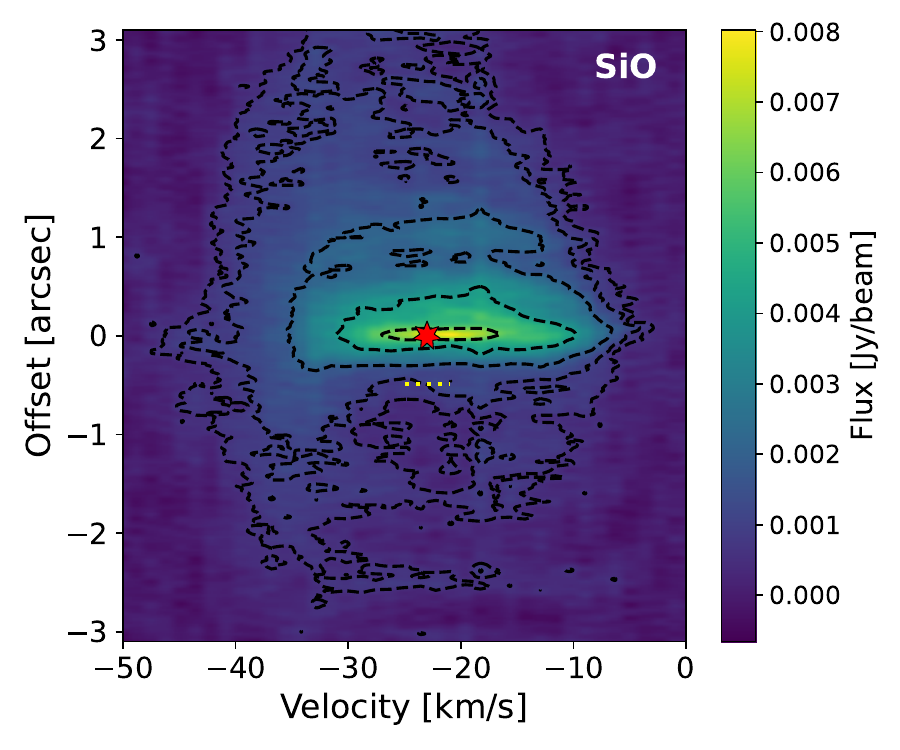}
\includegraphics[width=0.49\textwidth]{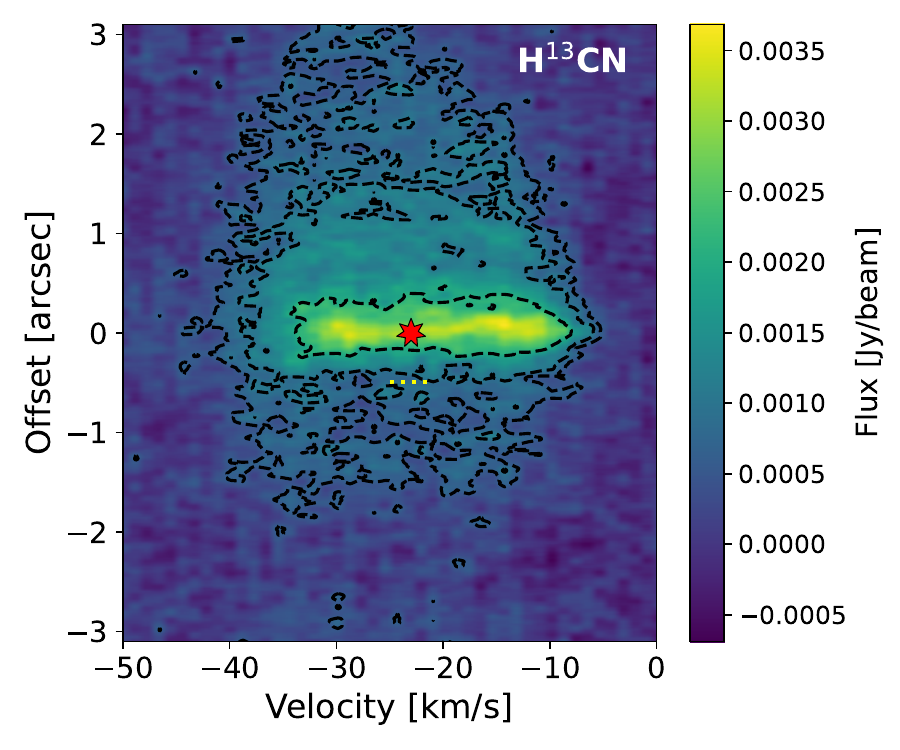}
\includegraphics[width=0.49\textwidth]{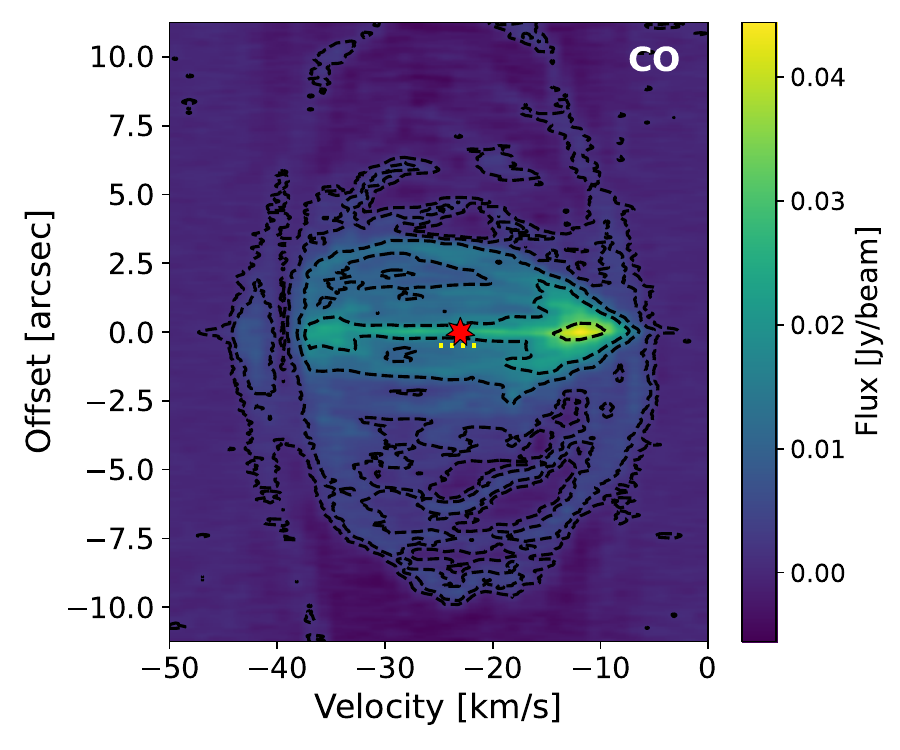}
\caption{Position-velocity diagrams of CS, SiO, H$^{13}$CN and CO, taken with the same slit that was used for SiN (Fig.~\ref{sin}). The black contours are at levels of 3, 5, 10, 20$\sigma$, except for CO, where they are at levels of 3, 10, 30, 50, 100$\sigma$. The position and LSR velocity of the AGB star are indicated by the red star and the horizontal dotted yellow line indicates the present offset of the F9 star. Note that the reduced emission at positive offsets for CO is the result of resolved out flux.
}
\label{CSPV}
\end{center}
\end{figure}

\begin{figure}[t]
\begin{center}
\includegraphics[width=\textwidth]{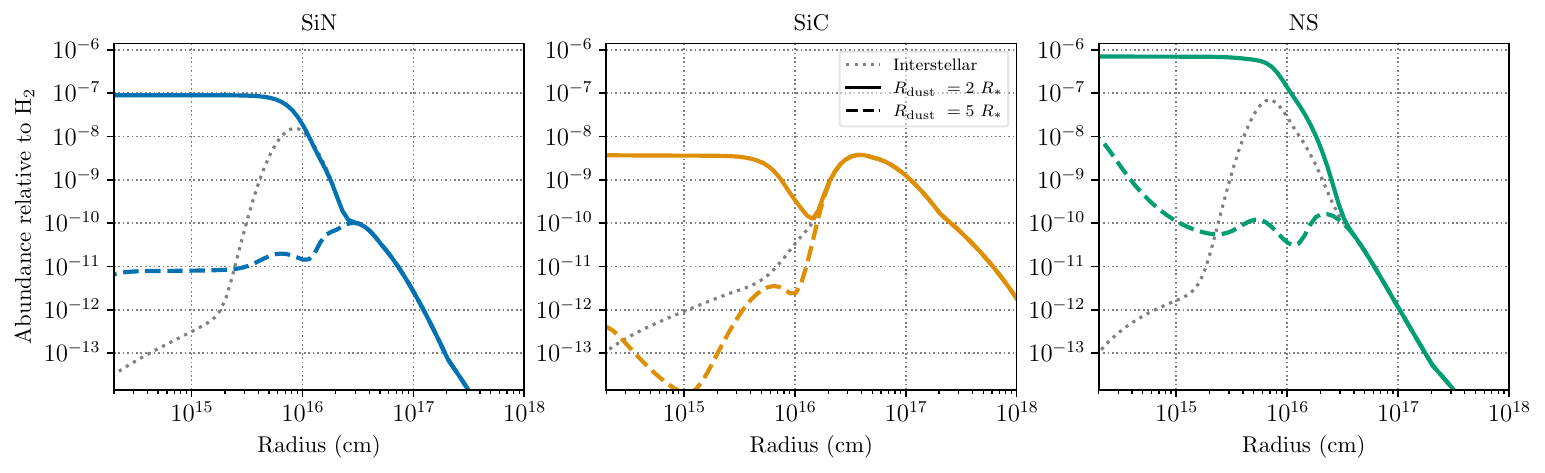}
\includegraphics[width=\textwidth]{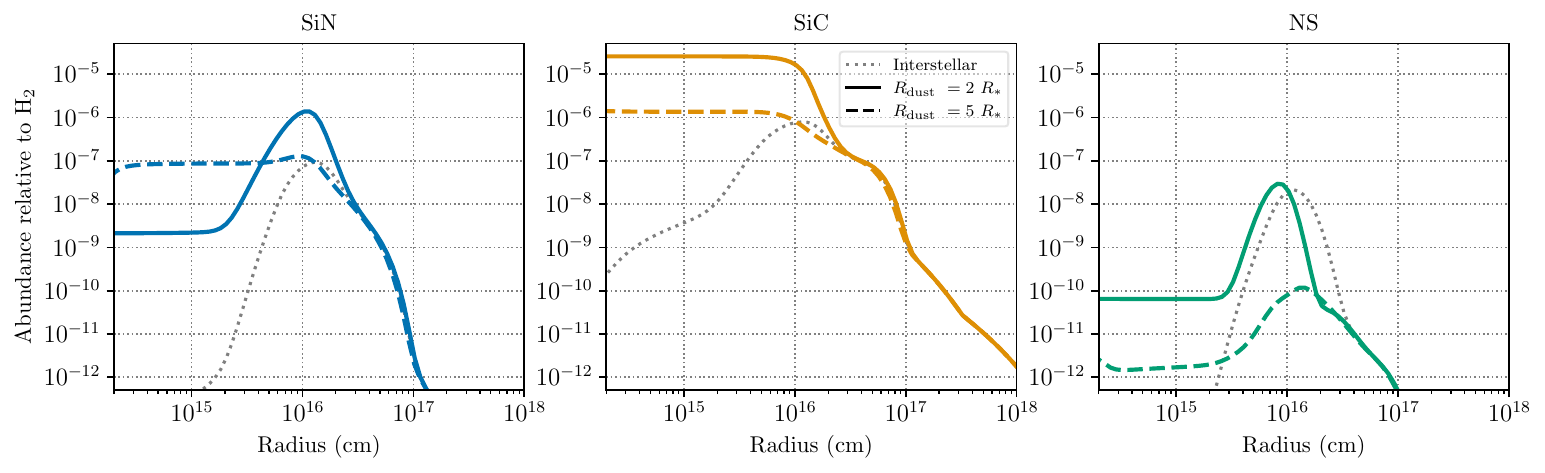}
\caption{\edits{Predicted abundances of SiN (\textit{left}), SiC (\textit{middle}) and NS (\textit{right}) based on chemical models for a stellar wind with a similar density as W~Aql. Plots show the predicted abundances in the absence of a companion (\textit{grey dotted lines}), and for when the effects of the companion are felt from $5R_\star$ (\textit{dashed coloured lines}) and $2R_\star$ (\textit{solid coloured lines}). Predictions for an oxygen-rich outflow are shown in the \textit{top} row and for a carbon-rich outflow in the \textit{bottom} row. Plotted models assume a 6000 K companion and are for a clumpy (two-component) outflow with full details given in \cite{Van-de-Sande2022}.}}
\label{fig:chemsin}
\end{center}
\end{figure}

\begin{figure}[t]
\begin{center}
\includegraphics[width=0.7\textwidth]{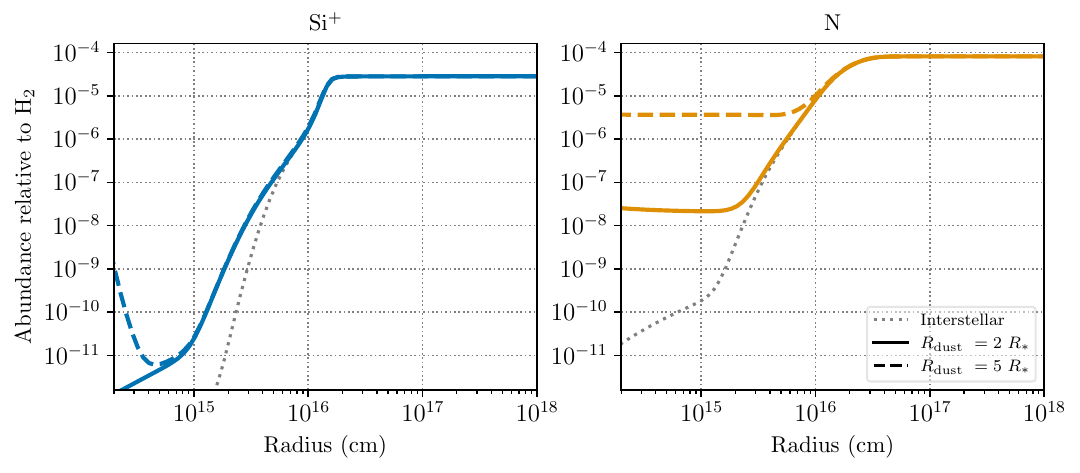}
\includegraphics[width=0.7\textwidth]{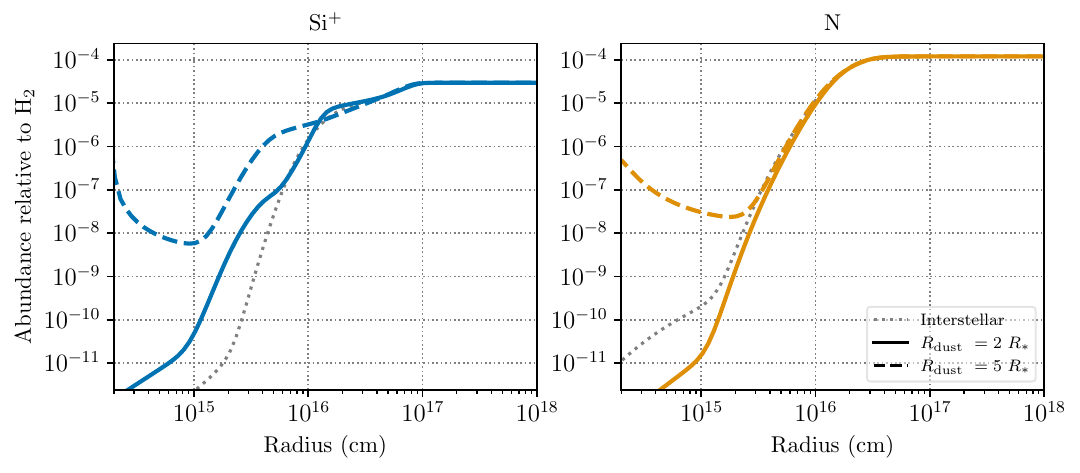}
\caption{\edits{Predicted abundances of \ce{Si+} (\textit{left}) and N (\textit{right}) based on chemical models for a stellar wind with a similar density as W~Aql. Plots show the predicted abundances in the absence of a companion (\textit{grey dotted lines}), and for when the effects of the companion are felt from $5R_\star$ (\textit{dashed coloured lines}) and $2R_\star$ (\textit{solid coloured lines}). Predictions for an oxygen-rich outflow are shown in the \textit{top} row and for a carbon-rich outflow in the \textit{bottom} row. Plotted models assume a 6000 K companion and are for a clumpy (two-component) outflow with full details given in \cite{Van-de-Sande2022}.}}
\label{fig:chemnsi}
\end{center}
\end{figure}

\begin{figure}[t]
\begin{center}
\includegraphics[width=0.49\textwidth]{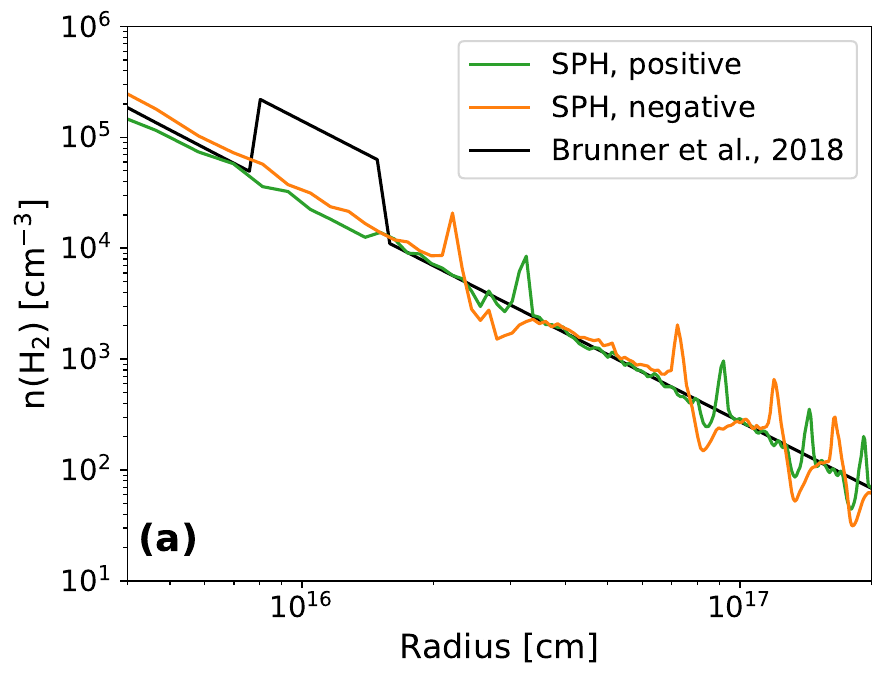}
\includegraphics[width=0.49\textwidth]{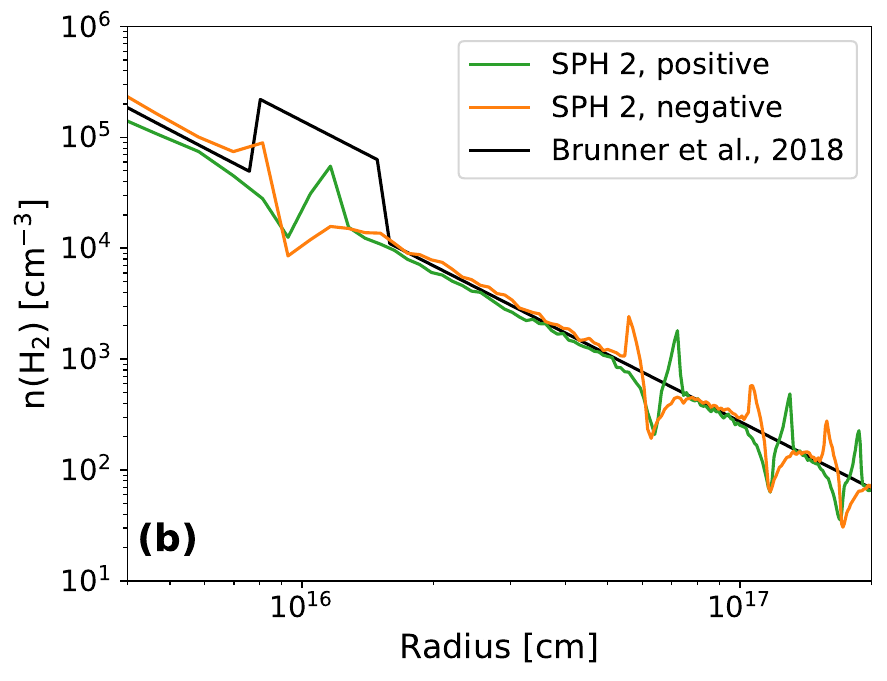}
\caption{\edits{Plots of number density in the hydrodynamic models along the $x$-axis (with $y=z=0$), showing number densities with increasing distance from the AGB star in both the positive (green) and negative (orange) $x$ directions (see Figures \ref{coarcs}c and \ref{faceonSPH}a for the definition of the axes), and compared with the spherically symmetric model of Brunner et al.~\cite{Brunner2018} (black). The innermost regions are excluded owing to limitations in the resolutions of our models. \textbf{(a)} The number density for our main hydrodynamic model (see Methods \ref{hydro} for details); \textbf{(b)} As for (a) but plotted for a second model with the orbital parameters derived in Methods \ref{sec:solution} and neglecting the more complex structures formed in the companion's wake. For this model, the location of the first higher-density circle agrees well with the location of the overdensity found from low-resolution ALMA observations \cite{Brunner2018}.}}
\label{fig:nh2}
\end{center}
\end{figure}

\begin{sidewaysfigure}[t]
\centering
\includegraphics[height=0.24\textheight]{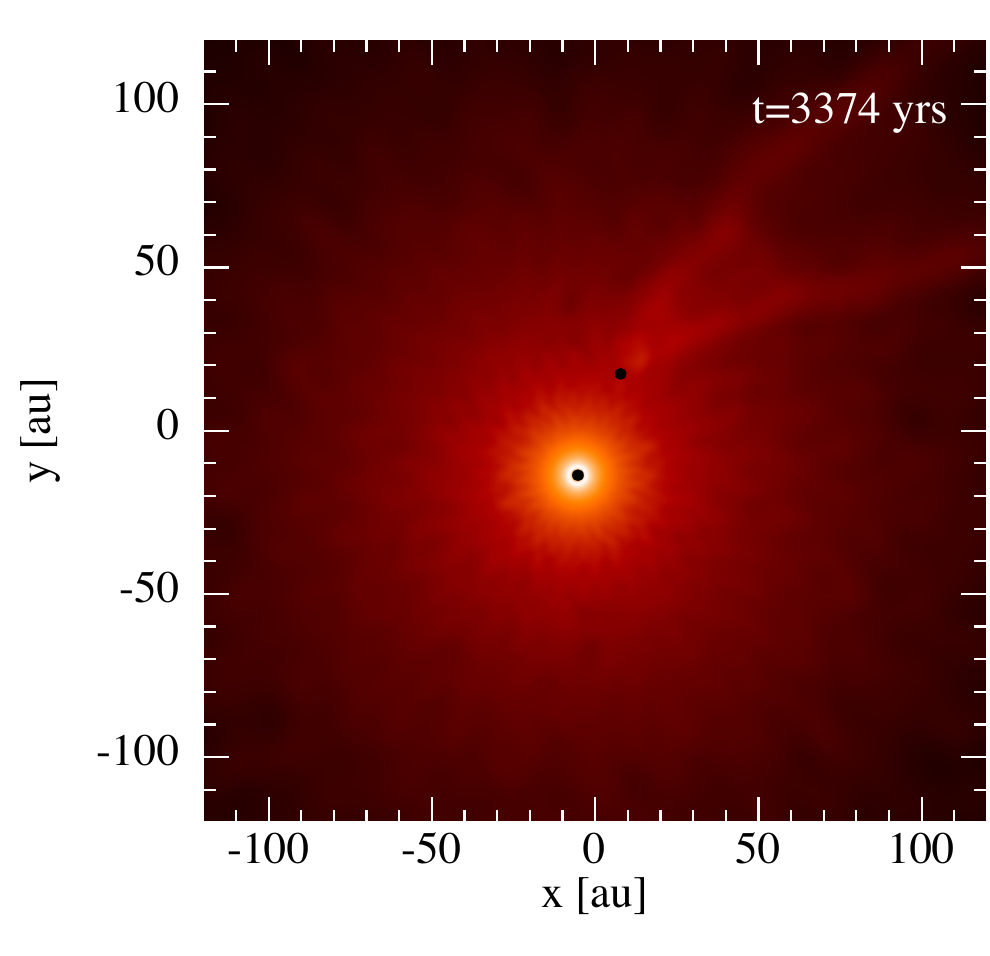}
\includegraphics[height=0.24\textheight]{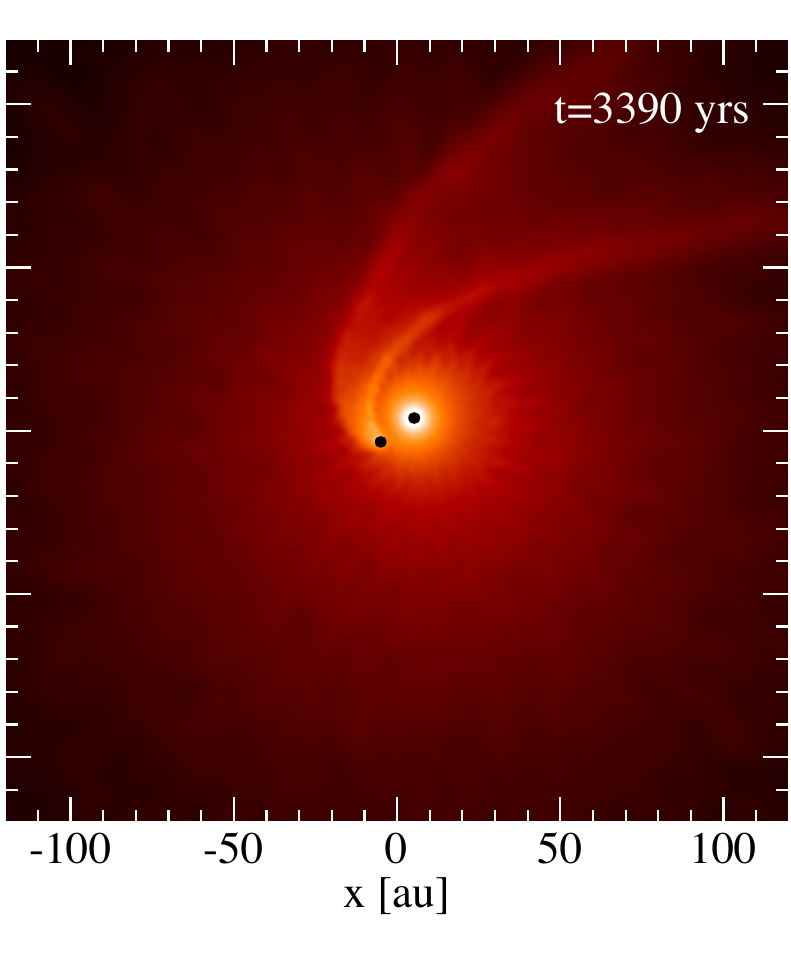}
\includegraphics[height=0.24\textheight]{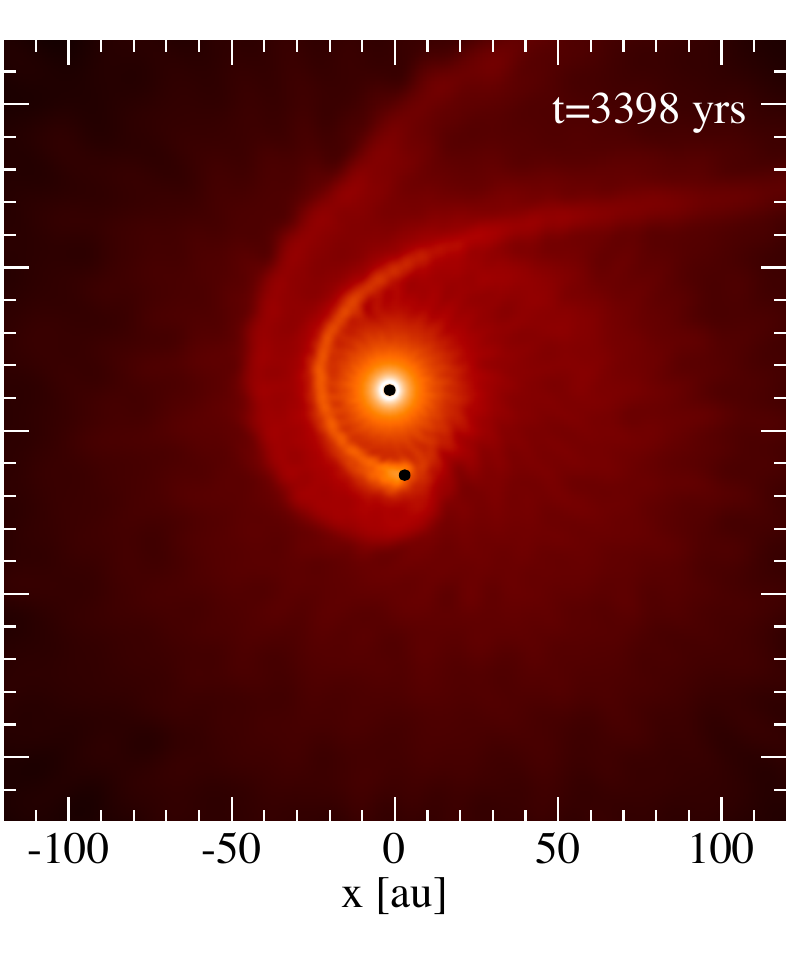}
\includegraphics[height=0.24\textheight]{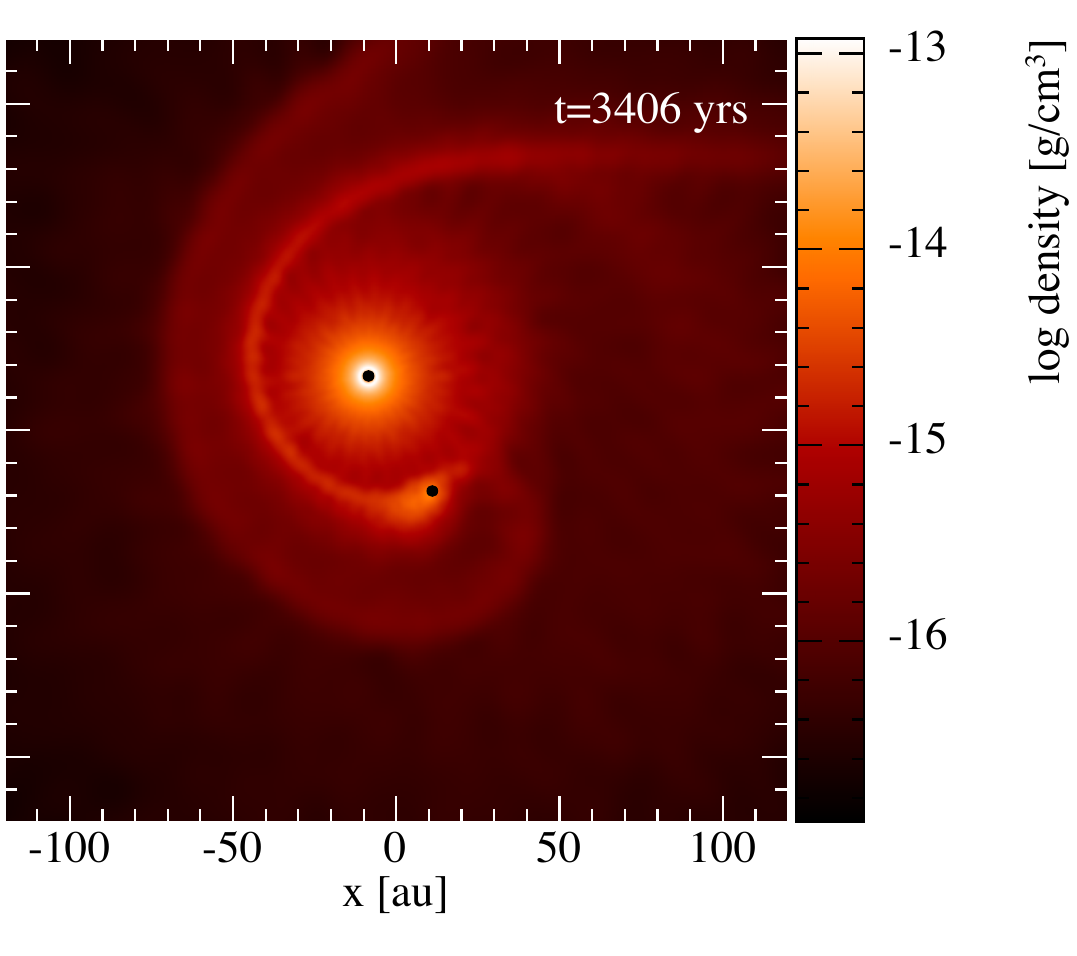}
\includegraphics[height=0.24\textheight]{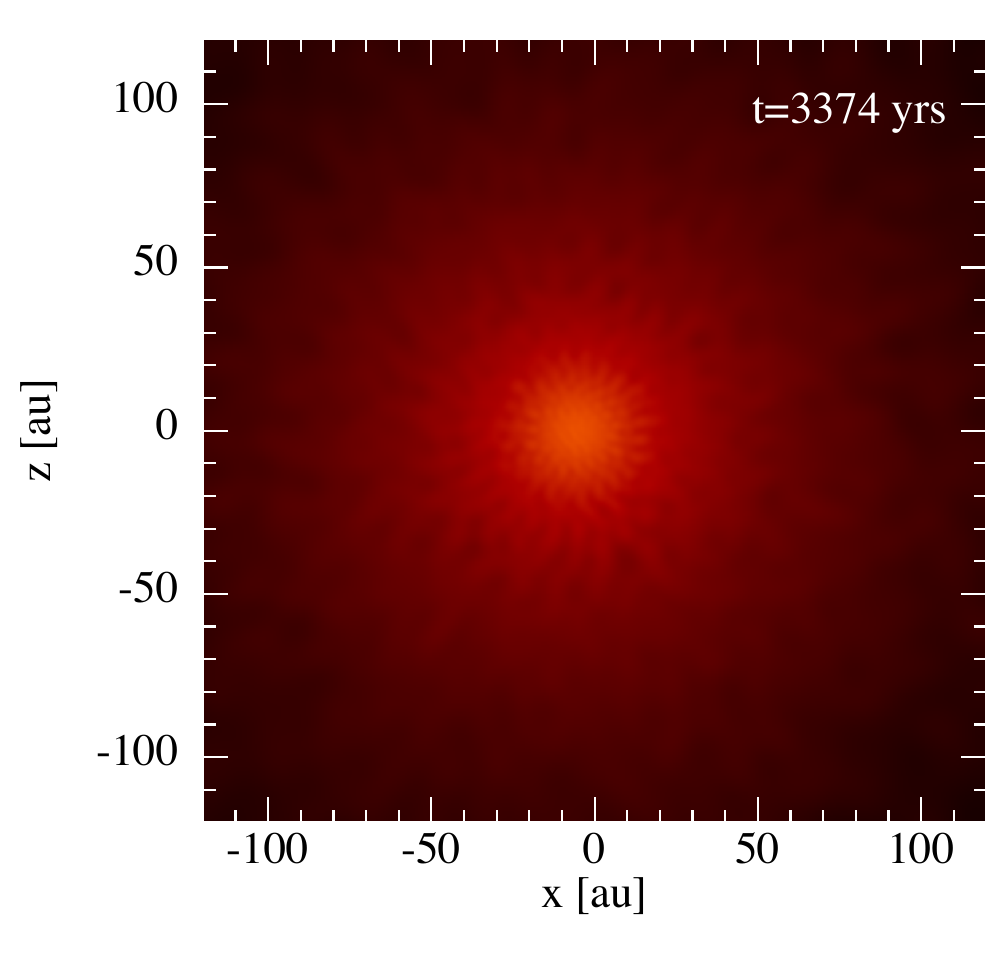}
\includegraphics[height=0.24\textheight]{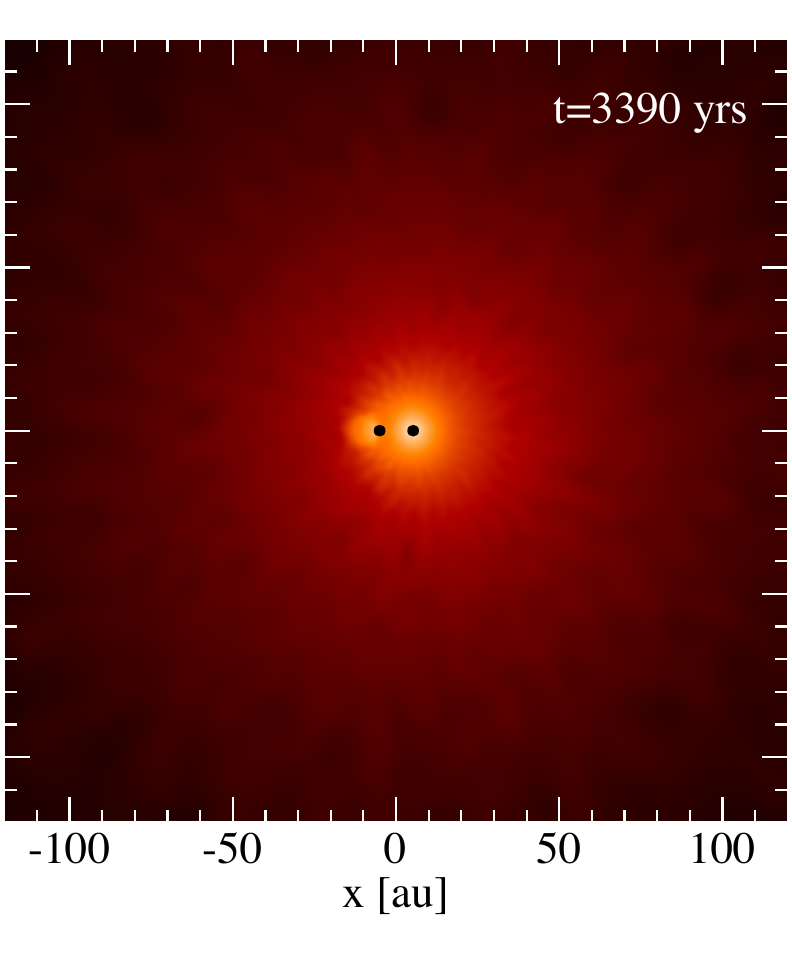}
\includegraphics[height=0.24\textheight]{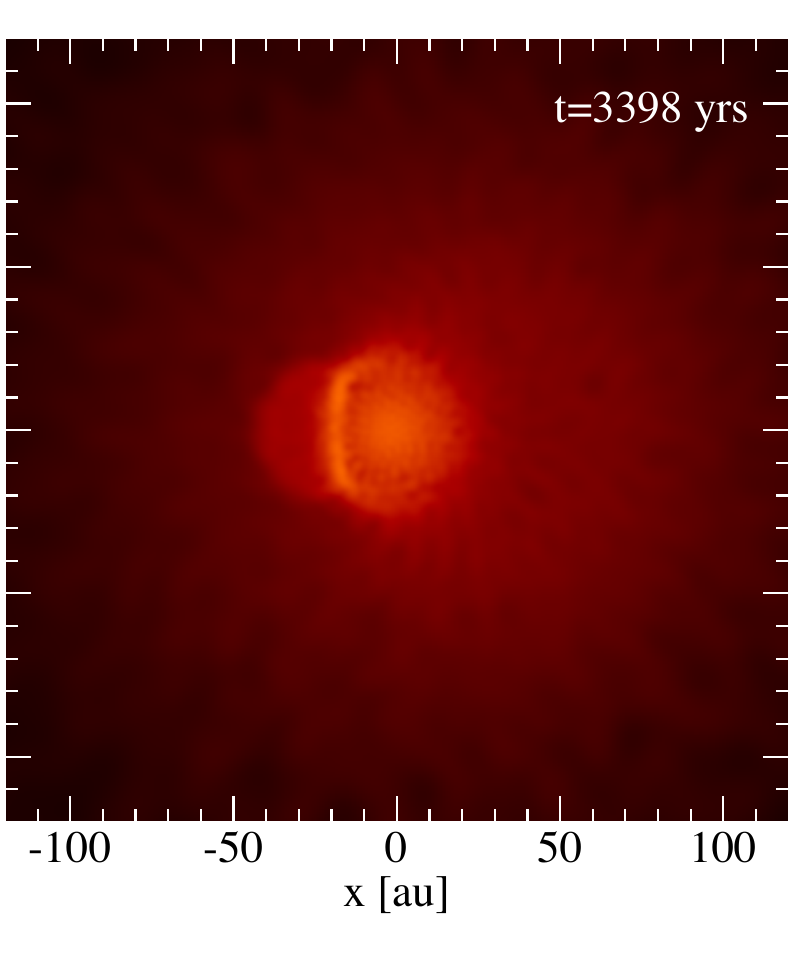}
\includegraphics[height=0.24\textheight]{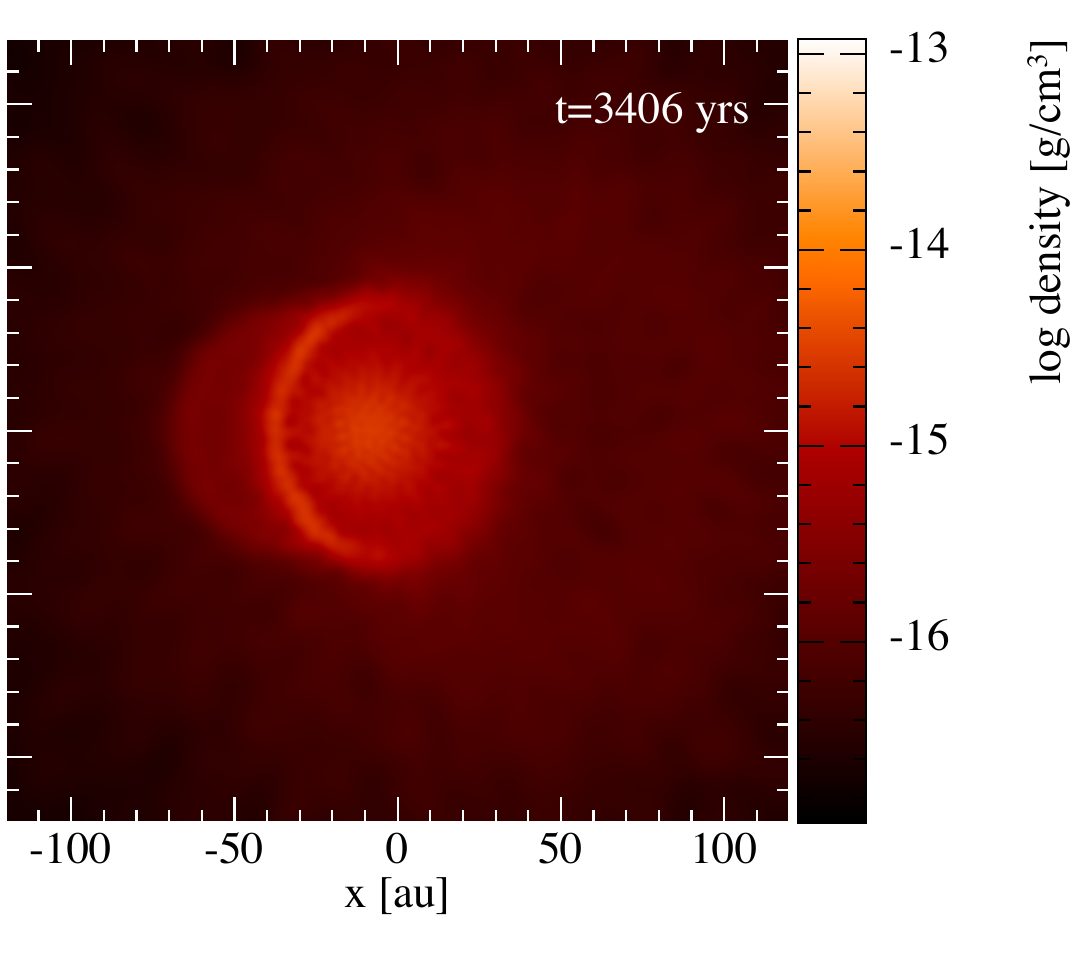}
\caption{\edits{Plots of the density in the inner regions of the hydrodynamic model at snapshots taken before, during and after the periastron passage. The time since the start of the simulation is show in the top right hand corner of each plot. The top row of plots show a slice through the orbital plane and the bottom row shows the same time steps but for a slice perpendicular to the orbital plane.}}
\label{fig:peri}
\end{sidewaysfigure}

\begin{figure}[t]
\begin{center}
\includegraphics[width=\textwidth]{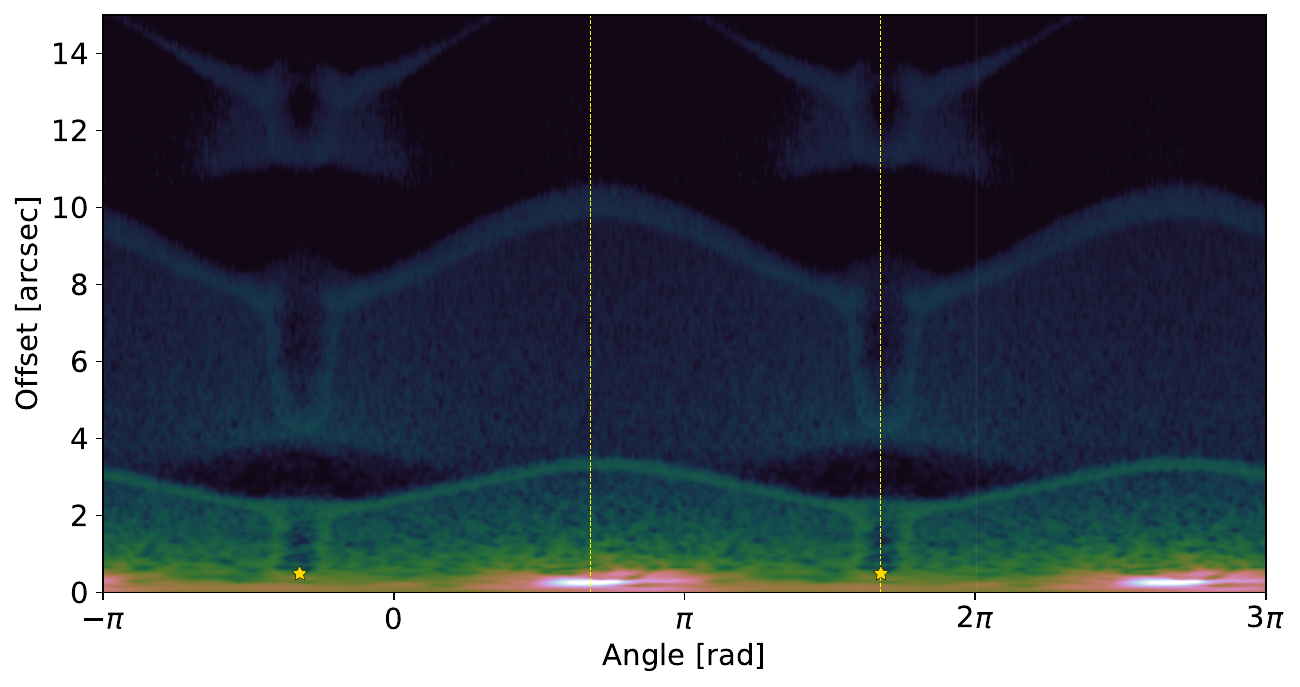}
\caption{Plot of the modelled radial emission distribution against angle for the central channel of CO generated from the MCFOST radiative transfer output of our Phantom model. One full revolution is \editstwo{shown in the centre (0 to $2\pi$) and half a revolution is shown on either side ($-\pi$ to 0 and $2\pi$ to $3\pi$) to show how the structures extend onwards, and} to match the equivalent plot constructed for the ALMA observations in Figs.~\ref{angrad} \editstwo{and \ref{angrad-no-lines}}. The location of the F9 star in the model is indicated by the yellow star and the yellow dotted line passes through both stars and is plotted in the central winding to guide the eye. Similar sinusoidal features are seen to those in the ALMA observations.}
\label{mcfostangrad}
\end{center}
\end{figure}

\begin{figure}[t]
\begin{center}
\includegraphics[width=\textwidth]{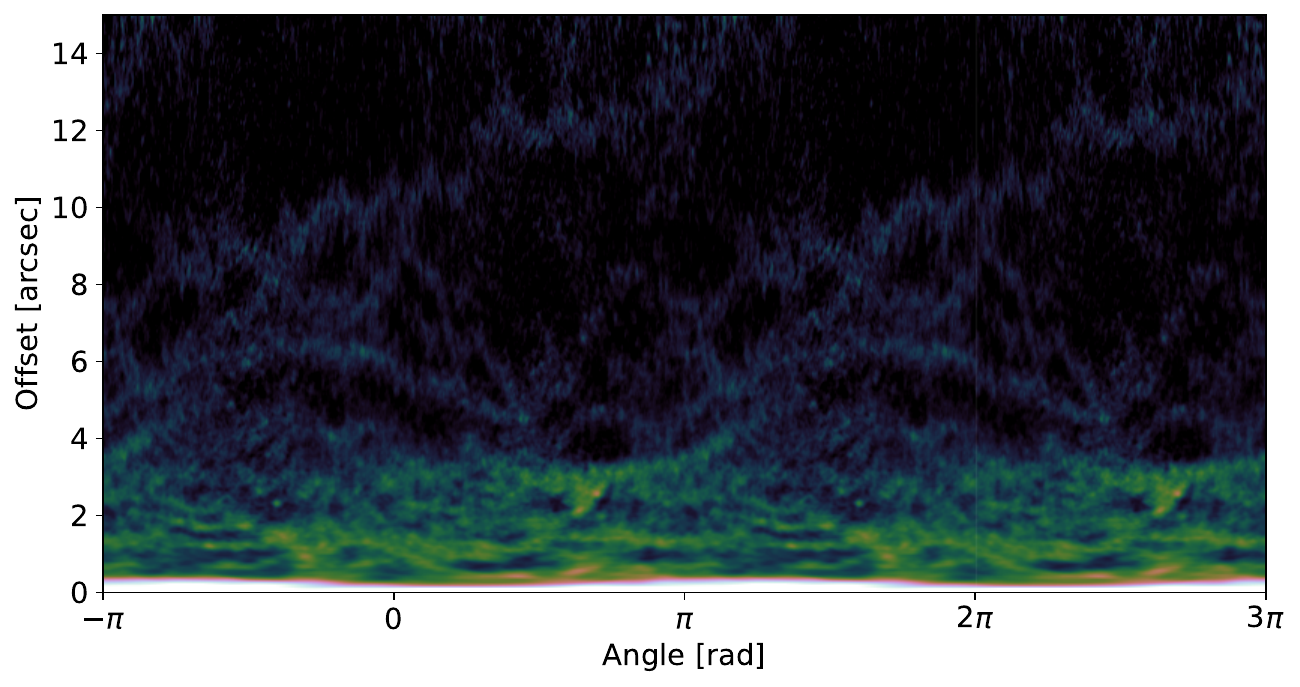}
\includegraphics[width=\textwidth]{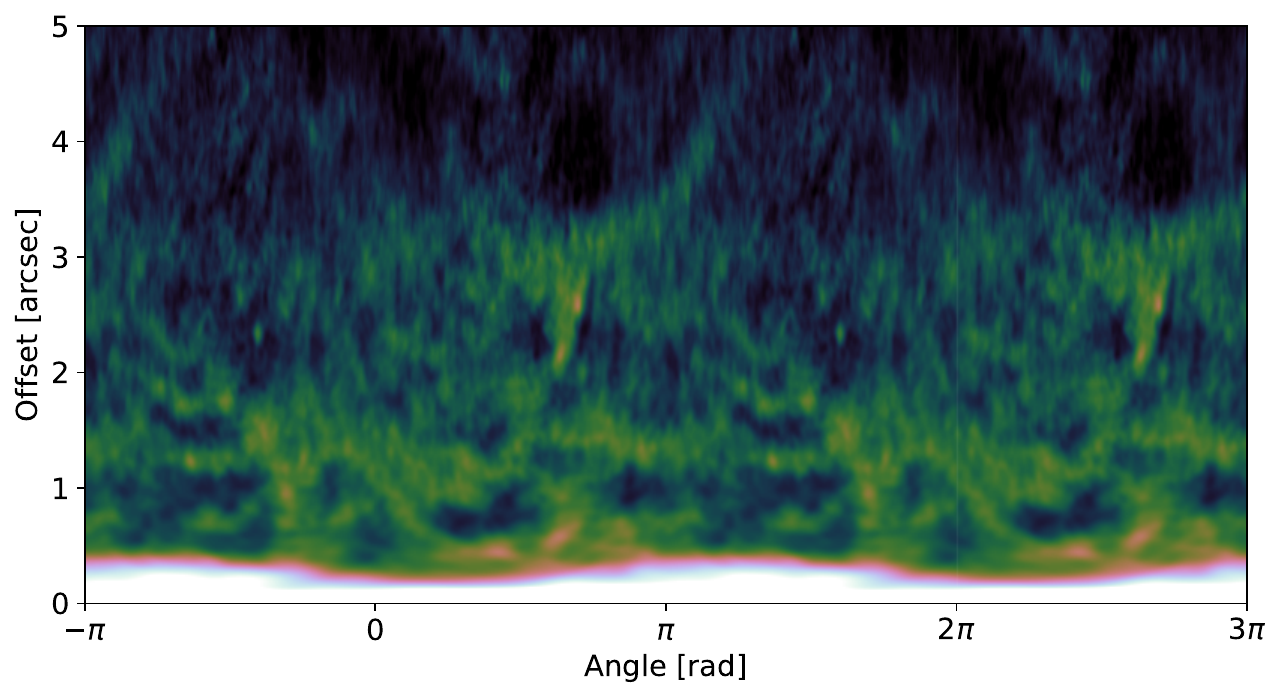}
\caption{\editstwo{As for Fig.~\ref{angrad} but without the additional annotations to highlight structures.}
}
\label{angrad-no-lines}
\end{center}
\end{figure}

\begin{figure}[t]
\begin{center}
\includegraphics[width=0.49\textwidth]{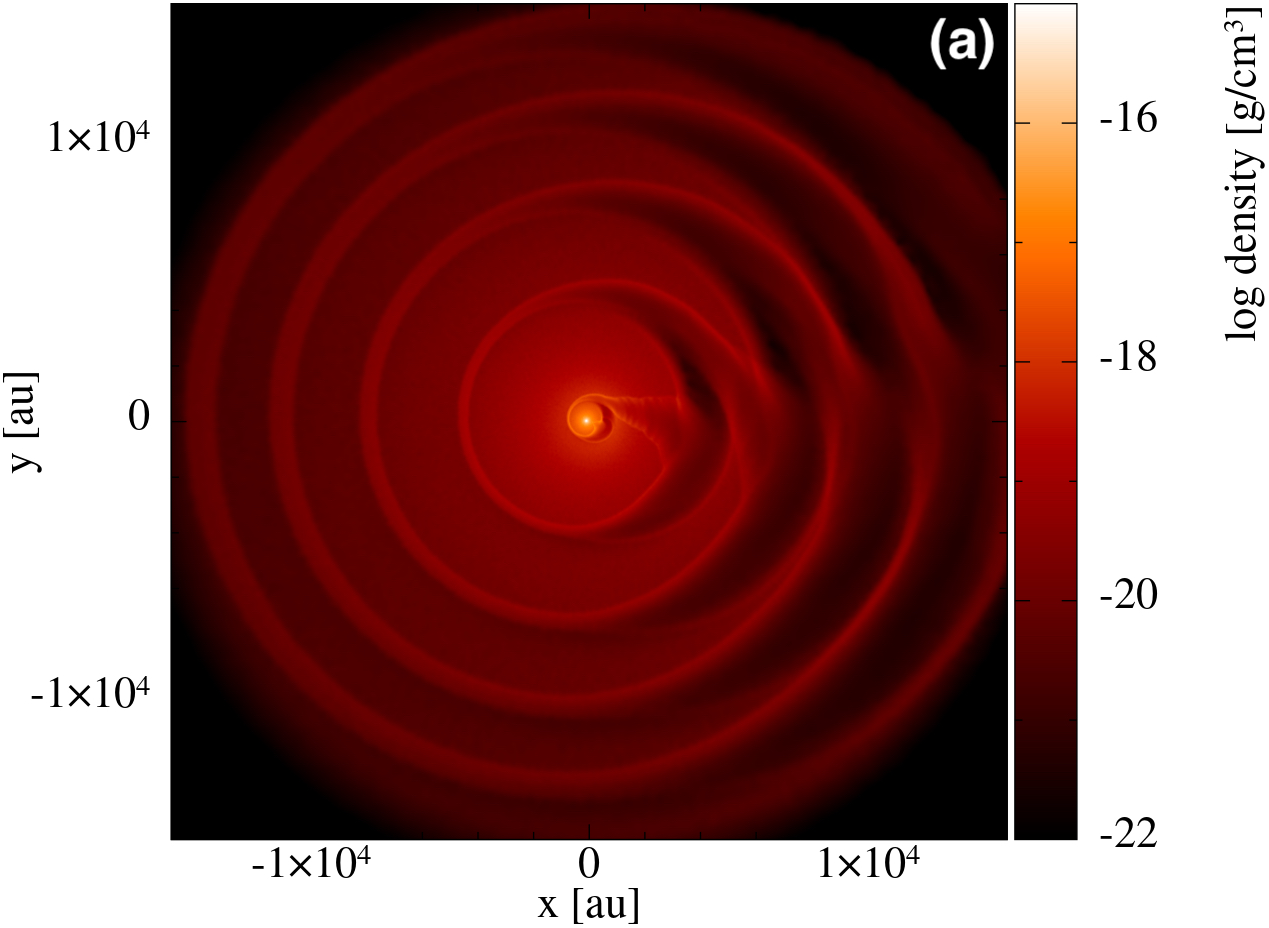}
\includegraphics[width=0.49\textwidth]{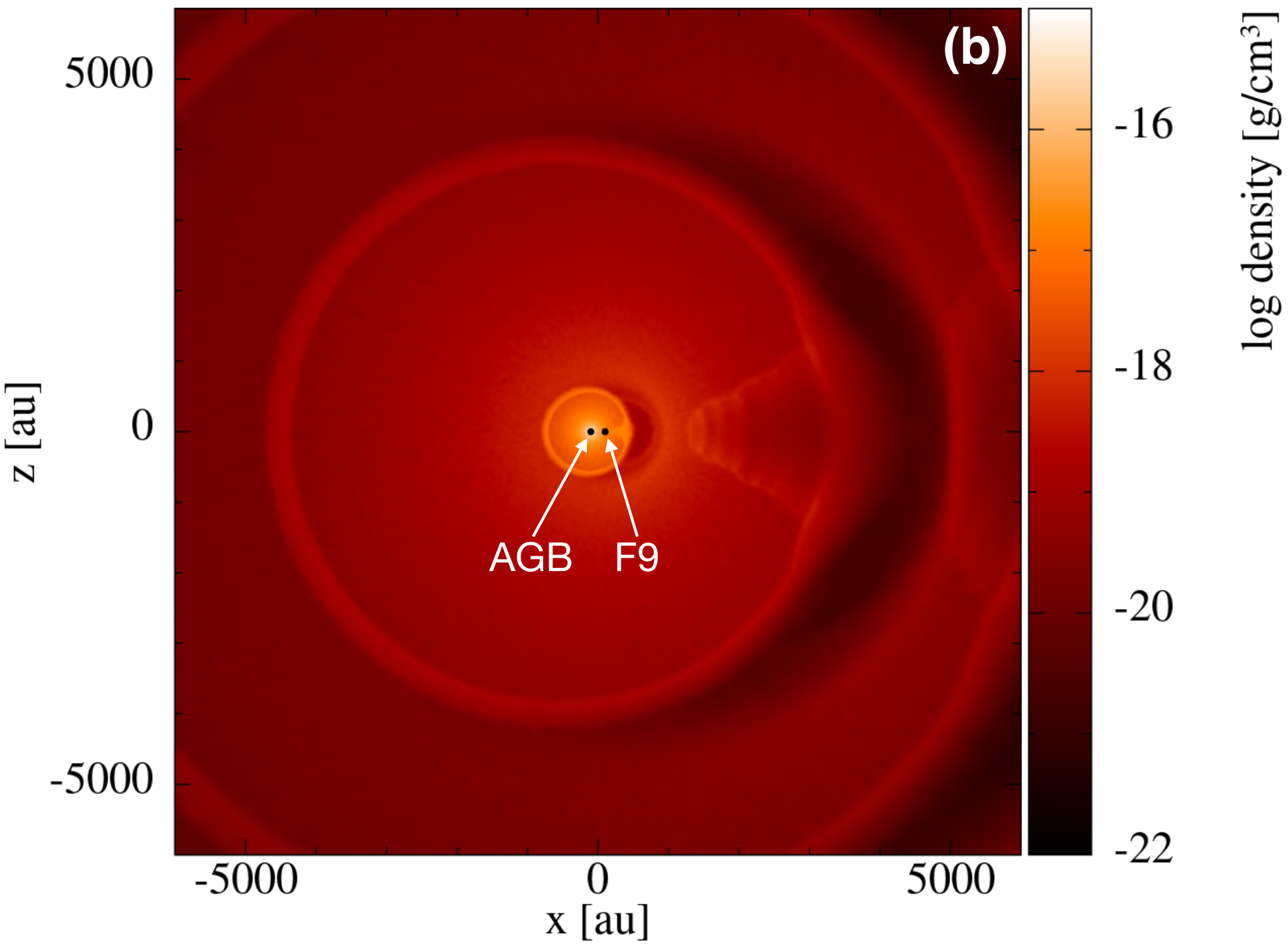}
\includegraphics[width=0.49\textwidth]{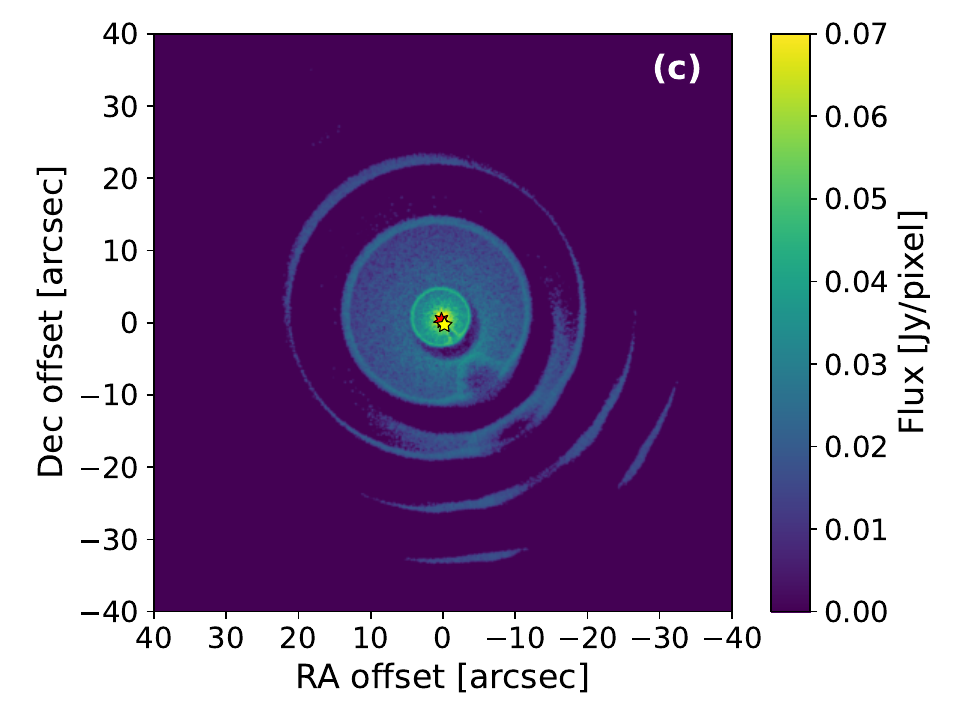}
\caption{
\textbf{(a)} Density distribution in a 2D slice through the orbital plane ($z=0$) from a 3D SPH model with masses $M_\mathrm{AGB}=1.6~\msol$ and $M_\mathrm{2}=1.06~\msol$, eccentricity $e=0.92$, and semimajor axis $a=125$~au; i.e. a face-on view of Fig. \ref{coarcs}c. See Methods \ref{hydro} for more details. \textbf{(b)} The central part of a slice perpendicular to the orbital plane of the same model, i.e. the central part of Fig. \ref{coarcs}c with the stars labelled and the $x$ and $y$ axes chosen to match the scale of Fig. \ref{coarcs}a.
\textbf{(c)} The full emission distribution predicted by MCFOST for the central channel at the LSR velocity, based on the SPH model, extending further than the field of view of our ALMA observations.
\label{faceonSPH}}
\end{center}
\end{figure}


\begin{figure}[t]
\begin{center}
\includegraphics[height=8cm]{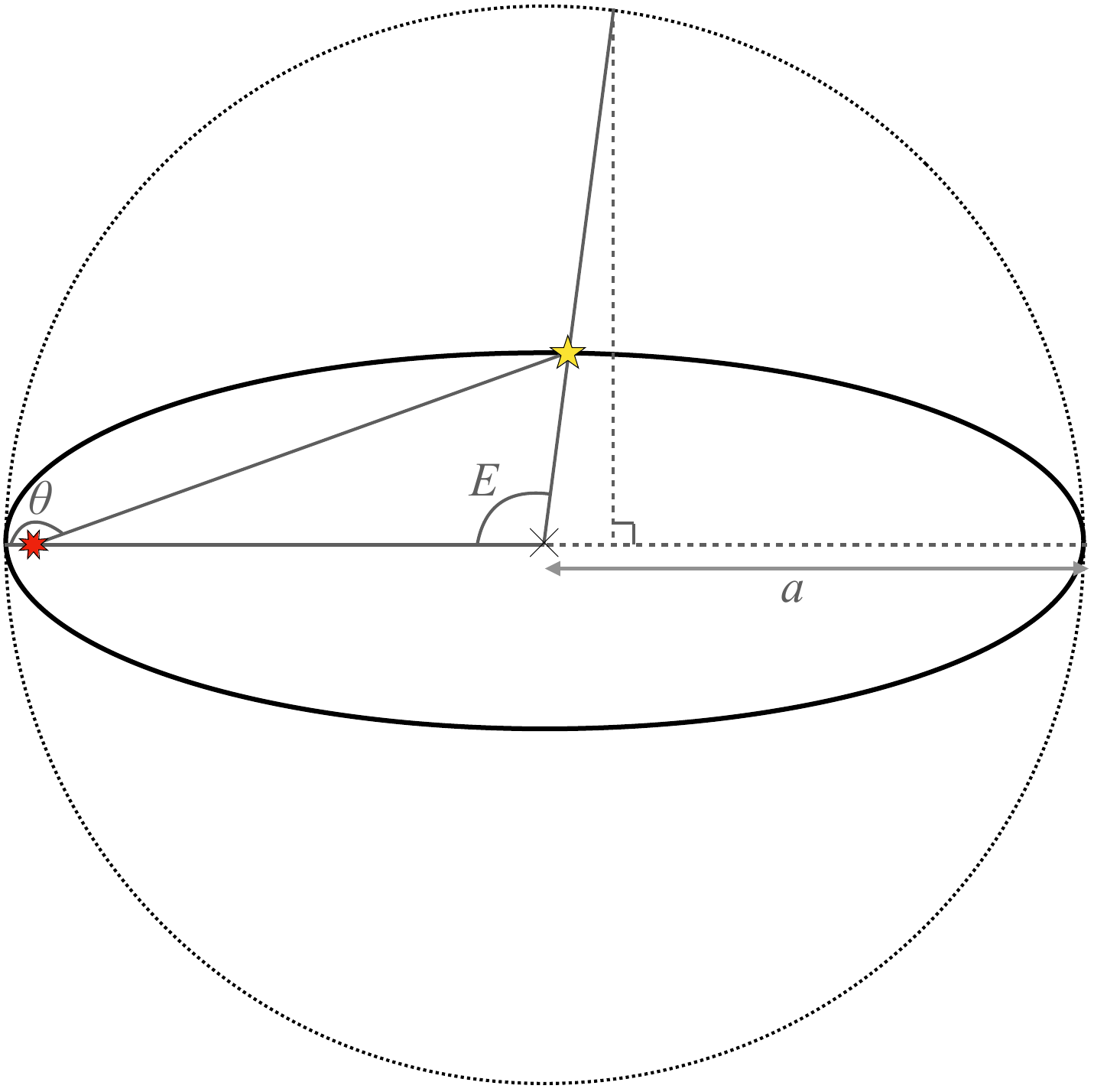}
\caption{A schematic view of the W~Aql system looking down onto the orbital plane in the frame of reference of the AGB star (red). The solid black ellipse shows a representative orbit of the F9 star (yellow) and the cross shows the centre of the orbital ellipse.  The semimajor axis, $a$, and the angles $\theta$ and $E$ are also shown, with the dotted outer circle having a radius equal to the semimajor axis.}
\label{schematic}
\end{center}
\end{figure}

\end{document}